\documentclass[english,final,numbers]{aipproc}
\layoutstyle{8x11single}

\usepackage{babel}
\usepackage{graphicx}
\usepackage{natbib}

\usepackage{amsmath}
\usepackage{amsfonts}
\usepackage{amssymb}
\usepackage{dcolumn}
\usepackage{ifpdf}

\newcolumntype{e}[1]{D{.}{.}{#1}}

\begin{document}

		\title{The Blackholic energy and the canonical Gamma-Ray Burst IV: the ``long'', `` genuine short'' and ``fake - disguised short'' GRBs\thanks{Part I, Part II and Part III of these Lecture notes have been published respectively in \textit{COSMOLOGY AND GRAVITATION: X$^{th}$ Brazilian School of Cosmology and Gravitation; 25$^{th}$ Anniversary (1977-2002)}, M. Novello, S.E. Perez Bergliaffa (eds.), \textit{AIP Conf. Proc.}, \textbf{668}, 16 (2003), see Ref. \citep{2003AIPC..668...16R}, in \textit{COSMOLOGY AND GRAVITATION: XI$^{th}$ Brazilian School of Cosmology and Gravitation}, M. Novello, S.E. Perez Bergliaffa (eds.), \textit{AIP Conf. Proc.}, \textbf{782}, 42 (2005), see Ref. \citep{2005AIPC..782...42R}, and in \textit{COSMOLOGY AND GRAVITATION: XII$^{th}$ Brazilian School of Cosmology and Gravitation}, M. Novello, S.E. Perez Bergliaffa (eds.), \textit{AIP Conf. Proc.}, \textbf{910}, 55 (2007), see Ref. \citep{2007AIPC..910...55R}.}}

\author{Remo Ruffini}{address={ICRANet, Piazzale della Repubblica 10, 65122 Pescara, Italy.}, altaddress={Dip. di Fisica and ICRA, Universit\`a di Roma ``La Sapienza'', Piazzale Aldo Moro 5, 00185 Roma, Italy.}}

\author{Alexey G. Aksenov}{address={Institute for Theoretical and Experimental Physics, B. Cheremushkinskaya, 25, 117218 Moscow, Russia}}

\author{Maria Grazia Bernardini}{address={Dip. di Fisica and ICRA, Universit\`a di Roma ``La Sapienza'', Piazzale Aldo Moro 5, 00185 Roma, Italy.}, altaddress={ICRANet, Piazzale della Repubblica 10, 65122 Pescara, Italy.}}

\author{Carlo Luciano Bianco}{address={ICRANet, Piazzale della Repubblica 10, 65122 Pescara, Italy.}, altaddress={Dip. di Fisica and ICRA, Universit\`a di Roma ``La Sapienza'', Piazzale Aldo Moro 5, 00185 Roma, Italy.}}

\author{Letizia Caito}{address={Dip. di Fisica and ICRA, Universit\`a di Roma ``La Sapienza'', Piazzale Aldo Moro 5, 00185 Roma, Italy.}, altaddress={ICRANet, Piazzale della Repubblica 10, 65122 Pescara, Italy.}}

\author{Pascal Chardonnet}{address={Universit\'e de Savoie, LAPTH - LAPP, BP 110, F-74941 Annecy-le-Vieux Cedex, France.}, altaddress={ICRANet, Piazzale della Repubblica 10, 65122 Pescara, Italy.}}

\author{Maria Giovanna Dainotti}{address={Dip. di Fisica and ICRA, Universit\`a di Roma ``La Sapienza'', Piazzale Aldo Moro 5, 00185 Roma, Italy.}, altaddress={ICRANet, Piazzale della Repubblica 10, 65122 Pescara, Italy.}}

\author{Gustavo De Barros}{address={Dip. di Fisica and ICRA, Universit\`a di Roma ``La Sapienza'', Piazzale Aldo Moro 5, 00185 Roma, Italy.}, altaddress={ICRANet, Piazzale della Repubblica 10, 65122 Pescara, Italy.}}

\author{Roberto Guida}{address={Dip. di Fisica and ICRA, Universit\`a di Roma ``La Sapienza'', Piazzale Aldo Moro 5, 00185 Roma, Italy.}, altaddress={ICRANet, Piazzale della Repubblica 10, 65122 Pescara, Italy.}}

\author{Luca Izzo}{address={Dip. di Fisica and ICRA, Universit\`a di Roma ``La Sapienza'', Piazzale Aldo Moro 5, 00185 Roma, Italy.}, altaddress={ICRANet, Piazzale della Repubblica 10, 65122 Pescara, Italy.}}

\author{Barbara Patricelli}{address={Dip. di Fisica and ICRA, Universit\`a di Roma ``La Sapienza'', Piazzale Aldo Moro 5, 00185 Roma, Italy.}, altaddress={ICRANet, Piazzale della Repubblica 10, 65122 Pescara, Italy.}}

\author{Luis Juracy Rangel Lemos}{address={Dip. di Fisica and ICRA, Universit\`a di Roma ``La Sapienza'', Piazzale Aldo Moro 5, 00185 Roma, Italy.}, altaddress={ICRANet, Piazzale della Repubblica 10, 65122 Pescara, Italy.}}

\author{Michael Rotondo}{address={Dip. di Fisica and ICRA, Universit\`a di Roma ``La Sapienza'', Piazzale Aldo Moro 5, 00185 Roma, Italy.}, altaddress={ICRANet, Piazzale della Repubblica 10, 65122 Pescara, Italy.}}

\author{Jorge Armando Rueda Hernandez}{address={Dip. di Fisica and ICRA, Universit\`a di Roma ``La Sapienza'', Piazzale Aldo Moro 5, 00185 Roma, Italy.}, altaddress={ICRANet, Piazzale della Repubblica 10, 65122 Pescara, Italy.}}

\author{Gregory Vereshchagin}{address={ICRANet, Piazzale della Repubblica 10, 65122 Pescara, Italy.}}

\author{She-Sheng Xue}{address={ICRANet, Piazzale della Repubblica 10, 65122 Pescara, Italy.}}

\begin{abstract}
We report some recent developments in the understanding of GRBs based on the theoretical framework of the ``fireshell'' model, already presented in the last three editions of the ``Brazilian School of Cosmology and Gravitation''. After recalling the basic features of the ``fireshell model'', we emphasize the following novel results: 1) the interpretation of the X-ray flares in GRB afterglows as due to the interaction of the optically thin fireshell with isolated clouds in the CircumBurst Medium (CBM); 2) an interpretation as ``fake - disguised'' short GRBs of the GRBs belonging to the class identified by Norris \& Bonnell; we present two prototypes, GRB 970228 and GRB 060614; both these cases are consistent with an origin from the final coalescence of a binary system in the halo of their host galaxies with particularly low CBM density $n_{cbm} \sim 10^{-3}$ particles/cm$^3$; 3) the first attempt to study a genuine short GRB with the analysis of GRB 050509B, that reveals indeed still an open question; 4) the interpretation of the GRB-SN association in the case of GRB 060218 via the ``induced gravitational collapse'' process; 5) a first attempt to understand the nature of the ``Amati relation'', a phenomenological correlation between the isotropic-equivalent radiated energy of the prompt emission $E_{iso}$ with the cosmological rest-frame $\nu F_{\nu}$ spectrum peak energy $E_{p,i}$. In addition, recent progress on the thermalization of the electron-positron plasma close to their formation phase, as well as the structure of the electrodynamics of Kerr-Newman Black Holes are presented. An outlook for possible explanation of high-energy phenomena in GRBs to be expected from the AGILE and the Fermi satellites are discussed. As an example of high energy process, the work by Enrico Fermi dealing with ultrarelativistic collisions is examined. It is clear that all the GRB physics points to the existence of overcritical electrodynamical fields. In this sense we present some progresses on a unified approach to heavy nuclei and neutron stars cores, which leads to the existence of overcritical fields under the neutron star crust.
\end{abstract}



	\maketitle
	
	\section{Introduction}

Gamma-Ray Bursts (GRBs) represent very likely ``the'' most extensive computational, theoretical and observational effort ever carried out successfully in physics and astrophysics. The extensive campaign of observation from space based X-ray and $\gamma$-ray observatory, such as the \emph{Vela}, CGRO, BeppoSAX, HETE-II, INTEGRAL, \emph{Swift}, Agile, GLAST, R-XTE, \emph{Chandra}, XMM satellites, have been matched by complementary observations in the radio wavelength (e.g. by the VLA) and in the optical band (e.g. by VLT, Keck, REM). The very fortunate situation occurs that these data can be confronted with a mature theoretical development.

We outline how this progress leads to the confirmation of three interpretation paradigms for GRBs we proposed in July 2001 \citep{2001ApJ...555L.107R,2001ApJ...555L.113R,2001ApJ...555L.117R}. The outcome of this analysis points to the existence of a ``canonical'' GRB, originating from a variety of different initial astrophysical scenarios. The communality of these GRBs appears to be that they all are emitted in the process of formation of a black hole with a negligible value of its angular momentum. The following sequence of events appears to be canonical: the gravitational collapse to a black hole, the vacuum polarization process in the dyadosphere with the creation of the optically thick self accelerating electron-positron plasma; the engulfment of baryonic mass during the plasma expansion; adiabatic expansion of the optically thick ``fireshell'' of electron-positron-baryon plasma up to the transparency; the interaction of the accelerated baryonic matter with the CircumBurst Medium (CBM). This leads to the canonical GRB composed of a proper GRB (P-GRB), emitted at the moment of transparency, followed by an extended afterglow. The sole parameters in this scenario are the total energy of the dyadosphere $E_{dya}$, the fireshell baryon loading $M_B$ defined by the dimensionless parameter $B\equiv M_Bc^2/E_{dya}$, and the CBM filamentary distribution around the source. In the limit $B\to 0$ the total energy is radiated in the P-GRB with a vanishing contribution in the extended afterglow. We refer globally to this model as the ``fireshell'' model.

The increase of observational details on many different GRBs, gained in the last two year observational campaign, is showing very clearly an evolution in the understanding of this basic scenario. This has allowed to explore GRBs with different baryon loading and very different CBM properties. A major new result has been obtained in the understanding of the origin of flares, which is traced back to the interaction of the fireshell with isolated CBM clouds (see below, section \emph{GRB 060607A: a complete analysis of the prompt emission and X-ray flares}).

Another key result, presented in these lectures, has been the understanding of a new GRB class, which was pioneered by \citet{2006ApJ...643..266N}, characterized by ``an occasional softer extended emission lasting tenths of seconds after an initial spikelike emission''. From the ``fireshell'' model it clearly follows that these sources explore a new range of parameters and mark a fundamental step in the understanding of the astrophysical nature of the GRB progenitor systems. The new class is shown to be characterized by GRBs occurring in a particularly low density CBM, $n_{cbm} \sim 10^{-3}$ particles/cm$^3$, typical of galactic halos. The progenitors must therefore necessarily be binary sources which have migrated from their birth location in a star forming region to a low density region within the galactic halo, where the final merging occurs. If such sources did explode in a CBM with average density $n_{cbm} \sim 1$ particle/cm$^3$, they would look exactly like what is called long GRBs. In this sense, we decided to call them ``fake'' or ``disguised'' short GRBs (see below, section \emph{The Norris \& Bonnel kind of sources: the new class of ``fake - disguised'' short GRBs}).

This leads, in turn, to a novel interpretation of the traditional GRB classification \citep{1992grbo.book..161K,1992AIPC..265..304D} in ``short'' GRBs (with a $T_{90}$ duration lasting less than $\sim 2$ s) and ``long'' ones (with a $T_{90}$ duration lasting more than $\sim 2$ s up to $\sim 1000$ s). The duration of long GRBs is shown to be just due to the instrumental noise threshold and not to represent any intrinsic GRB feature. On the other hand, there is an increasing evidence that the majority of the sources classified as short GRBs are actually ``fake - disguised'' short ones. No Supernova can possibly be related to these GRBs (see below, section \emph{The ``fireshell'' model and GRB progenitors}). However, some issues remains open (see below, section \emph{Open issues in current theoretical models}). Moreover, the quest for the first clear identification of a ``genuine'' short GRB is also still open (see below, section \emph{The search for a ``genuine short'' GRB: the case of GRB 050509B}).

The GRBs associated with SN do necessarily form a different class, of weakest and more numerous sources, originating also from binary systems, formed by a neutron star, close to its critical mass, and a companion star, evolved out of the main sequence, via the ``induced gravitational collapse'' process (see below, section \emph{GRBs and SNe: the induced gravitational collapse}).

The detection of GRBs up to very high redshifts \citep[up to $z = 6.7$, see Ref.][]{2008arXiv0810.2314G}, their high observed rate of one every few days, and the progress in the theoretical understanding of these sources all make them useful as cosmological tools, complementary to supernovae Ia, which are observed only up to $z = 1.7$ \citep{2001ARA&A..39...67L,2001ApJ...560...49R}. One of the hottest topics on GRBs is the possible existence of empirical relations between GRB observables \citep{2002A&A...390...81A,2004ApJ...616..331G,2004ApJ...609..935Y,2005ApJ...633..611L,2006MNRAS.370..185F,2008MNRAS.391..577A}, which may lead, if confirmed, to using GRBs as tracers of models of universe. The first empirical relation, discovered when analyzing the \emph{BeppoSAX} so-called ``long'' bursts with known redshift, was the ``Amati relation'' \citep{2002A&A...390...81A}. It was found that the isotropic-equivalent radiated energy of the prompt emission $E_{iso}$ is correlated with the cosmological rest-frame $\nu F_{\nu}$ spectrum peak energy $E_{p,i}$: $E_{p,i}\propto (E_{iso})^{a}$, with $a = 0.52 \pm 0.06$ \citep{2002A&A...390...81A}. The existence of the Amati relation has been confirmed by studying a sample of GRBs discovered by Swift, with $a = 0.49^{+0.06}_{-0.05}$ \citep{2006ApJ...636L..73S,2006MNRAS.372..233A}. We present a first attempt to understand the nature of such a phenomenological correlation (see below, section \emph{Theoretical background for GRBs' empirical correlations}).

All these theoretical and phenomenological approach has motivated significant progresses on:

a) The electron-positron plasma thermalization. We solved numerically relativistic Boltzmann equations with collisional integrals representing two-body as well as three-body interactions, in particular Compton, Moller and Bhabha scattering, pair creation and annihilation, relativistic bremsstrahlung, three photon creation and annihilation, double Compton scattering and the corresponding processes where protons participate. This allowed determination of characteristic timescales of thermalization as well as clarified the role of binary and triple interactions in reaching thermal equilibrium (see below, section \emph{Thermalization process of electron-positron plasma with baryon loading}).

b) The electrodynamics of neutron stars and black holes. We present a unified treatment of nuclear density cores recovering the classical results of neutral atoms with heavy nuclei with mass number $A\approx 10^2$-$10 ^6$ and extrapolating these results to massive nuclear density cores with $A\approx(m_{Planck}/m_n)^3 \sim 10^{57}$. The treatment is approached by solving the relativistic Thomas-Fermi equation describing a system of $N_n$ neutrons, $N_p$ protons and $N_e$ electrons in beta equilibrium. In order to show the stability of such cores under the competing effects of self gravity and Coulomb repulsion, we have started to take the gravitational field into duly account, and put the issue within the framework of general relativity. We expect that starting from such configuration, gravitational collapse would lead to Dyadosphere (electron-positron-photon plasma) in the Reissner-Nordstr\"m geometry, or Dyadotorus in the Kerr-Newmann geometry. We present in some details the analysis of the Dyadotorus (see below, section \emph{Critical electric Fields on the surface of massive cores and Dyadotorus of the Kerr-Newman Geometry}).

c) In view of the new data from the Fermi and AGILE satellites, an analysis of the GRB radiation over $1$ MeV. It is by now clear that our pure thermal emission previously considered, and which has been fundamental in expressing the average CBM density, is not appropriate to the description of this high-energy component. In parallel, we are currently examining how Fermi ideas \citep{1950PThPh...5..570F} have been further developed in large data analysis procedures at CERN and other accelerators all over the world (see below, section \emph{Selected processes originating high-energy emission}).

	\section{Brief reminder of the \textit{fireshell} model}

\begin{figure}
\centering
\includegraphics[width=0.35\hsize]{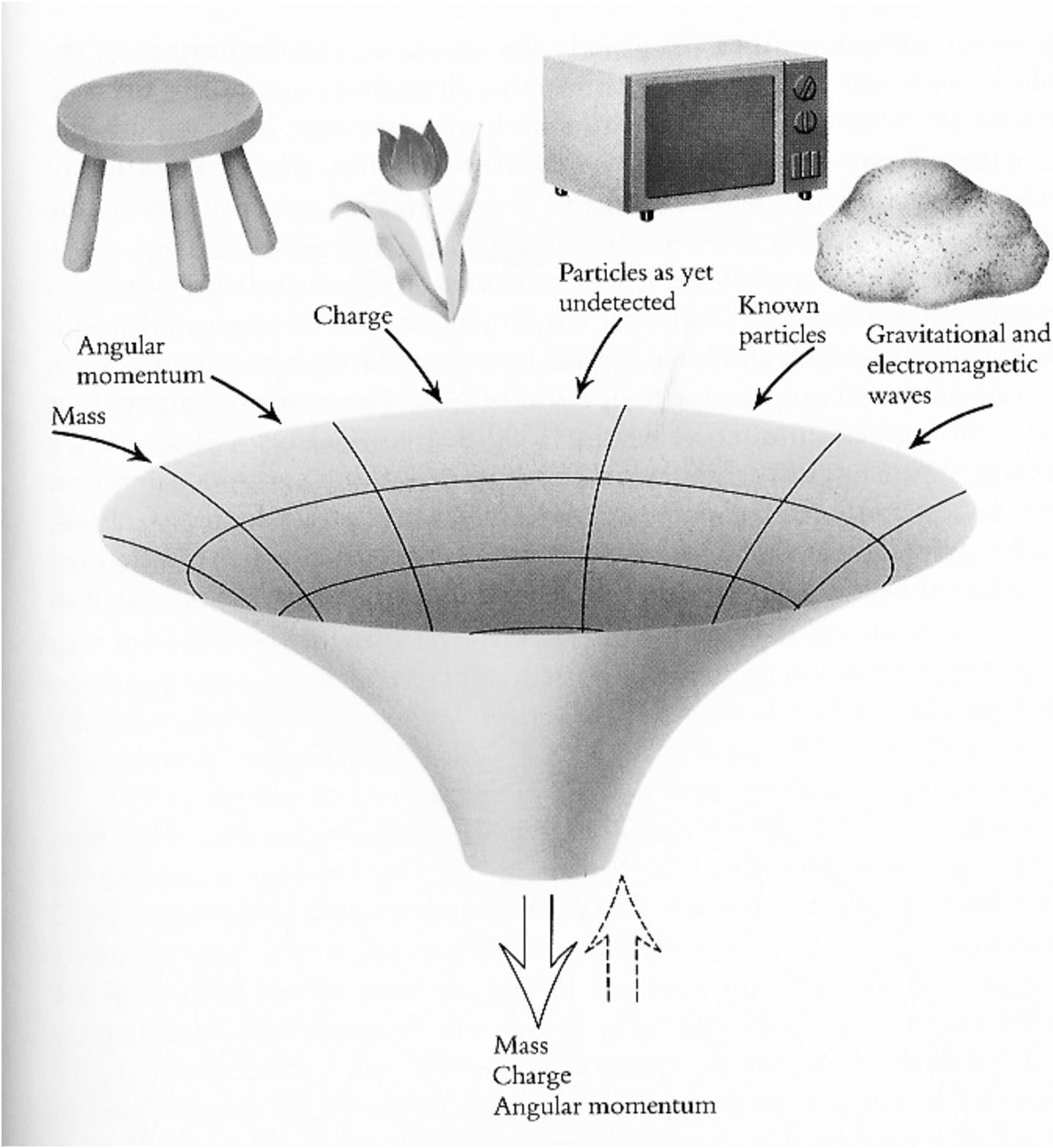}
\includegraphics[width=0.65\hsize]{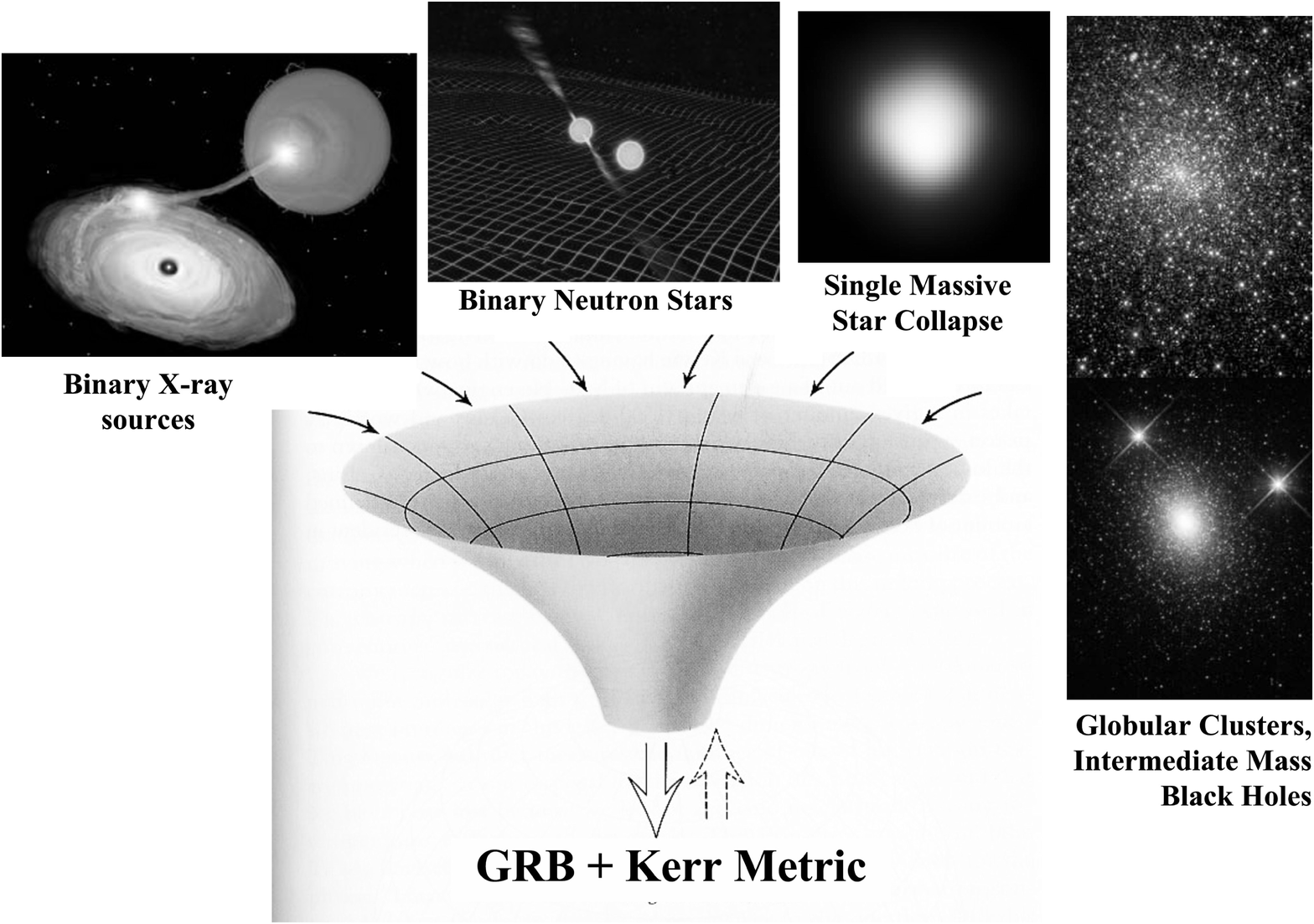}
\caption{\textbf{Left:} The Black Hole Uniqueness theorem \citep[see e.g. Ref.][]{1971PhT....24a..30R}. \textbf{Right:} The GRB uniqueness.}
\label{BH_Uniqueness_1}
\end{figure}

The black hole uniqueness theorem \citep[see Left panel in Fig. \ref{BH_Uniqueness_1} and e.g. Ref.][]{1971PhT....24a..30R} is at the very ground of the fact that it is possible to explain the different Gamma-Ray Burst (GRB) features with a single theoretical model, over a range of energies spanning over $6$ orders of magnitude. The fundamental point is that, independently of the fact that the progenitor of the gravitational collapse is represented by merging binaries composed by neutron stars and white dwarfs in all possible combinations, or by a single process of gravitational collapse, or by the process of ``induced'' gravitational collapse, the formed black hole is totally independent from the initial conditions and reaches the standard configuration of a Kerr-Newman black hole (see Right panel in Fig. \ref{BH_Uniqueness_1}). It is well known that pair creation by vacuum polarization process can occur in a Kerr-Newman black hole \citep{1975PhRvL..35..463D,PhysRep}.

\begin{figure}
\includegraphics[width=\hsize,clip]{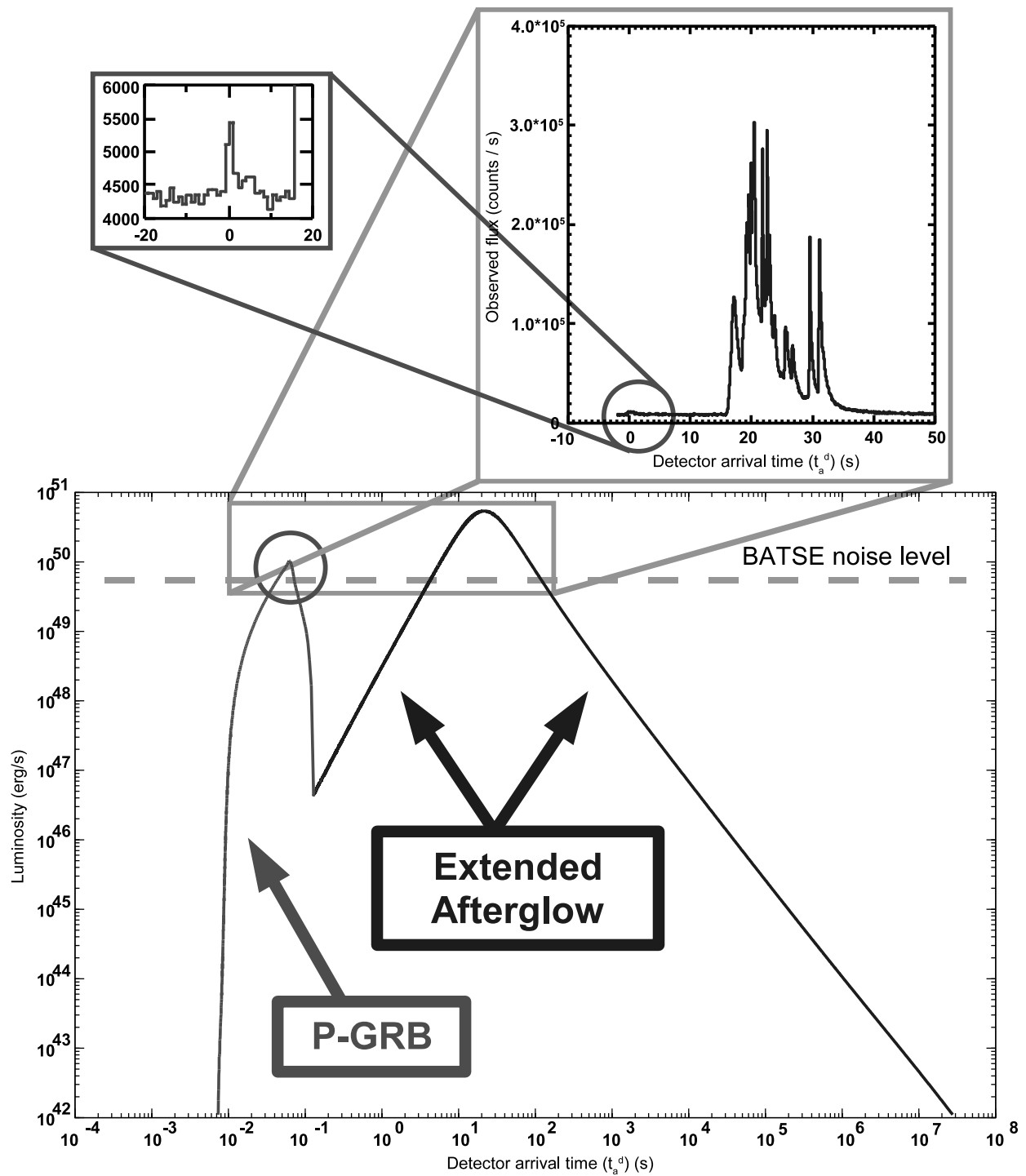}
\caption{The ``canonical GRB'' light curve theoretically computed for GRB 991216. The prompt emission observed by BATSE is identified with the peak of the extended afterglow, while the small precursor is identified with the P-GRB. For this source we have $E_{e^\pm}^{tot} = 4.83\times 10^{53}$ ergs, $B\simeq 3.0\times 10^{-3}$ and $\langle n_{cbm} \rangle \sim 1.0$ particles/cm$^3$. Details in \citet{2001ApJ...555L.113R,2002ApJ...581L..19R,2007AIPC..910...55R}.}
\label{canonical_991216_fig}
\end{figure}

We consequently assume, within the fireshell model, that all GRBs originate from an optically thick $e^\pm$ plasma with total energy $E_{tot}^{e^\pm}$ in the range $10^{49}$--$10^{54}$ ergs and a temperature $T$ in the range $1$--$4$ MeV \citep{1998A&A...338L..87P}. Such an $e^\pm$ plasma has been widely adopted in the current literature \citep[see e.g. Refs.][and references therein]{2005RvMP...76.1143P,2006RPPh...69.2259M}. After an early expansion, the $e^\pm$-photon plasma reaches thermal equilibrium with the engulfed baryonic matter $M_B$ described by the dimensionless parameter $B=M_{B}c^{2}/E_{tot}^{e^\pm}$, that must be $B < 10^{-2}$ \citep{1999A&A...350..334R,2000A&A...359..855R}. As the optically thick fireshell composed by $e^\pm$-photon-baryon plasma self-accelerate to ultrarelativistic velocities, it finally reaches the transparency condition. A flash of radiation is then emitted. This is the P-GRB \citep{2001ApJ...555L.113R}. Different current theoretical treatments of these early expansion phases of GRBs are compared and contrasted in \citet{brvx06} and \citet{2008AIPC.1065..219R}. The amount of energy radiated in the P-GRB is only a fraction of the initial energy $E_{tot}^{e^\pm}$. The remaining energy is stored in the kinetic energy of the optically thin baryonic and leptonic matter fireshell that, by inelastic collisions with the CBM, gives rise to a multi-wavelength emission. This is the extended afterglow. It presents three different regimes: a rising part, a peak and a decaying tail.
We therefore define a ``canonical GRB'' light curve with two sharply different components (see Fig.~\ref{canonical_991216_fig}) \citep{2001ApJ...555L.113R,2007AIPC..910...55R,2007A&A...474L..13B,2008AIPC..966...12B,2008AIPC.1000..305B}: 1) the P-GRB and 2) the extended afterglow. What is usually called ``Prompt emission'' in the current literature mixes the P-GRB with the raising part and the peak of the extended afterglow. Such an unjustified mixing of these components, originating from different physical processes, leads to difficulties in the current models of GRBs, and can as well be responsible for some of the intrinsic scatter observed in the Amati relation \citep[][, see also below, section \emph{Theoretical background for GRBs' empirical correlations}]{2006MNRAS.372..233A,2008A&A...487L..37G}.

\subsection{The optically thick phase}

\begin{figure}
\begin{minipage}{\hsize}
\centering
\includegraphics[width=0.85\hsize]{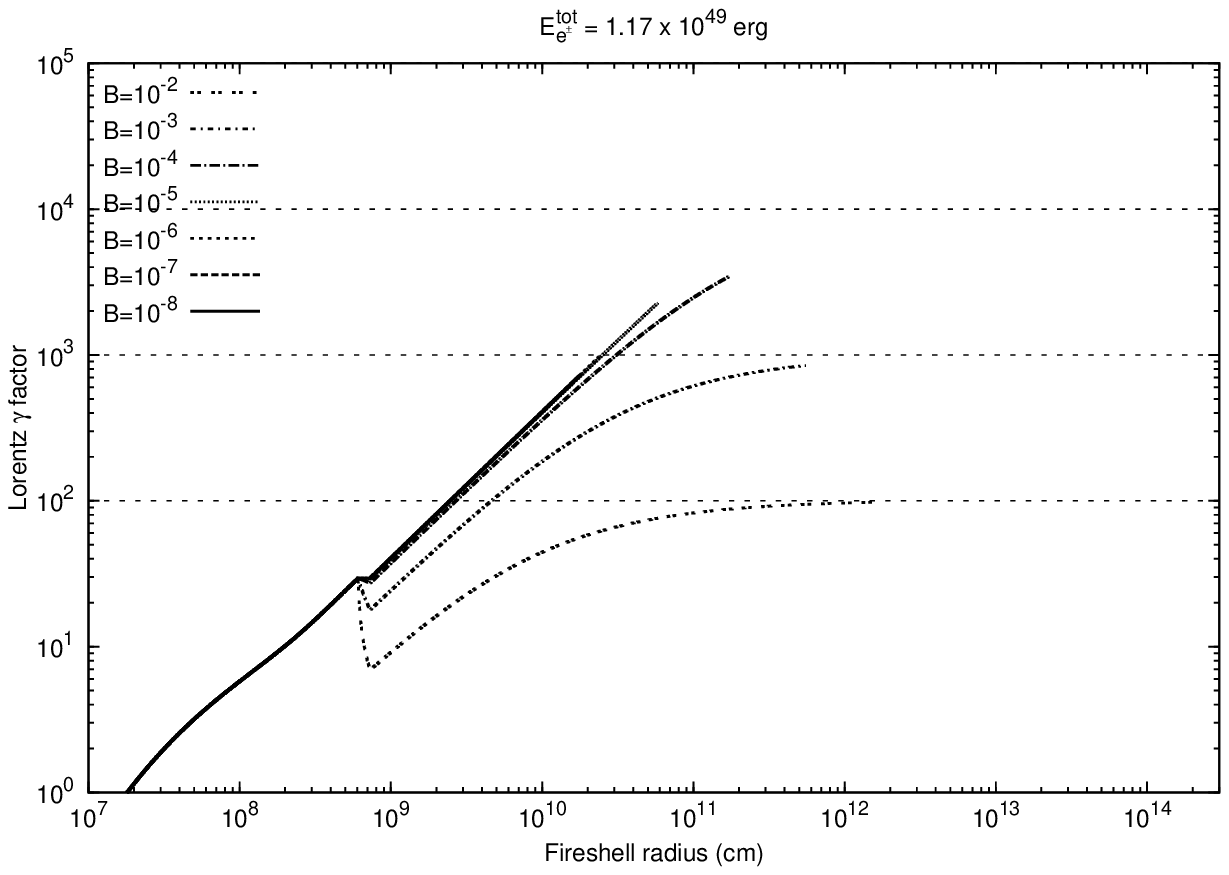}\\
\includegraphics[width=0.85\hsize]{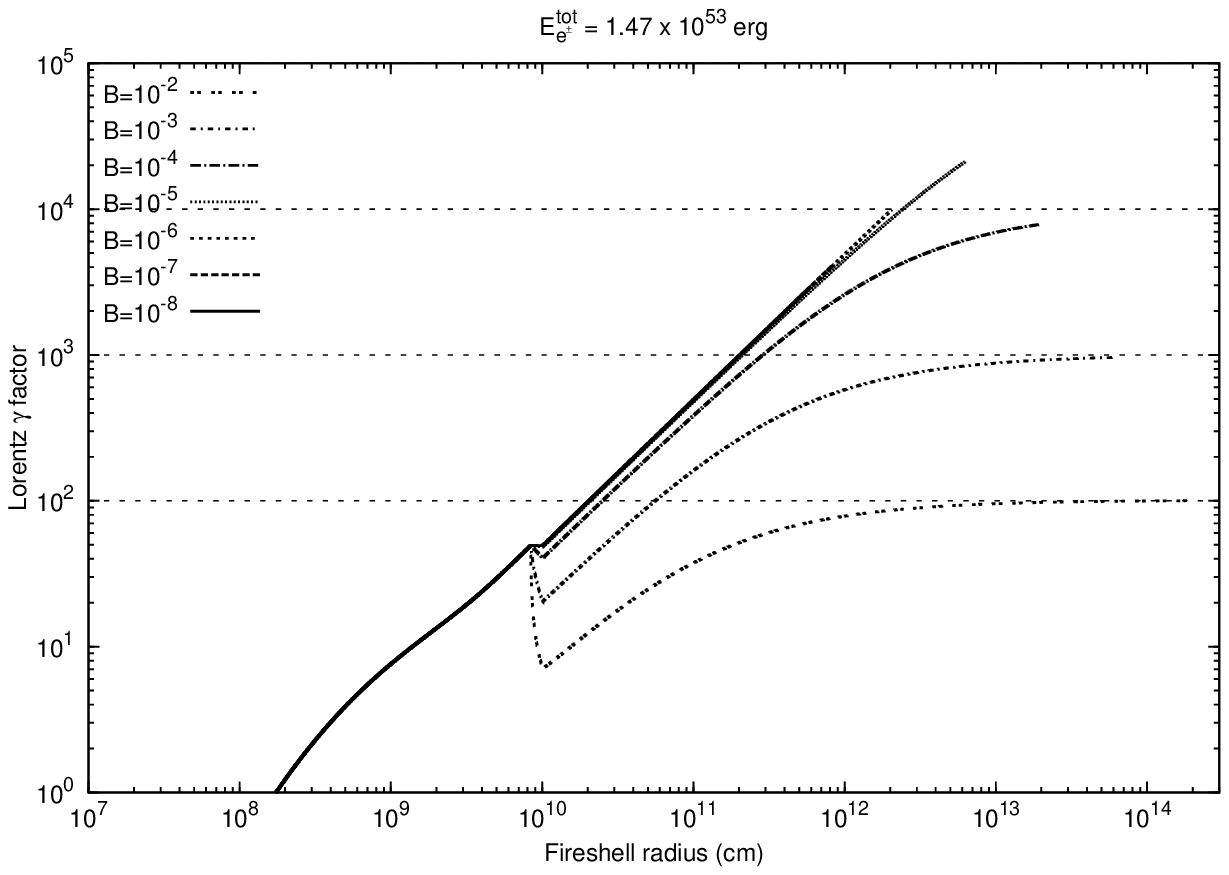}
\end{minipage}
\caption{The Lorentz $\gamma$ factor of the expanding fireshell is plotted as a function of its external radius for $7$ different values of the fireshell baryon loading $B$, ranging from $B=10^{-8}$ and $B=10^{-2}$, and two selected limiting values of the total energy $E_{e^\pm}^{tot}$ of the $e^\pm$ plasma: $E_{e^\pm}^{tot} = 1.17\times 10^{49}$ ergs (upper panel) and $E_{e^\pm}^{tot} = 1.47\times 10^{53}$ ergs (lower panel). The asymptotic values $\gamma \to 1/B$ are also plotted (dashed horizontal lines). The lines are plotted up to when the fireshell transparency is reached.}
\label{MultiGamma}
\end{figure}

In Fig.~\ref{MultiGamma} we present the evolution of the optically thick fireshell Lorentz $\gamma$ factor as a function of the external radius for $7$ different values of the fireshell baryon loading $B$ and two selected limiting values of the total energy $E_{e^\pm}^{tot}$ of the $e^\pm$ plasma. We can identify three different eras:
\begin{enumerate}
\item \textbf{Era I:} The fireshell is made only of electrons, positrons and photons in thermodynamical equilibrium (the ``pair-electromagnetic pulse'', or PEM pulse for short). It self-accelerate and begins its expansion into vacuum, because the environment has been cleared by the black hole collapse. The Lorentz $\gamma$ factor increases with radius and the dynamics can be described by the energy conservation and the condition of adiabatic expansion \citep{1999A&A...350..334R,brvx06}:
\begin{align}
& T^{0\nu}_{\phantom{0\nu},\nu} = 0\\
& \frac{\epsilon_\circ}{\epsilon} = \left(\frac{V}{V_\circ}\right)^{\Gamma} = \left(\frac{{\cal V}\gamma}{{\cal V}_\circ\gamma_\circ}\right)^{\Gamma}
\end{align}
where $T^{\mu\nu}$ is the energy-momentum tensor of the $e^+e^-$ plasma (assumed to be a perfect fluid), $\epsilon$ is its internal energy density, $V$ and ${\cal V}$ are its volumes in the co-moving and laboratory frames respectively, $\Gamma$ is the thermal index and the quantities with and without the $\circ$ subscript are measured at two different times during the expansion.\\
\item \textbf{Era II:} The fireshell impacts with the non-collapsed bayonic remnants and engulfs them. The Lorentz $\gamma$ factor drops. The dynamics of this era can be described by imposing energy and momentum conservation during the fully inelastic collision between the fireshell and the baryonic remnant. For the fireshell solution to be still valid, it must be $B \lesssim 10^{-2}$ \citep{2000A&A...359..855R}.\\
\item \textbf{Era III:} The fireshell is now made of electrons, positrons, baryons and photons in thermodynamical equilibrium (the ``pair-electromagnetic-baryonic pulse'', or PEMB pulse for short). It self-accelerate again and the Lorentz $\gamma$ factor increases again with radius up to when the transparency condition is reached, going to an asymptotic value $\gamma_{asym} = 1/B$. If $B \sim 10^{-2}$ the transparency condition is reached when $\gamma \sim \gamma_{asym}$. On the other hand, when $B < 10^{-2}$, the transparency condition is reached much before $\gamma$ reaches its asymptotic value \citep[][, see next section for details]{2000A&A...359..855R,2001ApJ...555L.113R}. In this era the dynamical equations are the same of the first one, together with the baryon number conservation:
\begin{align}
& T^{0\nu}_{\phantom{0\nu},\nu} = 0\\
& \frac{\epsilon_\circ}{\epsilon} = \left(\frac{V}{V_\circ}\right)^{\Gamma} = \left(\frac{{\cal V}\gamma}{{\cal V}_\circ\gamma_\circ}\right)^{\Gamma}\\
& \frac{n^\circ_B}{n_B} = \frac{V}{V_\circ} = \frac{{\cal V}\gamma}{{\cal V}_\circ\gamma_\circ}
\end{align}
where $n_B$ is the baryon number density in the fireshell. In this era it starts to be crucial the contribution of the rate equation to describe the annihilation of the $e^+e^-$ pairs:
\begin{equation}
\frac{\partial}{\partial t} N_{e^\pm} = - N_{e^\pm} \frac{1}{\cal V}\frac{\partial{\cal V}}{\partial t} + \overline{\sigma v} \frac{1}{\gamma^2}\left(N_{e^\pm}^2(T)-N_{e^\pm}^2\right)
\end{equation}
where $N_{e^\pm}$ is the number of $e^+e^-$ pairs and $N_{e^\pm}(T)$ is the number of $e^+e^-$ pairs at thermal equilibrium at temperature $T$ \citep{2000A&A...359..855R,brvx06}.
\end{enumerate}
In the ``fireball'' model in the current literature the baryons are usually considered present in the plasma since the beginning. In other words, in the fireball dynamics there is only one era corresponding to the Era III above \citep{1993ApJ...415..181M,1990ApJ...365L..55S,1993MNRAS.263..861P,brvx06}. Moreover, the rate equation is usually neglected, and this affects the reaching of the transparency condition. A detailed comparison between the different approaches is reported in \citet{brvx06}.

\subsection{The transparency point}

At the transparency point, the value of the $B$ parameter rules the ratio between the energetics of the P-GRB and the kinetic energy of the baryonic and leptonic matter giving rise to the extended afterglow. It rules as well the time separation between the corresponding peaks \citep{2001ApJ...555L.113R,2008AIPC.1065..219R}.

\begin{figure}
\centering
\includegraphics[width=0.7\hsize]{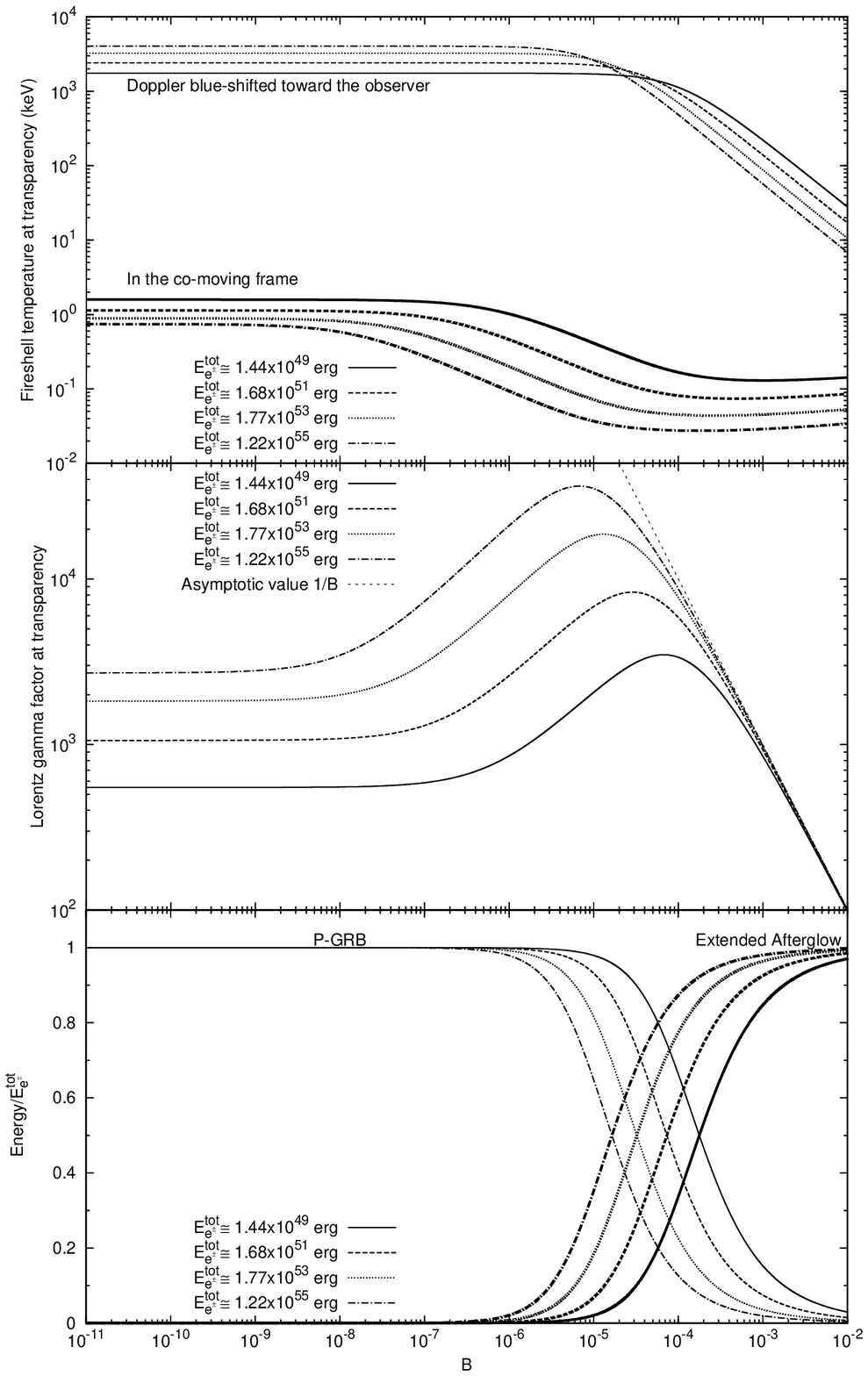}
\caption{At the fireshell transparency point, for $4$ different values of $E^{tot}_{e^\pm}$, we plot as a function of $B$: \textbf{(Above)} The fireshell temperature in the co-moving frame $T_\circ^{com}$ (thicker lines) and the one Doppler blue-shifted along the line of sight toward the observer in the source cosmological rest frame $T_\circ^{obs}$ (thinner lines); \textbf{(Middle)} The fireshell Lorentz gamma factor $\gamma_\circ$ together with the asymptotic value $\gamma_\circ = 1/B$; \textbf{(Below)} The energy radiated in the P-GRB (thinner lines, rising when $B$ decreases) and the one converted into baryonic kinetic energy and later emitted in the extended afterglow (thicker lines, rising when $B$ increases), in units of $E_{e^\pm}^{tot}$.}
\label{ftemp-fgamma-bcross}
\end{figure}

We have recently shown \citep{2009PhRvD..79d3008A} that a thermal spectrum still occurs in presence of $e^\pm$ pairs and baryons. By solving the rate equation we have evaluated the evolution of the temperature during the fireshell expansion, all the way up to when the transparency condition is reached \citep{1999A&A...350..334R,2000A&A...359..855R}. In the upper panel of Fig.~\ref{ftemp-fgamma-bcross} we plot, as a function of $B$, the fireshell temperature $T_\circ$ at the transparency point, i.e. the temperature of the P-GRB radiation. The plot is drawn for four different values of $E_{e^\pm}^{tot}$ in the interval $[10^{49}, 10^{55}]$ ergs, well encompassing GRBs' observed isotropic energies. We plot both the value in the co-moving frame $T_\circ^{com}$ and the one Doppler blue-shifted toward the observed $T_\circ^{obs} = (1+\beta_\circ) \gamma_\circ T_\circ^{com}$, where $\beta_\circ$ is the fireshell speed at the transparency point in units of $c$ \citep{2000A&A...359..855R}.

In the middle panel of Fig.~\ref{ftemp-fgamma-bcross} we plot, as a function of $B$, the fireshell Lorentz gamma factor at the transparency point $\gamma_\circ$. The plot is drawn for the same four different values of $E_{e^\pm}^{tot}$ of the upper panel. Also plotted is the asymptotic value $\gamma_\circ = 1/B$, which corresponds to the condition when the entire initial internal energy of the plasma $E_{e^\pm}^{tot}$ has been converted into kinetic energy of the baryons \citep{2000A&A...359..855R}. We see that such an asymptotic value is approached for $B \to 10^{-2}$. We see also that, if $E_{e^\pm}^{tot}$ increases, the maximum values of $\gamma_\circ$ are higher and they are reached for lower values of $B$.

\begin{figure}
\includegraphics[width=\hsize]{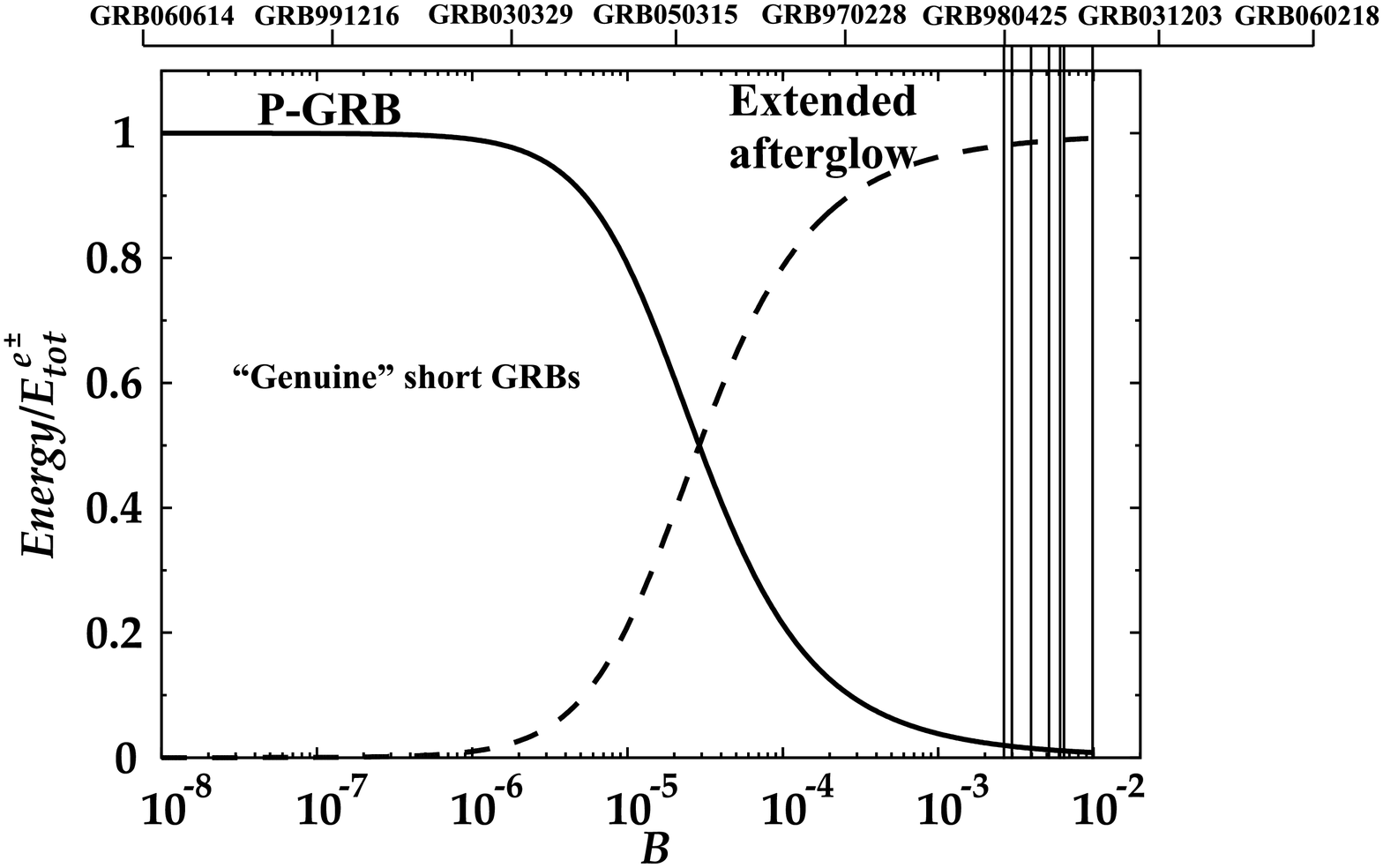}
\caption{Here the energies emitted in the P-GRB (solid line) and in the extended afterglow (dashed line), in units of the total energy of the plasma, are plotted as functions of the $B$ parameter for a typical value of $E^{tot}_{e^\pm} \sim 10^{53}$ erg (see lower panel of Fig. \ref{ftemp-fgamma-bcross}). When $B \lesssim 10^{-5}$, the P-GRB becomes predominant over the extended afterglow, giving rise to a ``genuine'' short GRB. In the figure are also marked the values of the $B$ parameters corresponding to some GRBs we analyzed, all belonging to the class of long GRBs.}
\label{f2}
\end{figure}

In the lower panel of Fig.~\ref{ftemp-fgamma-bcross} we plot, as a function of $B$, the total energy radiated at the transparency point in the P-GRB and the one converted into baryonic and leptonic kinetic energy and later emitted in the extended afterglow. The plot is drawn for the same four different values of $E_{e^\pm}^{tot}$ of the upper panel and middle panels. We see that for $B \lesssim 10^{-5}$ the total energy emitted in the P-GRB is always larger than the one emitted in the extended afterglow. In the limit $B \rightarrow 0$ it gives rise to a ``genuine'' short GRB (see also Fig. \ref{f2}). On the other hand, for $3.0\times 10^{-4} \lesssim B < 10^{-2}$ the total energy emitted in the P-GRB is always smaller than the one emitted in the extended afterglow. If it is not below the instrumental threshold and if $n_{cbm}\sim 1$ particle/cm$^3$ (see Fig. \ref{canonical_canonical}), the P-GRB can be observed in this case as a small pulse preceding the main GRB event (which coincides with the peak of the extended afterglow), i.e. as a GRB ``precursor'' \citep{2001ApJ...555L.113R,2003AIPC..668...16R,2008AIPC.1000..305B,2008AIPC.1065..219R,2007A&A...474L..13B,2008AIPC.1065..223B}. 

Particularly relevant for the new era of the \emph{Agile} and \emph{GLAST} satellites is that for $B < 10^{-3}$ the P-GRB emission has an observed temperature up to $10^{3}$ keV or higher. This high-energy emission has been unobservable by the \emph{Swift} satellite.

\subsection{The optically thin phase}

The dynamics of the optically thin fireshell of baryonic matter propagating in the CBM can be obtained from the relativistic conservation laws of energy and momentum \citep[see e.g. Ref.][]{2005ApJ...620L..23B}:
\begin{equation}
\left\{\begin{aligned}
dE_{\mathrm{int}} & = \left(\gamma - 1\right) dM_{\mathrm{cbm}} c^2\\
d\gamma & = - \frac{{\gamma}^2 - 1}{M} dM_{\mathrm{cbm}}\\
dM & = \frac{1-\varepsilon}{c^2}dE_{\mathrm{int}}+dM_\mathrm{cbm}\\
dM_\mathrm{cbm} & = 4\pi m_p n_\mathrm{cbm} r^2 dr \end{aligned}\right.
\label{Taub_Eq}
\end{equation}
where $\gamma$, $E_\mathrm{int}$ and $M$ are the pulse Lorentz gamma factor, internal energy and mass-energy respectively, $n_\mathrm{cbm}$ is the CBM number density, $m_p$ is the proton mass, $\varepsilon$ is the emitted fraction of the energy developed in the collision with the CBM and $M_\mathrm{cbm}$ is the amount of CBM mass swept up within the radius $r$: $M_\mathrm{cbm}=m_pn_\mathrm{cbm}(4\pi/3)(r^3-{r_\circ}^3)$, where $r_\circ$ is the starting radius of the shock front.

In both our approach and in the other ones in the current literature \citep[see e.g. Refs.][]{1999PhR...314..575P,1999ApJ...512..699C,2005ApJ...620L..23B,2005ApJ...633L..13B,2007AIPC..910...55R} such conservations laws are used. The main difference is that in the current literature an ultra-relativistic approximation, following the \citet{1976PhFl...19.1130B} self-similar solution, is widely adopted, leading to a simple constant-index power-law relations between the Lorentz $\gamma$ factor of the optically thin ``fireshell'' and its radius:
\begin{equation}
\gamma\propto r^{-a}\, ,
\label{gr0}
\end{equation}
with $a=3$ in the fully radiative case and $a=3/2$ in the adiabatic case \citep{1999PhR...314..575P,2005ApJ...633L..13B}. On the contrary, we use the exact solutions of the equations of motion of the fireshell \citep{2004ApJ...605L...1B,2005ApJ...620L..23B,2005ApJ...633L..13B,2006ApJ...644L.105B,2007AIPC..910...55R}. In the adiabatic regime ($\varepsilon = 0$) we get:
\begin{equation}
\gamma^2=\frac{\gamma_\circ^2+2\gamma_\circ\left(M_\mathrm{cbm}/M_B\right)+\left(M_\mathrm{cbm}/M_B\right)^2}{1+2\gamma_\circ\left(M_\mathrm{cbm}/M_B\right)+\left(M_\mathrm{cbm}/M_B\right)^2}\, ,
\label{gamma_ad}
\end{equation}
where $\gamma_\circ$ and $M_B$ are respectively the values of the Lorentz gamma factor and of the mass of the accelerated baryons at the beginning of the afterglow phase. In the fully radiative regime ($\varepsilon = 1$), instead, we have:
\begin{equation}
\gamma=\frac{1+\left(M_\mathrm{cbm}/M_B\right)\left(1+\gamma_\circ^{-1}\right)\left[1+\left(1/2\right)\left(M_\mathrm{cbm}/M_B\right)\right]}{\gamma_\circ^{-1}+\left(M_\mathrm{cbm}/M_B\right)\left(1+\gamma_\circ^{-1}\right)\left[1+\left(1/2\right)\left(M_\mathrm{cbm}/M_B\right)\right]}\, .
\label{gamma_rad}
\end{equation}

A detailed comparison between the equations used in the two approaches has been presented by \citet{2004ApJ...605L...1B,2005ApJ...620L..23B,2005ApJ...633L..13B,2006ApJ...644L.105B}. In particular, \citet{2005ApJ...633L..13B} show that the regime represented in Eq.(\ref{gr0}) is reached only asymptotically when $\gamma_\circ \gg \gamma \gg 1$ in the fully radiative regime and $\gamma_\circ^2 \gg \gamma^2 \gg 1$ in the adiabatic regime, where $\gamma_\circ$ the initial Lorentz gamma factor of the optically thin fireshell.

\begin{figure}
\centering
\includegraphics[width=0.75\hsize,clip]{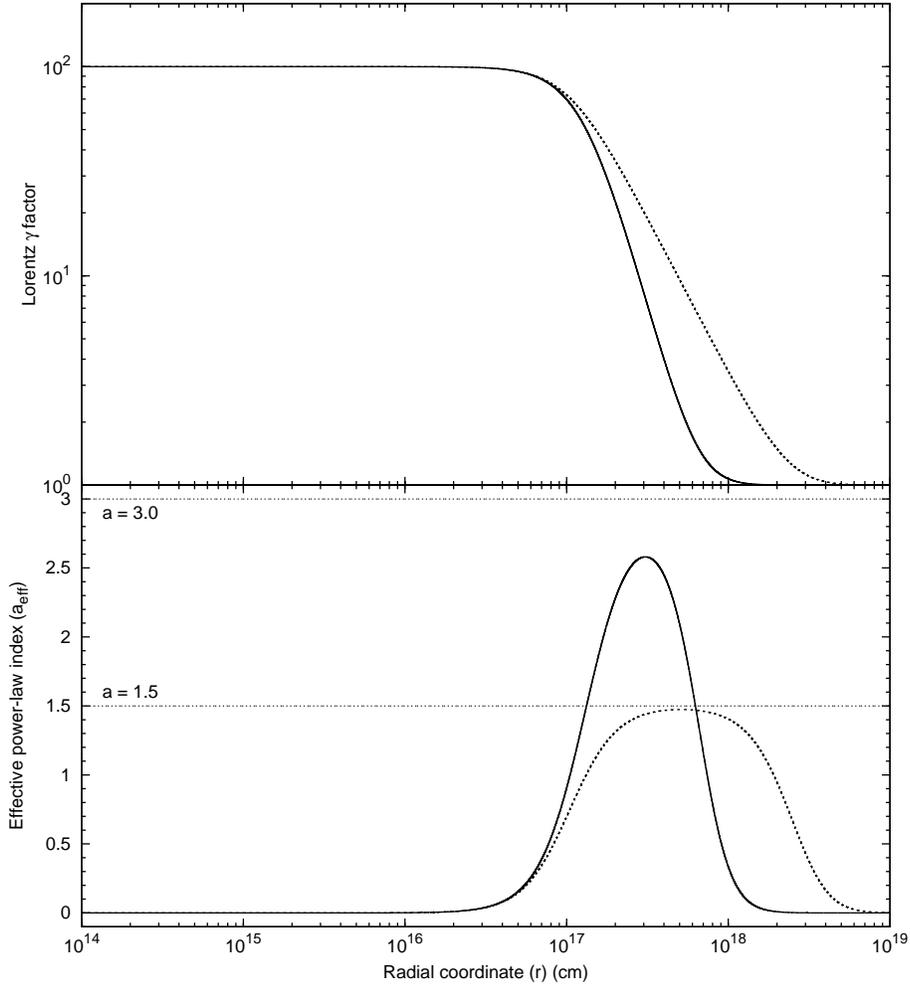}
\caption{In the upper panel, the analytic behavior of the Lorentz $\gamma$ factor during the afterglow era is plotted versus the radial coordinate of the expanding optically thin fireshell in the fully radiative case (solid line) and in the adiabatic case (dotted line) starting from $\gamma_\circ = 10^2$ and the same initial conditions as GRB 991216 \citep{2005ApJ...633L..13B}. In the lower panel are plotted the corresponding values of the ``effective'' power-law index $a_{eff}$ (see Eq.(\ref{eff_a})), which is clearly not constant, is highly varying and systematically lower than the constant values $3$ and $3/2$ purported in the current literature (horizontal thin dotted lines).}
\label{gdir_a_comp_rad-ad_mg11}
\end{figure}

In Fig.~\ref{gdir_a_comp_rad-ad_mg11} we show the differences between the two approaches. In the upper panel there are plotted the exact solutions for the fireshell dynamics in the fully radiative and adiabatic cases. In the lower panel we plot the corresponding ``effective'' power-law index $a_{eff}$, defined as the index of the power-law tangent to the exact solution \citep{2005ApJ...633L..13B}:
\begin{equation}
a_{eff} = - \frac{d\ln\gamma}{d\ln r}\, .
\label{eff_a}
\end{equation}
Such an ``effective'' power-law index of the exact solution smoothly varies from $0$ to a maximum value which is always smaller than $3$ or $3/2$, in the fully radiative and adiabatic cases respectively, and finally decreases back to $0$ (see Fig.~\ref{gdir_a_comp_rad-ad_mg11}).

\subsection{Extended afterglow luminosity and spectrum}

The extended afterglow luminosity in the different energy bands is governed by two quantities associated to the environment. Within the fireshell model, these are the effective CBM density profile, $n_{cbm}$, and the ratio between the effective emitting area $A_{eff}$ and the total area $A_{tot}$ of the expanding baryonic and leptonic shell, ${\cal R}= A_{eff}/A_{tot}$. This last parameter takes into account the CBM filamentary structure \citep{2004IJMPD..13..843R,2005IJMPD..14...97R} and the possible occurrence of a fragmentation in the shell \citep{2007A&A...471L..29D}.

Within the ``fireshell'' model, in addition to the determination of the baryon loading, it is therefore possible to infer a detailed description of the CBM, its average density and its porosity and filamentary structure, all the way from the black hole horizon to distance $r \lesssim 10^{17}$ cm. This corresponds to the prompt emission. This description is lacking in the traditional model based on the synchrotron emission. The attempt to use the internal shock model for the prompt emission \citep[see e.g. Refs.][and references therein]{1994ApJ...430L..93R,2005RvMP...76.1143P,2006RPPh...69.2259M} only applies to regions where $r > 10^{17}$ cm \citep{2008MNRAS.384...33K}.

In our hypothesis, the emission from the baryonic and leptonic matter shell is spherically symmetric. This allows us to assume, in a first approximation, a modeling of thin spherical shells for the CBM distribution and consequently to consider just its radial dependence \citep{2002ApJ...581L..19R}. For simplicity, and in order to have an estimate of the energetics, the emission process is postulated to be thermal in the co-moving frame of the shell \citep{2004IJMPD..13..843R}. We are currently examining a departure from this basic mechanism by taking into account inverse Compton effects due to the electron collisions with the thermal photons. The observed GRB non-thermal spectral shape is due to the convolution of an infinite number of thermal spectra with different temperatures and different Lorentz and Doppler factors. Such a convolution is to be performed over the surfaces of constant arrival time of the photons at the detector \citep[EQuiTemporal Surfaces, EQTSs; see e.g. Ref.][]{2005ApJ...620L..23B} encompassing the whole observation time \citep{2005ApJ...634L..29B}.

\begin{figure}
\centering
\includegraphics[width=\hsize]{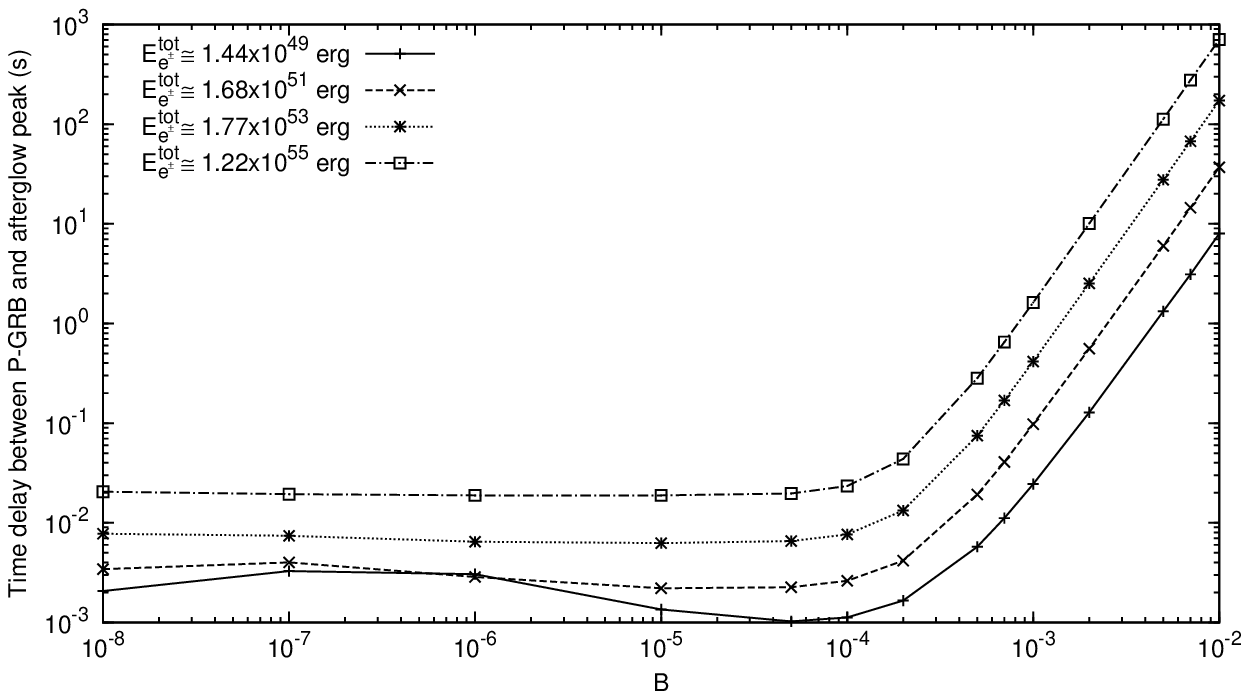}
\caption{For $4$ different values of $E^{tot}_{e^\pm}$, we plot as a function of $B$ the arrival time separation $\Delta t_a$ between the P-GRB and the peak of the extended afterglow (i.e. the ``quiescent time between the ``precursor'' and the main GRB event), measured in the source cosmological rest frame. This computation has been performed assuming a constant CBM density $n_{cbm}=1.0$ particles/cm$^3$. The points represents the actually numerically computed values, connected by straight line segments.}
\label{dta}
\end{figure}

In Fig.~\ref{dta} we plot, as a function of $B$, the arrival time separation $\Delta t_a$ between the P-GRB and the peak of the extended afterglow measured in the cosmological rest frame of the source. Such a time separation $\Delta t_a$ is the ``quiescent time'' between the precursor (i.e. the P-GRB) and the main GRB event (i.e. the peak of the extended afterglow). The plot is drawn for the same four different values of $E_{e^\pm}^{tot}$ of Fig.~\ref{ftemp-fgamma-bcross}. The arrival time of the peak of the extended afterglow emission depends on the detailed profile of the CBM density. In this plot it has been assumed a constant CBM density $n_{cbm}=1.0$ particles/cm$^3$. We can see that, for $3.0\times 10^{-4} \lesssim B < 10^{-2}$, which is the condition for P-GRBs to be ``precursors'' (see above), $\Delta t_a$ increases both with $B$ and with $E_{e^\pm}^{tot}$. We can have $\Delta t_a > 10^2$ s and, in some extreme cases even $\Delta t_a \sim 10^3$ s. For $B \lesssim 3.0\times 10^{-4}$, instead, $\Delta t_a$ presents a behavior which qualitatively follows the opposite of $\gamma_\circ$ (see middle panel of Fig.~\ref{ftemp-fgamma-bcross}).

\begin{figure}
\centering
\includegraphics[width=\hsize]{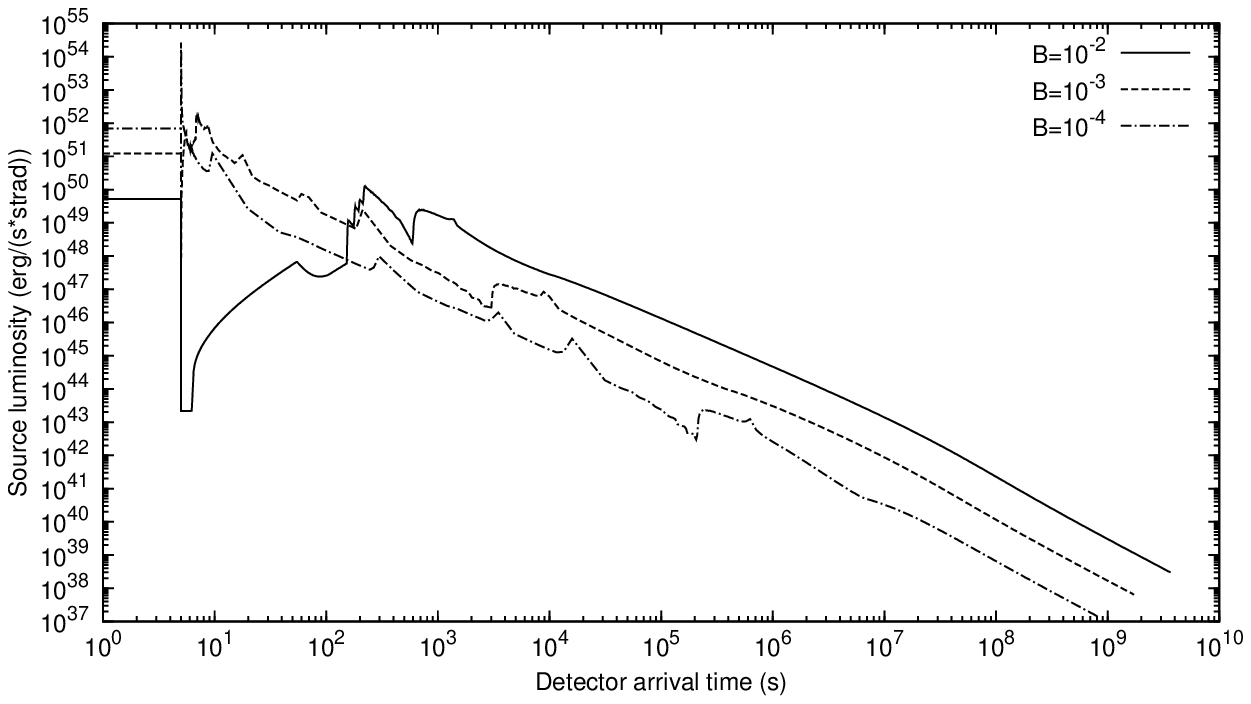}
\caption{We plot three theoretical extended afterglow bolometric light curves together with the corresponding P-GRB peak luminosities (the horizontal segments). The computations have been performed assuming the same $E^{tot}_{e^\pm}$ and CBM structure of GRB 991216 and three different values of $B$. The P-GRBs have been assumed to have the same duration in the three cases, i.e. $5$ s. For $B$ decreasing, the extended afterglow light curve squeezes itself on the P-GRB.}
\label{multi_b}
\end{figure}

Finally, in Fig.~\ref{multi_b} we present three theoretical extended afterglow bolometric light curves together with the corresponding P-GRB peak luminosities for three different values of $B$. The duration of the P-GRBs has been assumed to be the same in the three cases (i.e. $5$ s). The computations have been performed assuming the same $E^{tot}_{e^\pm}$ and the same detailed CBM density profile of GRB 991216 \citep{2003AIPC..668...16R}. In this picture we clearly see how, for $B$ decreasing, the extended afterglow light curve ``squeezes'' itself on the P-GRB and the P-GRB peak luminosity increases.

The ``prompt emission'' light curves of many GRBs present a small pulse preceding the main GRB event, with a lower peak flux and separated by this last one by a quiescent time. The nature of such GRB ``precursors'' is one of the most debated issues in the current literature \citep[see e.g. Refs.][]{2008ApJ...685L..19B,2008AIPC.1065..219R}. Already in 2001 \citep{2001ApJ...555L.113R}, within the ``fireshell'' model, we proposed that GRB ``precursors'' are the P-GRBs emitted when the fireshell becomes transparent, and we gave precise estimates of the time sequence and intensities of the P-GRB and of the extended afterglow, recalled above.

The radiation viewed in the co-moving frame of the accelerated baryonic matter is assumed to have a thermal spectrum and to be produced by the interaction of the CBM with the front of the expanding baryonic shell \citep{2004IJMPD..13..843R}. In \citet{2005ApJ...634L..29B} it is shown that, although the instantaneous spectrum in the co-moving frame of the optically thin fireshell is thermal, the shape of the final instantaneous spectrum in the laboratory frame is non-thermal. In fact, as explained in \citet{2004IJMPD..13..843R}, the temperature of the fireshell is evolving with the co-moving time and, therefore, each single instantaneous spectrum is the result of an integration of hundreds of thermal spectra with different temperature over the corresponding EQTS. This calculation produces a non thermal instantaneous spectrum in the observer frame \citep{2005ApJ...634L..29B}. Another distinguishing feature of the GRBs spectra which is also present in these instantaneous spectra is the hard to soft transition during the evolution of the event \citep{1997ApJ...479L..39C,1999PhR...314..575P,2000ApJS..127...59F,2002A&A...393..409G}. In fact the peak of the energy distributions $E_p$ drift monotonically to softer frequencies with time \citep{2005ApJ...634L..29B}. This feature explains the change in the power-law low energy spectral index \citep{1993ApJ...413..281B} $\alpha$ which at the beginning of the prompt emission of the burst ($t_a^d=2$ s) is $\alpha=0.75$, and progressively decreases for later times \citep{2005ApJ...634L..29B}. In this way the link between $E_p$ and $\alpha$ identified in \citet{1997ApJ...479L..39C} is explicitly shown.

The time-integrated observed GRB spectra show a clear power-law behavior. Within a different framework \citep[see e.g. Ref.][ and references therein]{1983ASPRv...2..189P} it has been argued that it is possible to obtain such power-law spectra from a convolution of many non power-law instantaneous spectra monotonically evolving in time. This result was recalled and applied to GRBs \citep{1999ARep...43..739B} assuming for the instantaneous spectra a thermal shape with a temperature changing with time. It was shown that the integration of such energy distributions over the observation time gives a typical power-law shape possibly consistent with GRB spectra.

\begin{figure}
\includegraphics[width=\hsize,clip]{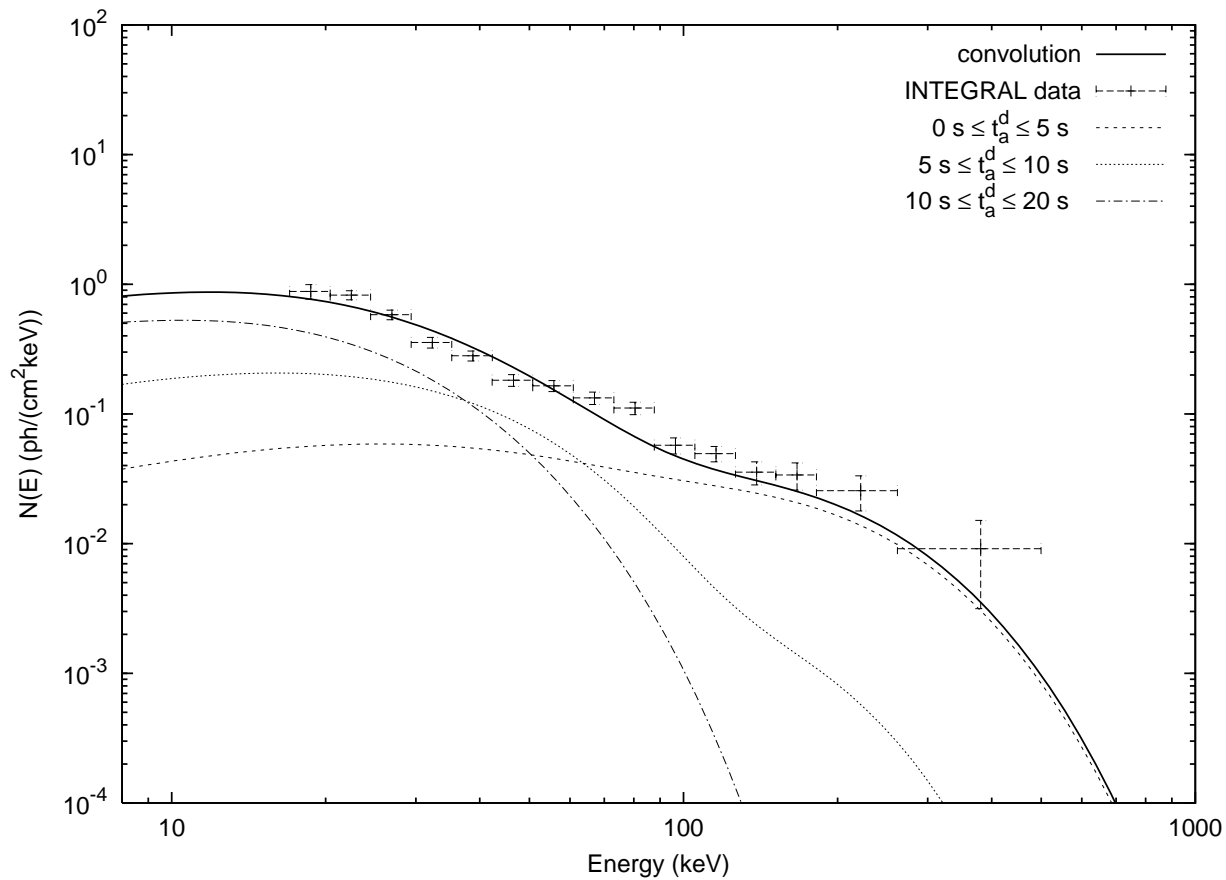}
\caption{Three theoretically predicted time-integrated photon number spectra $N(E)$, computed for GRB 031203 \citep{2005ApJ...634L..29B}, are here represented for $0 \le t_a^d \le 5$ s, $5 \le t_a^d \le 10$ s and $10 \le t_a^d \le 20$ s (dashed and dotted curves), where $t_a^d$ is the photon arrival time at the detector \citep{2001ApJ...555L.107R,2005ApJ...634L..29B}. The hard to soft behavior is confirmed. Moreover, the theoretically predicted time-integrated photon number spectrum $N(E)$ corresponding to the first $20$ s of the ``prompt emission'' (black bold curve) is compared with the data observed by INTEGRAL \citep{2004Natur.430..646S}. This curve is obtained as a convolution of 108 instantaneous spectra, which are enough to get a good agreement with the observed data. Details in \citet{2005ApJ...634L..29B}.}
\label{031203_spettro}
\end{figure}

Our specific quantitative model is more complicated than the one considered by \citet{1999ARep...43..739B}: the instantaneous spectrum here is not a black body. Each instantaneous spectrum is obtained by an integration over the corresponding EQTS: \citep{2004ApJ...605L...1B,2005ApJ...620L..23B} it is itself a convolution, weighted by appropriate Lorentz and Doppler factors, of $\sim 10^6$ thermal spectra with variable temperature. Therefore, the time-integrated spectra are not plain convolutions of thermal spectra: they are convolutions of convolutions of thermal spectra \citep{2004IJMPD..13..843R,2005ApJ...634L..29B}. In Fig.~\ref{031203_spettro} we present the photon number spectrum $N(E)$ time-integrated over the $20$ s of the whole duration of the prompt event of GRB 031203 observed by INTEGRAL \citep{2004Natur.430..646S}: in this way we obtain a typical non-thermal power-law spectrum which results to be in good agreement with the INTEGRAL data \citep{2004Natur.430..646S,2005ApJ...634L..29B} and gives a clear evidence of the possibility that the observed GRBs spectra are originated from a thermal emission \citep{2005ApJ...634L..29B}

Before closing, we like to mention that, using the diagrams represented in Figs.~\ref{ftemp-fgamma-bcross}-\ref{dta}, in principle one can compute the two free parameters of the fireshell model, namely $E^{tot}_{e^\pm}$ and $B$, from the ratio between the total energies of the P-GRB and of the extended afterglow and from the temporal separation between the peaks of the corresponding bolometric light curves. None of these quantities depends on the cosmological model. Therefore, one can in principle use this method to compute the GRBs' intrinsic luminosity and make GRBs the best cosmological distance indicators available today. The increase of the number of observed sources, as well as the more accurate knowledge of their CBM density profiles, will possibly make viable this procedure to test cosmological parameters, in addition to the Amati relation \citep{2008MNRAS.391..577A,2008A&A...487L..37G}.

	\section{GRB 060607A: a complete analysis of the prompt emission and X-ray flares.}

GRB 060607A is a very distant \citep[$z=3.082$, see Ref.][]{2006GCN..5237....1L} and energetic event \citep[$E_{iso}\sim 10^{53}$ erg, see Ref.][]{2007A&A...469L..13M}. Its BAT light curve shows a double-peaked structure with a duration of $T_{90}=(100\pm5)$ s \citep{2006GCN..5242....1T}. The time-integrated spectrum over the $T_{90}$ is best fit with a simple power-law model with an index $\Gamma= 1.45\pm0.08$ (Guidorzi, private communication). The XRT light curve shows a prominent flaring activity (at least three flares) superimposed to the normal afterglow decay \citep{2006GCN..5240....1P}.

The GRB 060607A main peculiarity is that the peak of the near-infrared (NIR) afterglow has been observed with the REM robotic telescope \citep{2007A&A...469L..13M}. Interpreting the NIR light curve as corresponding to the afterglow onset as predicted by the fireball forward shock model \citep{1999ApJ...520..641S,2006RPPh...69.2259M}, it is possible to infer the initial Lorentz gamma factor of the emitting system that results to be $\Gamma_\circ \sim 400$ \citep{2007A&A...469L..13M,2007arXiv0710.0727C,2007MNRAS.378.1043J}. Moreover, these measurements seem to be consistent with an interstellar medium environment, ruling out the wind-like medium \citep{2007A&A...469L..13M,2007MNRAS.378.1043J}.

We analyze GRB 060607A within the fireshell model \citep{2001ApJ...555L.107R,2001ApJ...555L.113R,2007AIPC..910...55R}. We show that within this interpretation the N(E) spectrum of the prompt emission can be fitted in a satisfactory way by a convolution of thermal spectra as predicted by the model we applied \citep{2004IJMPD..13..843R,2005IJMPD..14...97R,2005ApJ...634L..29B}. The theoretical spectrum and light curve in the BAT energy band obtained are in good agreement with the observations, enforcing the plausibility of our approach. Moreover, in analogy with the case of GRB 011121 \citep{Venezia_Flares}, we propose an interpretation of the observed X-ray flares as produced by overdense CBM clouds, in analogy with the gamma-ray light curve.

In this preliminary analysis we deal only with the BAT and XRT observations, which are the basic contribution to the afterglow emission according to the fireshell model. We do not deal with the infrared emission that, on the contrary, is used in the current literature to estimate the dynamical quantities of the fireball in the forward external shock regime. Nevertheless, the initial value of Lorentz gamma factor we predict is compatible with the one deduced from the REM observations even under very different assumptions.

\subsection{GRB 060607A prompt emission}

\subsubsection{Light curves}

\begin{figure}
\begin{minipage}{\hsize}
\centering
\includegraphics[width=0.4\hsize]{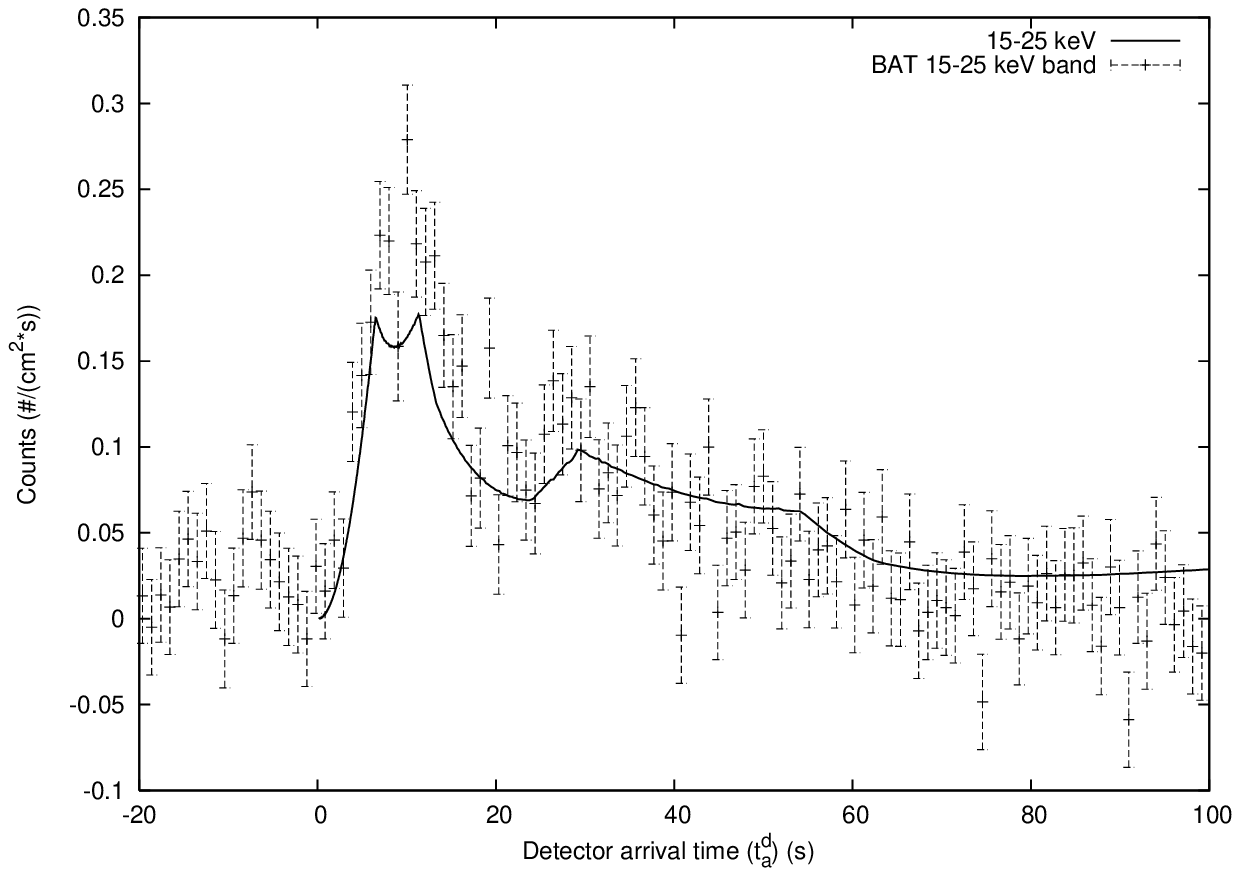}
\includegraphics[width=0.4\hsize]{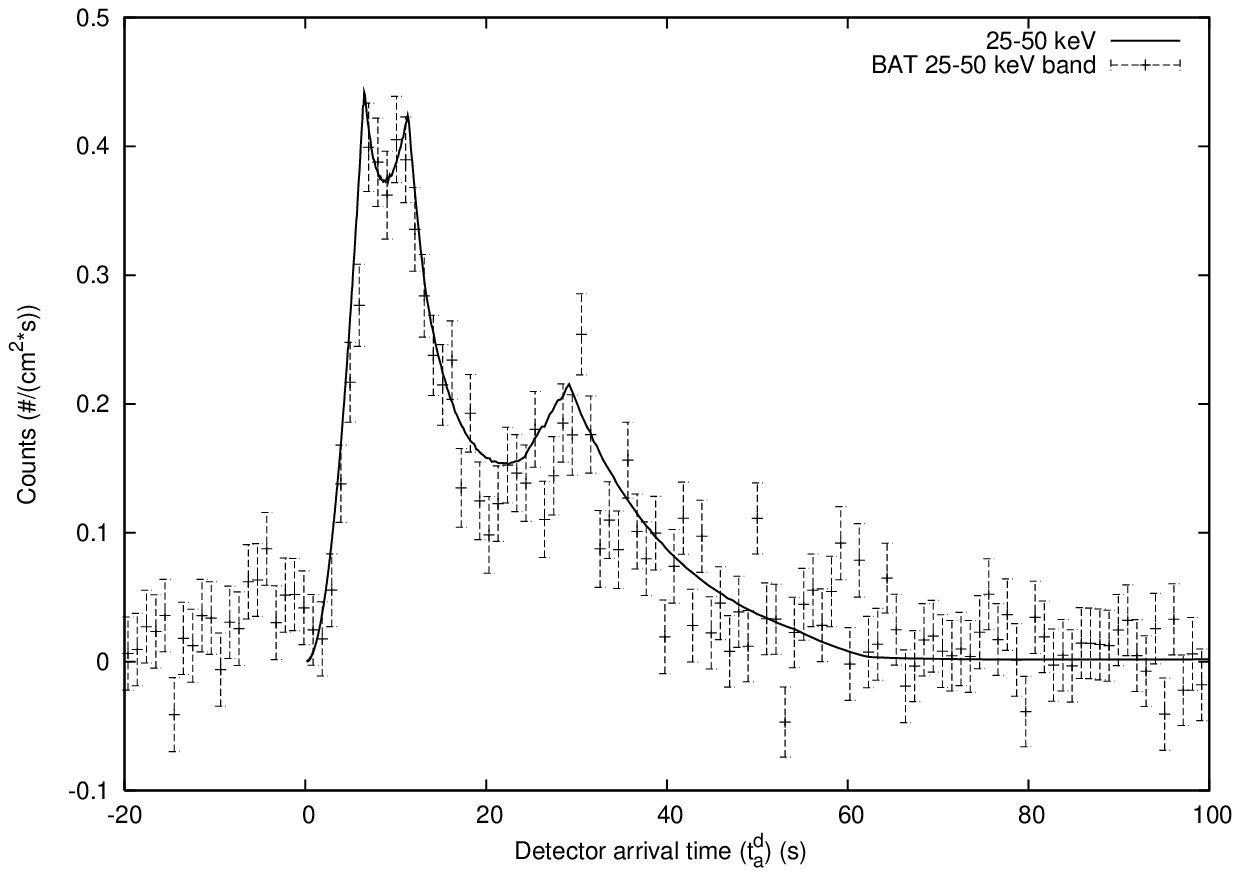}\\
\includegraphics[width=0.4\hsize]{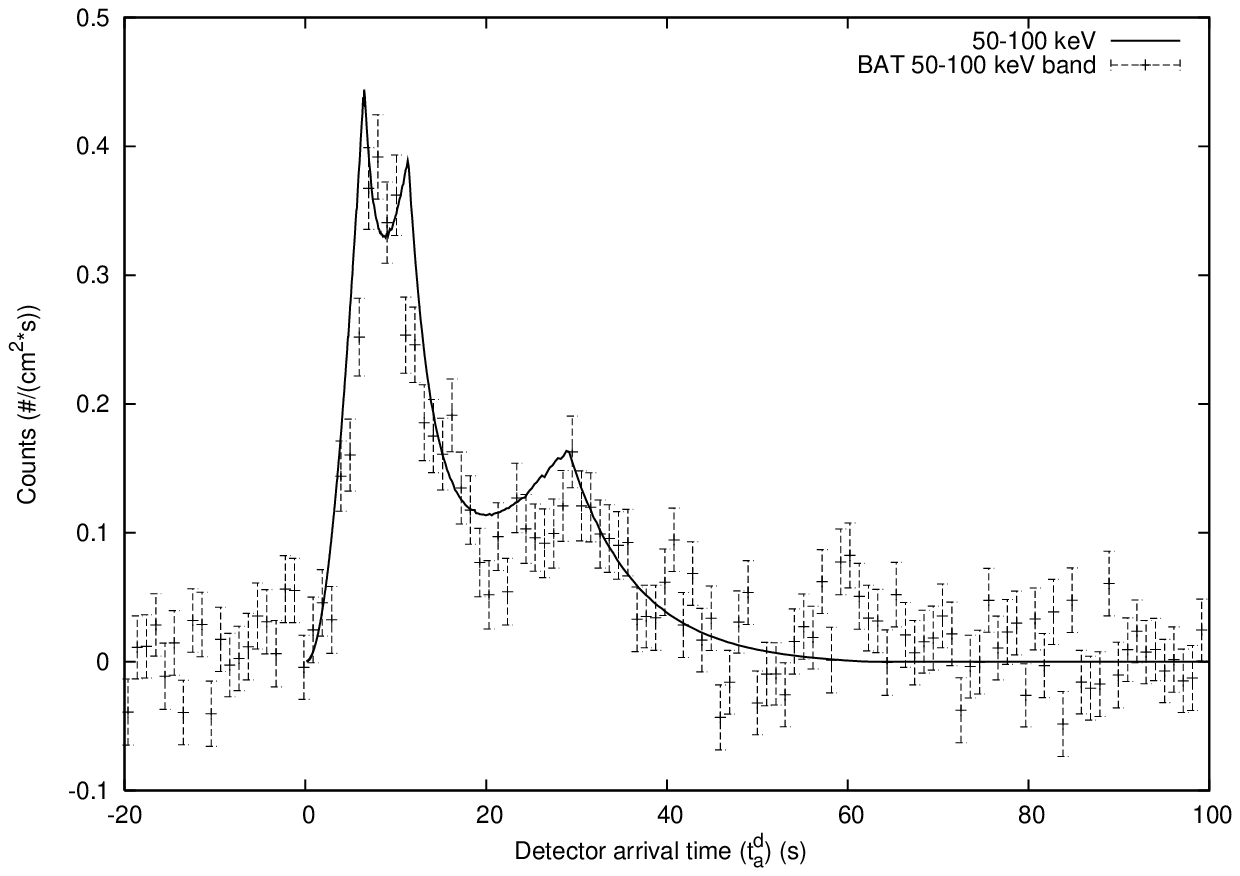}
\includegraphics[width=0.4\hsize]{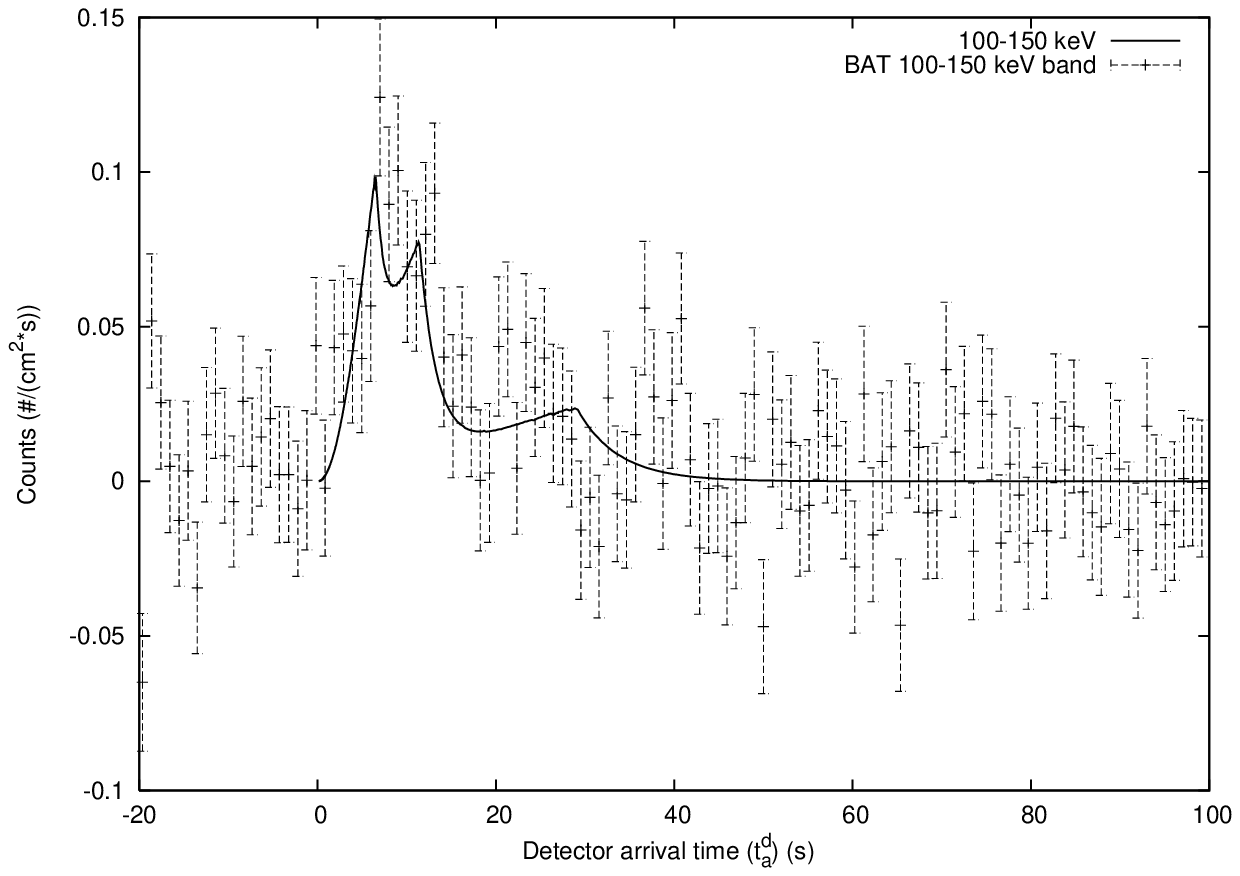}
\end{minipage}
\caption{\emph{Swift} BAT ($15$--$25$ keV, $25$--$50$ keV, $50$--$100$ keV, $100$--$150$ keV) light curves (points) compared with the theoretical ones (solid lines).}
\label{060607A_fit}
\end{figure}

\begin{figure}
\begin{minipage}{\hsize}
\centering
\includegraphics[width=0.7\hsize]{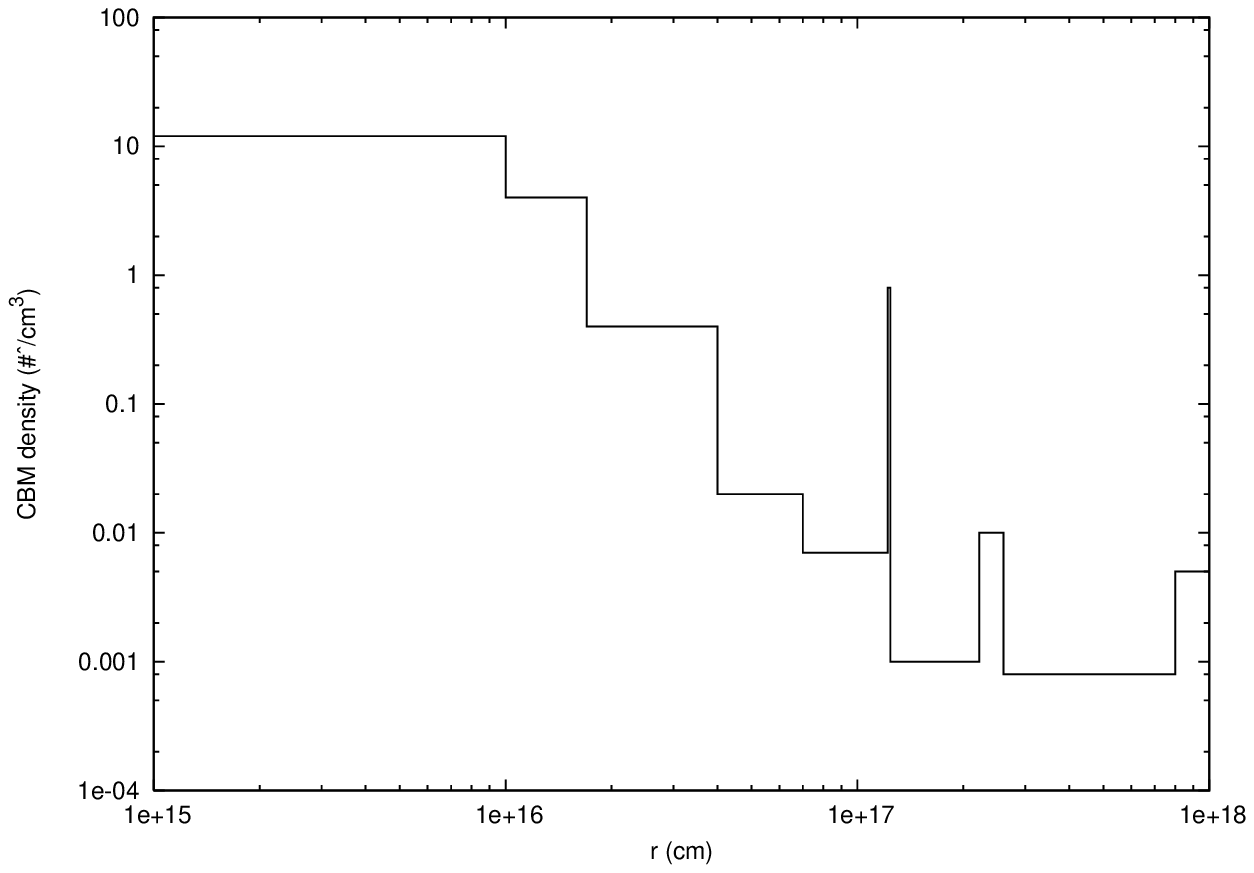}\\
\includegraphics[width=0.7\hsize]{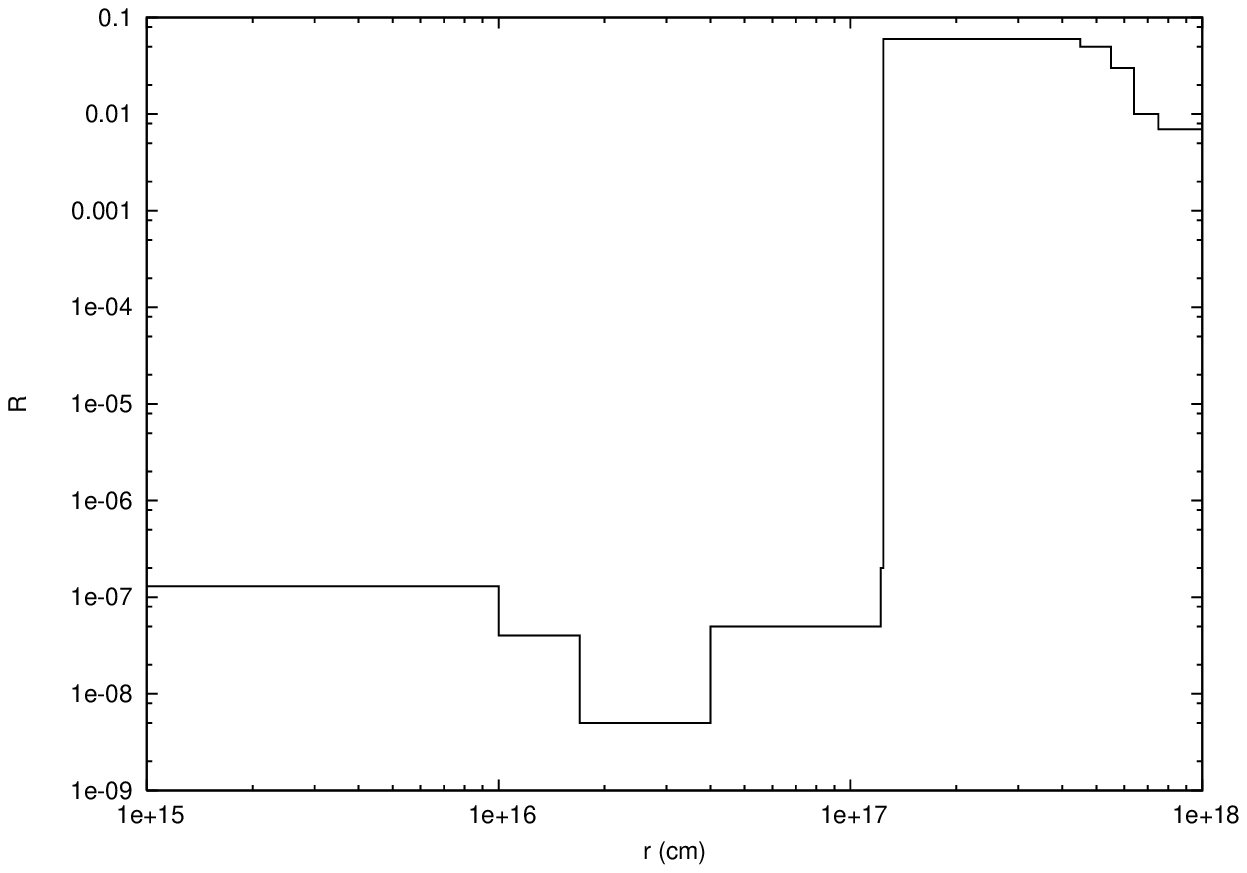}
\end{minipage}
\caption{Detailed structure of the CBM adopted: particle number density $n_{cbm}$ (upper panel) and fraction effective emitting area ${\cal R}$ (lower panel) versus distance from the progenitor. The two X-ray flares corresponds in the upper panel to the huge increases in the CBM density that departs from the roughly power-law decrease observed. In the lower panel, the X-ray flares produce an increase of the emitting area which is not real but due to the lack of a complete 3-dimensional treatment of the interaction between the fireshell and the CBM (see text).}
\label{060607A_CBM}
\end{figure}

In Fig.~\ref{060607A_fit} we present the theoretical fit of \emph{Swift} BAT light curves in different energy bands ($15$--$25$ keV, $25$--$50$ keV, $50$--$100$ keV, $100$--$150$ keV) of GRB 060607A. We identify the whole prompt emission with the peak of the extended afterglow emission, and the remaining part of the light curve with the decaying tail of the extended afterglow, according to our ``canonical GRB'' scenario \citep{2001ApJ...555L.113R,2007AIPC..910...55R}. The temporal variability of the light curves has been reproduced assuming overdense spherical CBM regions \citep{2002ApJ...581L..19R}. The detailed structure of the CBM adopted is presented in Fig. \ref{060607A_CBM}.

We therefore obtain for the two parameters characterizing the source in our model $E_{e^\pm}^{tot}=2.5\times 10^{53}$ erg and $B = 3.0\times 10^{-3}$. This implies an initial $e^\pm$ plasma with a total number of $e^{\pm}$ pairs $N_{e^\pm} = 2.6\times 10^{58}$ and an initial temperature $T = 1.7$ MeV. The theoretically estimated total isotropic energy emitted in the P-GRB is $E_{P-GRB}=1.9\% \, E_{e^\pm}^{tot}=4.7 \times 10^{51}$ erg, hence the P-GRB results to be undetectable if we assume a duration $\Delta t_{p-grb}\gtrsim 10$ s.

After the transparency point at $r_0 = 1.4\times 10^{14}$ cm from the progenitor, the initial Lorentz gamma factor of the fireshell is $\gamma_\circ = 328$. This value has been obtained adopting the exact solutions of the equations of motions of the fireshell \citep{2005ApJ...633L..13B} and using as initial conditions the two free parameters ($E_{e^\pm}^{tot}$ and $B$) estimated from the simultaneous analysis of the BAT and XRT light curves.

\subsubsection{Time-integrated spectra}

\begin{figure}
\begin{minipage}{\hsize}
\centering
\includegraphics[width=0.6\hsize]{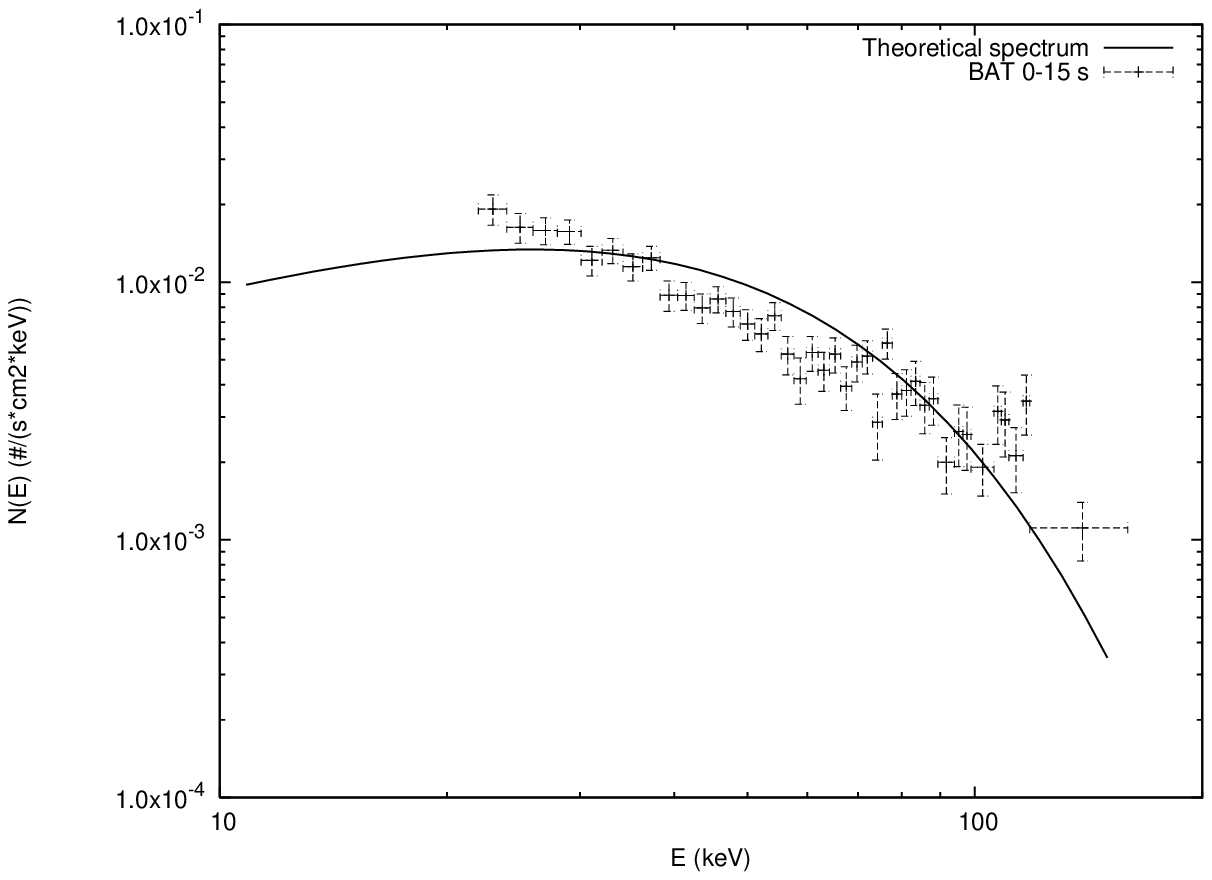}\\
\includegraphics[width=0.6\hsize]{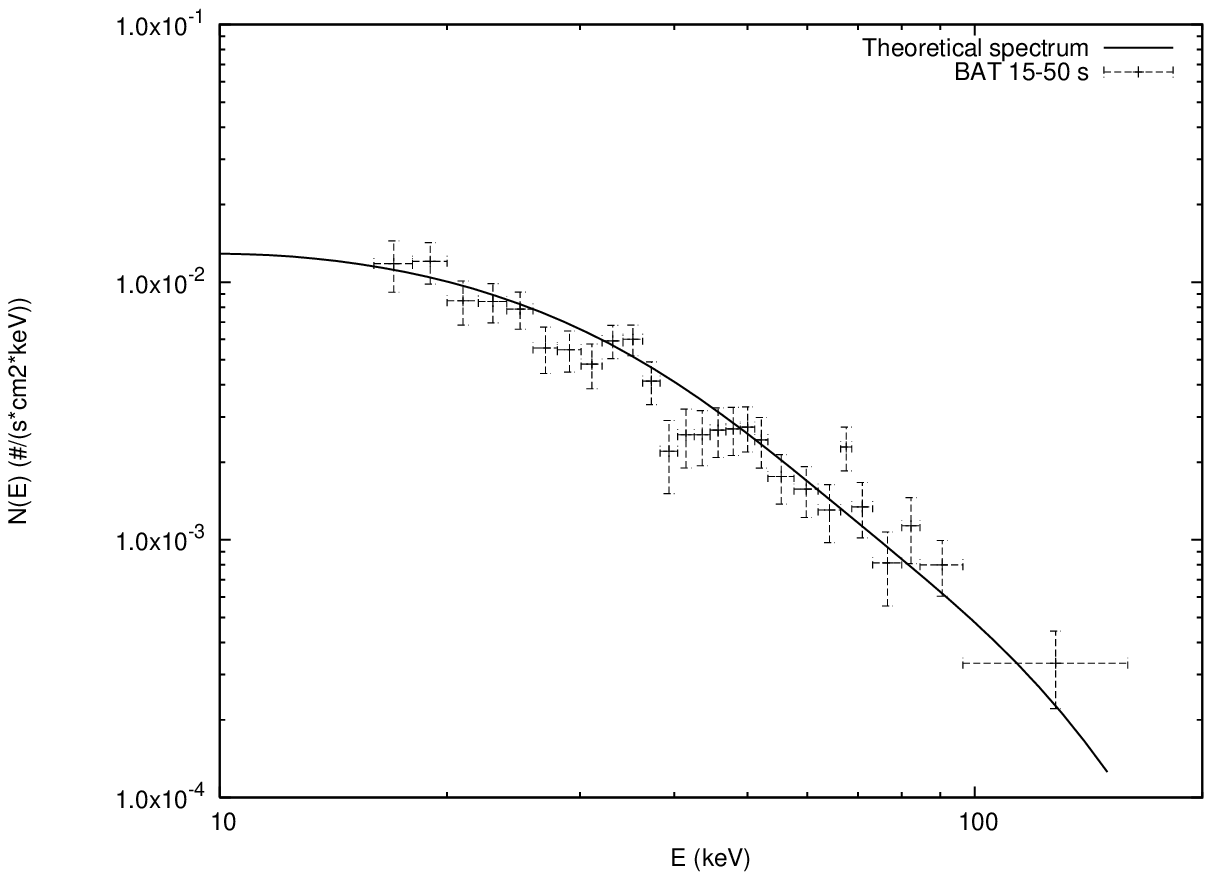}\\
\includegraphics[width=0.6\hsize]{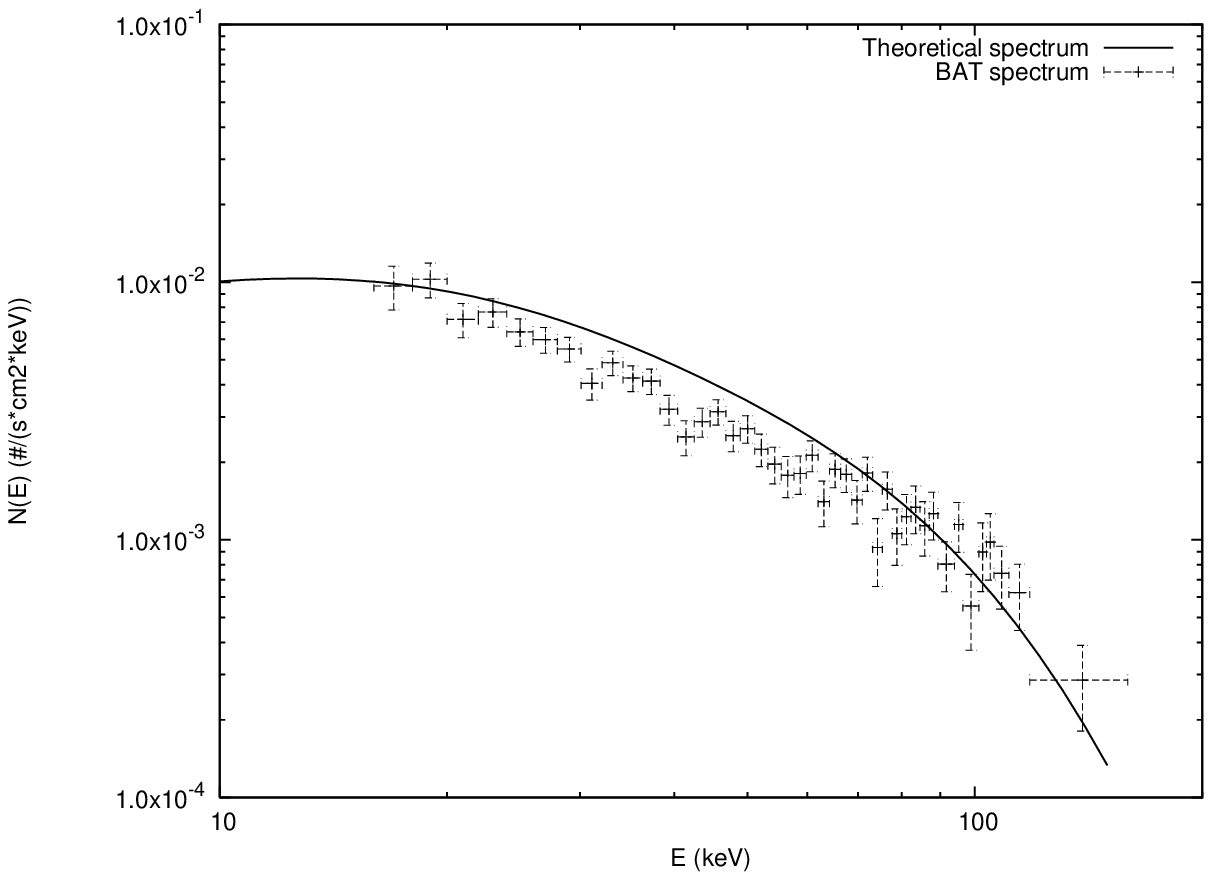}
\end{minipage}
\caption{Theoretically predicted time-integrated photon number spectrum $N(E)$ corresponding to the $0-15$ s (upper panel), $15-50$ s (middle panel), and to the whole duration ($T_{90}=100$ s, lower panel) of the prompt emission (solid lines) compared with the observed spectra integrated in the same intervals.}
\label{060607A_spettro}
\end{figure}

We turn now to the analysis of the GRB 060607A prompt emission time-integrated spectrum. As discussed in previous works \citep{2004IJMPD..13..843R,2005IJMPD..14...97R,2005ApJ...634L..29B}, even if the fireshell model assumes that the GRB spectrum is thermal in the comoving frame, the shape of the final spectrum in the laboratory frame is clearly non-thermal. In fact each single instantaneous spectrum is the result of a convolution of thermal spectra. In fact photons observed at the same arrival time are emitted at different comoving time \citep[the so called EQTS, see Refs.][]{2004ApJ...605L...1B,2005ApJ...620L..23B}, hence with different temperatures. This calculation produces a non-thermal instantaneous spectrum in the observer frame. This effect is enhanced if we calculate the time-integrated spectrum: we perform two different integrations, one on the observation time and one on the EQTS, and what we get is a typical non-thermal power-law spectrum which results to be in good agreement with the observations (see Fig. \ref{060607A_spettro}).

\subsection{The X-ray flares.}

\begin{figure}
\begin{minipage}{\hsize}
\centering
\includegraphics[width=0.68\hsize]{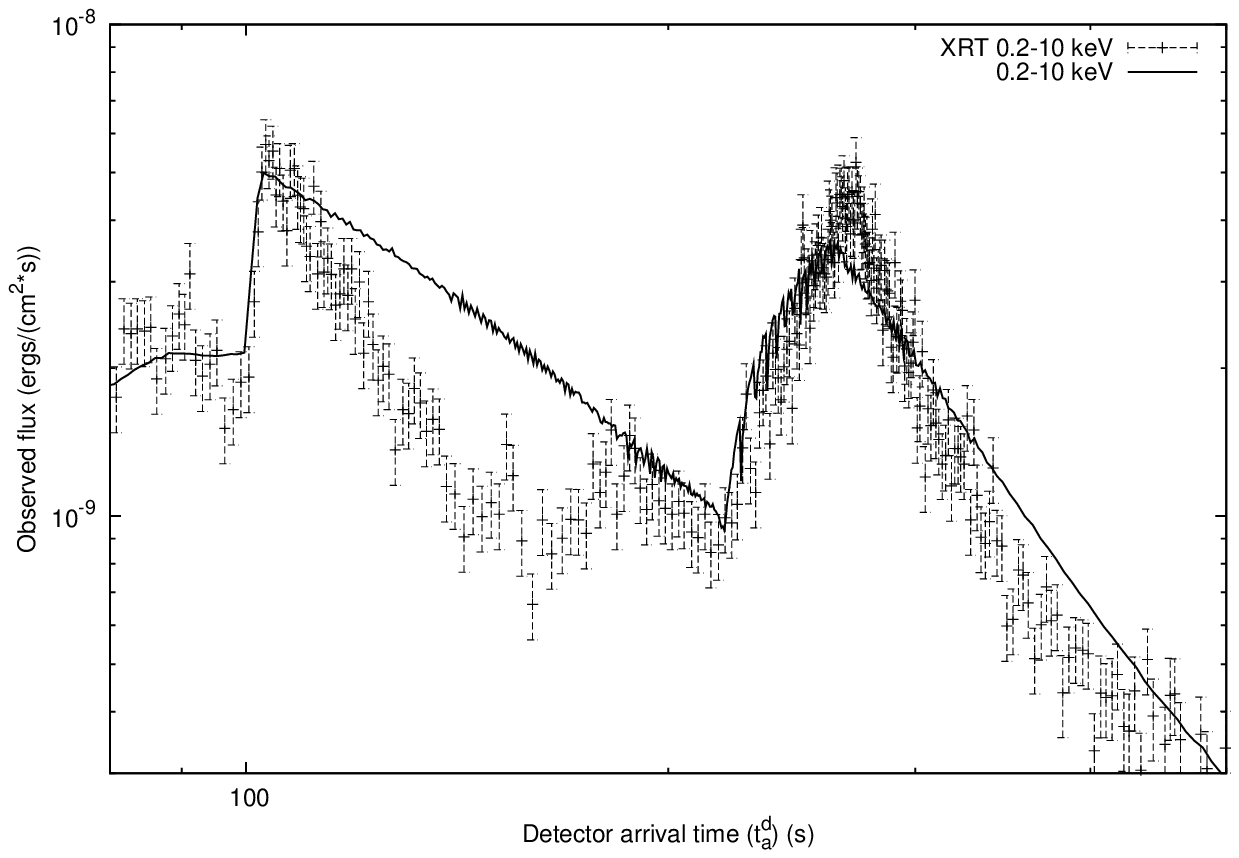}\\
\includegraphics[width=0.68\hsize]{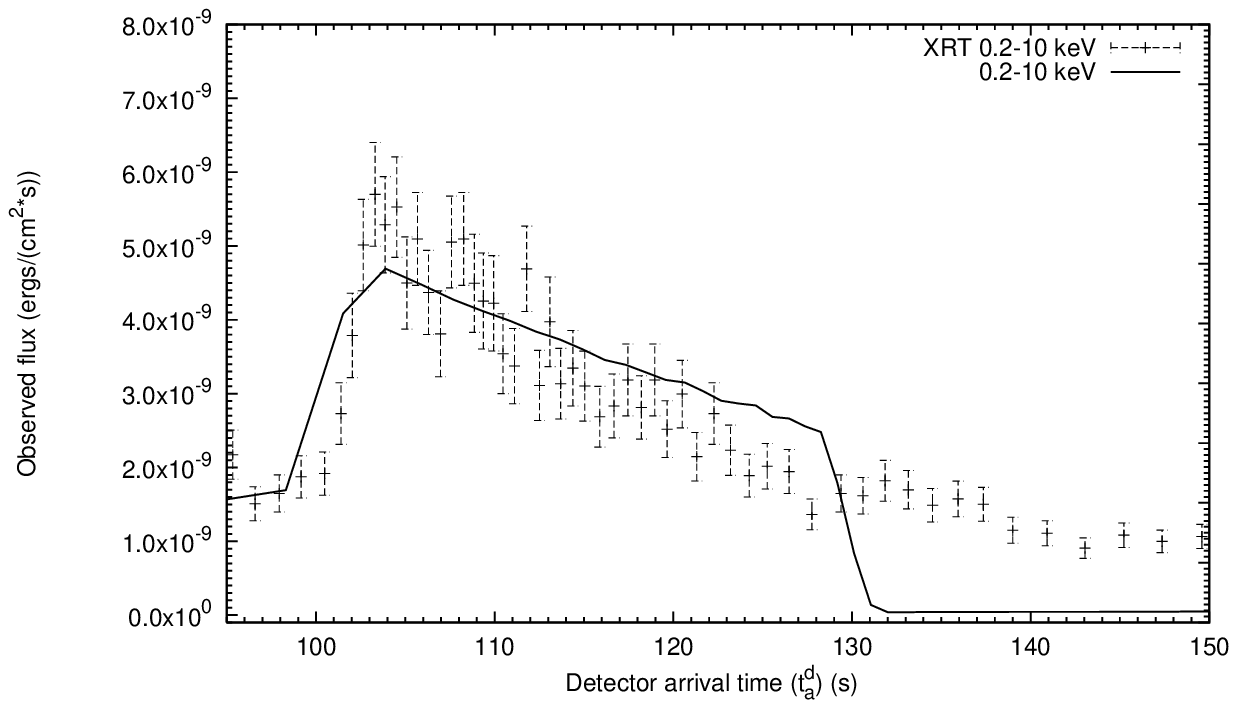}\\
\includegraphics[width=0.68\hsize]{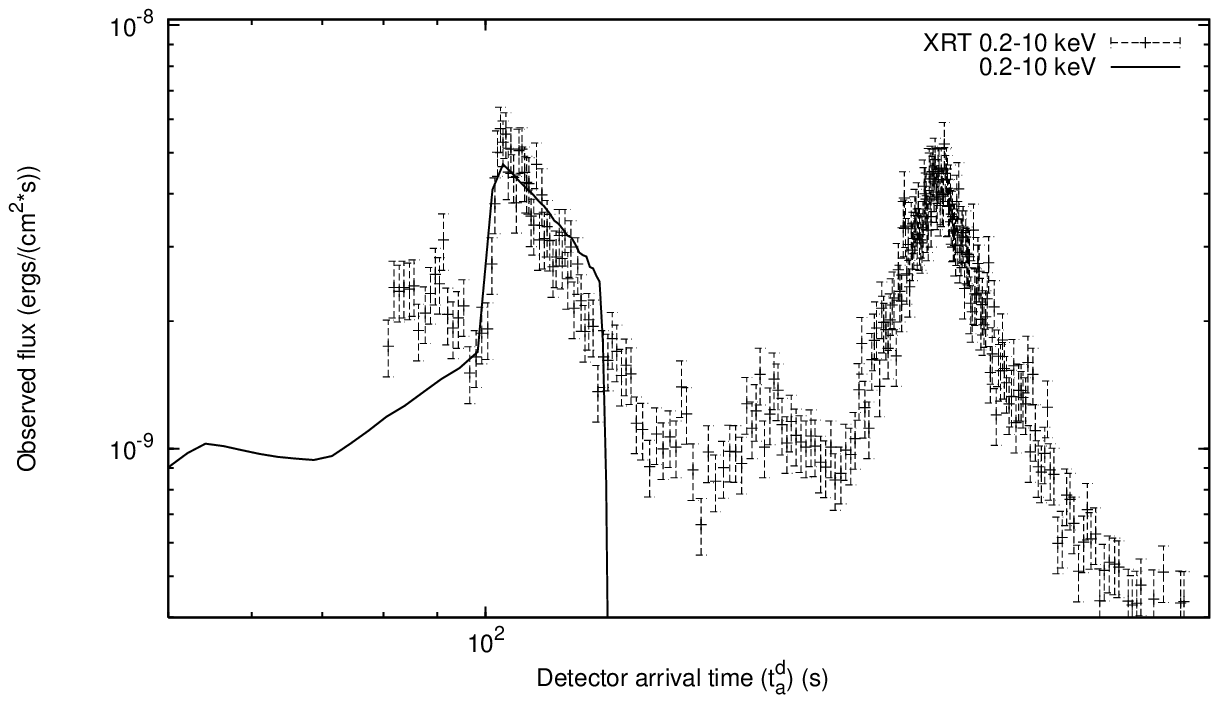}
\end{minipage}
\caption{\emph{Swift} XRT ($0.2$--$10$ keV) light curve compared with, respectively, the theoretical one obtained assuming the CBM distributed in spherical shells (upper panel), the theoretical one obtained imposing a finite transverse dimension for the CBM cloud (middle panel) and the same theoretical curve in logarithmic scale (lower panel).}
\label{060607A_flares}
\end{figure}

We analyze now the X-ray flares observed by \emph{Swift} XRT ($0.2-10$ keV) in the early part of the decaying phase of the extended afterglow. According to the fireshell model these flares have the same origin of the prompt emission, namely they are produced by the interaction of the fireshell with overdense CBM. As we can see in the upper panel in Fig. \ref{060607A_flares}, the result obtained is compatible with the observations only for the second flare but not for the first one since its duration is longer. This discrepancy is due to the simple modeling adopted, namely the CBM is arranged in spherical shells \citep{2002ApJ...581L..19R}. This approximetion fails when the visible area of the fireshell is comparable with the size of the CBM clouds. 

To solve this problem, following the results obtained for GRGB 011121 \citep{Venezia_Flares}, we tried to account for the three-dimensional structure of the CBM clouds by ``cutting'' the emission at a certain angle $\theta_{cloud}$ from the line of sight, corresponding to the transverse dimension of the CBM cloud, until the duration of the flare $\delta t/t_{tot}$ is compatible with the observation (see Fig. \ref{060607A_flares} middle and lower panels). It is worth to observe that with this procedure we kee the value of {\cal R} constant during the flare. Hence the increase in {\cal R} that we obtained in our previous analysis (see Fig. \ref{060607A_CBM}) is not real but it compensates the fact that spherical approximation is not valid at this stage.

This procedure affects the dynamics of the fireshell, so the light curve after the ``cut'' is meaningless. Nevertheless, it is a confirmation that it is possible to obtain arbitrarily short flares by the interaction with the CBM. 

\subsection{Conclusions}

We presented the analysis of GRB 060607A within the fireshell model \citep{2001ApJ...555L.107R,2001ApJ...555L.113R,2007AIPC..910...55R}. According to the ``canonical GRB'' scenario \citep{2001ApJ...555L.113R,2007AIPC..910...55R} we interpreted the whole prompt emission as the peak of the extended afterglow emission, and the remaining part of the light curve with the decaying tail of the extended afterglow. We found in this second case that the P-GRB is too faint to be detected, as we expected from our interpretation. The theoretical light curves obtained are well in agreement with the observations in all the \emph{Swift} BAT energy bands.

Furthermore, the initial Lorentz gamma factor of the fireshell, obtained adopting the exact solutions of its equations of motions \citep{2005ApJ...633L..13B} and as initial condition the free parameters of the fireshell estimated by the simultaneous analysis of the BAT and XRT light curves, is $\gamma_\circ = 328$. In this preliminary analysis we deal only with the BAT and XRT observations, which are the basic contribution to the afterglow emission according to the fireshell model. We do not deal with the infrared emission that, on the contrary, is used in the current literature to estimate the dynamical quantities of the fireball in the forward external shock regime. Nevertheless, the initial value of Lorentz gamma factor we predict is compatible with the one deduced from the REM observations even under very different assumptions, $\Gamma_\circ \sim 400$ \citep{2007A&A...469L..13M,2007arXiv0710.0727C,2007MNRAS.378.1043J}.

We investigated also the GRB 060607A prompt emission spectra integrated in different time intervals assuming a thermal spectrum in the comoving frame. The results obtained show clearly that, after the correct space-time transformations, both the instantaneous and the time-integrated spectra in the observer frame have nothing to do with a Planckian distribution, but they have a power-law shape, thus confirming our previous analyses \citep{2005ApJ...634L..29B,2006ApJ...645L.109R}.

Finally we analyzed the X-ray flares observed by \emph{Swift} XRT ($0.2-10$ keV) in the early part of the decaying phase of the extended afterglow. According to the fireshell model these flares have the same origin of the prompt emission, namely they are produced by the interaction of the fireshell with overdense CBM. We found that our theoretical light curve is not compatible with the observations since in such regime the one-dimensional approximation fails. Following the results obtained for GRGB 011121 \citep{Venezia_Flares}, we tried to account for the three-dimensional structure of the CBM clouds by ``cutting'' the emission at a certain angle $\theta_{cloud}$ from the line of sight, corresponding to the transverse dimension of the CBM cloud. We obtain in this way a flare whose duration $\delta t/t_{tot}$ is compatible with the observation.

	\section{The Norris \& Bonnel kind of sources: the new class of ``fake - disguised'' short GRBs}

\begin{figure}
\includegraphics[width=\hsize,clip]{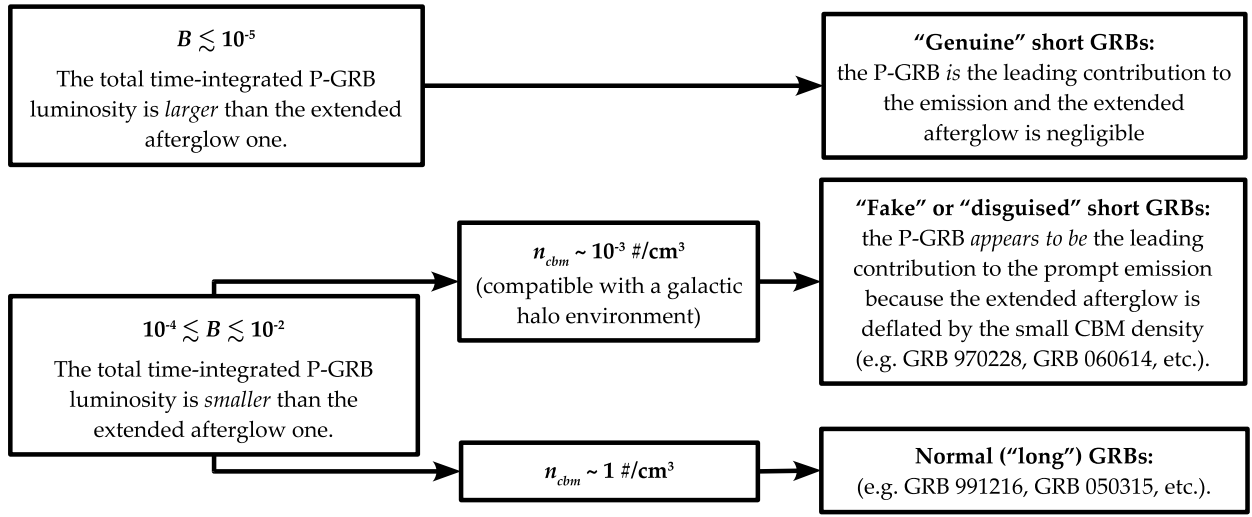}
\caption{A sketch summarizing the ``canonical GRB'' scenario.}
\label{canonical_canonical}
\end{figure}

We now present the theoretical understanding, within the fireshell model, of a new class of sources, pioneered by \citet{2006ApJ...643..266N}. This class is characterized by an occasional softer extended emission after an initial spikelike emission. The softer extended emission has a peak luminosity smaller than the one of the initial spikelike emission. This has misled the understanding of the correct role of the extended afterglow. As shown in the prototypical case of GRB 970228 \citep[see below and Ref.][]{2007A&A...474L..13B}, the initial spikelike emission can be identified with the P-GRB and the softer extended emission with the peak of the extended afterglow. Crucial is the fact that the time-integrated extended afterglow luminosity is much larger than the P-GRB one, and this fact unquestionably identifies GRB 970228 as a canonical GRB with $B > 10^{-4}$. The consistent application of the fireshell model allowed to compute the CBM porosity, filamentary structure and average density which, in that specific case, resulted to be $n_{cbm} \sim 10^{-3}$ particles/cm$^3$ \citep{2007A&A...474L..13B}. This explained the peculiarity of the low extended afterglow peak luminosity and of its much longer time evolution. These features are not intrinsic to the progenitor nor to the black hole, but they uniquely depend on the peculiarly low value of the CBM density, typical of galactic halos. If one takes the same total energy, baryon loading and CBM distribution as in GRB 970228, and rescales the CBM density profile by a constant numerical factor in order to raise its average value from $10^{-3}$ to $1$ particles/cm$^3$, he obtains a GRB with a much larger extended afterglow peak luminosity and a much reduced time scale. Such a GRB would appear a perfect traditional ``long'' GRB following the current literature \citep[see below and Ref.][]{2007A&A...474L..13B}. This has led us to expand the traditional classification of GRBs to three classes: ``genuine'' short GRBs, ``fake'' or ``disguised'' short GRBs, and all the remaining ``canonical'' ones \citep[see Fig.~\ref{canonical_canonical} and Ref.][]{RuCef}.

\subsection{GRB 970228 and a class of GRBs with an initial spikelike emission}

GRB 970228 was detected by the Gamma-Ray Burst Monitor (GRBM, $40$--$700$ keV) and Wide Field Cameras (WFC, $2$--$26$ keV) on board BeppoSAX on February $28.123620$ UT \citep{1998ApJ...493L..67F}. The burst prompt emission is characterized by an initial $5$ s strong pulse followed, after $30$ s, by a set of three additional pulses of decreasing intensity \citep{1998ApJ...493L..67F}. Eight hours after the initial detection, the NFIs on board BeppoSAX were pointed at the burst location for a first target of opportunity observation and a new X-ray source was detected in the GRB error box: this is the first ``afterglow'' ever detected \citep{1997Natur.387..783C}. A fading optical transient has been identified in a position consistent with the X-ray transient \citep{1997Natur.386..686V}, coincident with a faint galaxy with redshift $z=0.695$ \citep{2001ApJ...554..678B}. Further observations by the Hubble Space Telescope clearly showed that the optical counterpart was located in the outskirts of a late-type galaxy with an irregular morphology \citep{1997Natur.387R.476S}.

The BeppoSAX observations of GRB 970228 prompt emission revealed a discontinuity in the spectral index between the end of the first pulse and the beginning of the three additional ones \citep{1997Natur.387..783C,1998ApJ...493L..67F,2000ApJS..127...59F}. The spectrum during the first $3$ s of the second pulse is significantly harder than during the last part of the first pulse \citep{1998ApJ...493L..67F,2000ApJS..127...59F}, while the spectrum of the last three pulses appear to be consistent with the late X-ray afterglow \citep{1998ApJ...493L..67F,2000ApJS..127...59F}. This was soon recognized by \citet{1998ApJ...493L..67F,2000ApJS..127...59F} as pointing to an emission mechanism producing the X-ray afterglow already taking place after the first pulse.

The simultaneous occurrence of an extended afterglow with total time-integrated luminosity larger than the P-GRB one, but with a smaller peak luminosity, is indeed explainable in terms of a peculiarly small average value of the CBM density and not due to the intrinsic nature of the source. In this sense, GRBs belonging to this class are only ``fake'' or ``disguised'' short GRBs. We show that GRB 970228 is a very clear example of this situation. We identify the initial spikelike emission with the P-GRB, and the late soft bump with the peak of the extended afterglow. GRB 970228 shares the same morphology and observational features with the sources analyzed by \citet{2006ApJ...643..266N} as well as with e.g. GRB 050709 \citep{2005Natur.437..855V}, GRB 050724 \citep{2006A&A...454..113C} and GRB 060614 \citep[see next section and Ref.][]{2006Natur.444.1044G}. Therefore, we propose GRB 970228 as a prototype for this new GRB class.

\subsubsection{The analysis of GRB 970228 prompt emission}

\begin{figure}
\includegraphics[width=\hsize,clip]{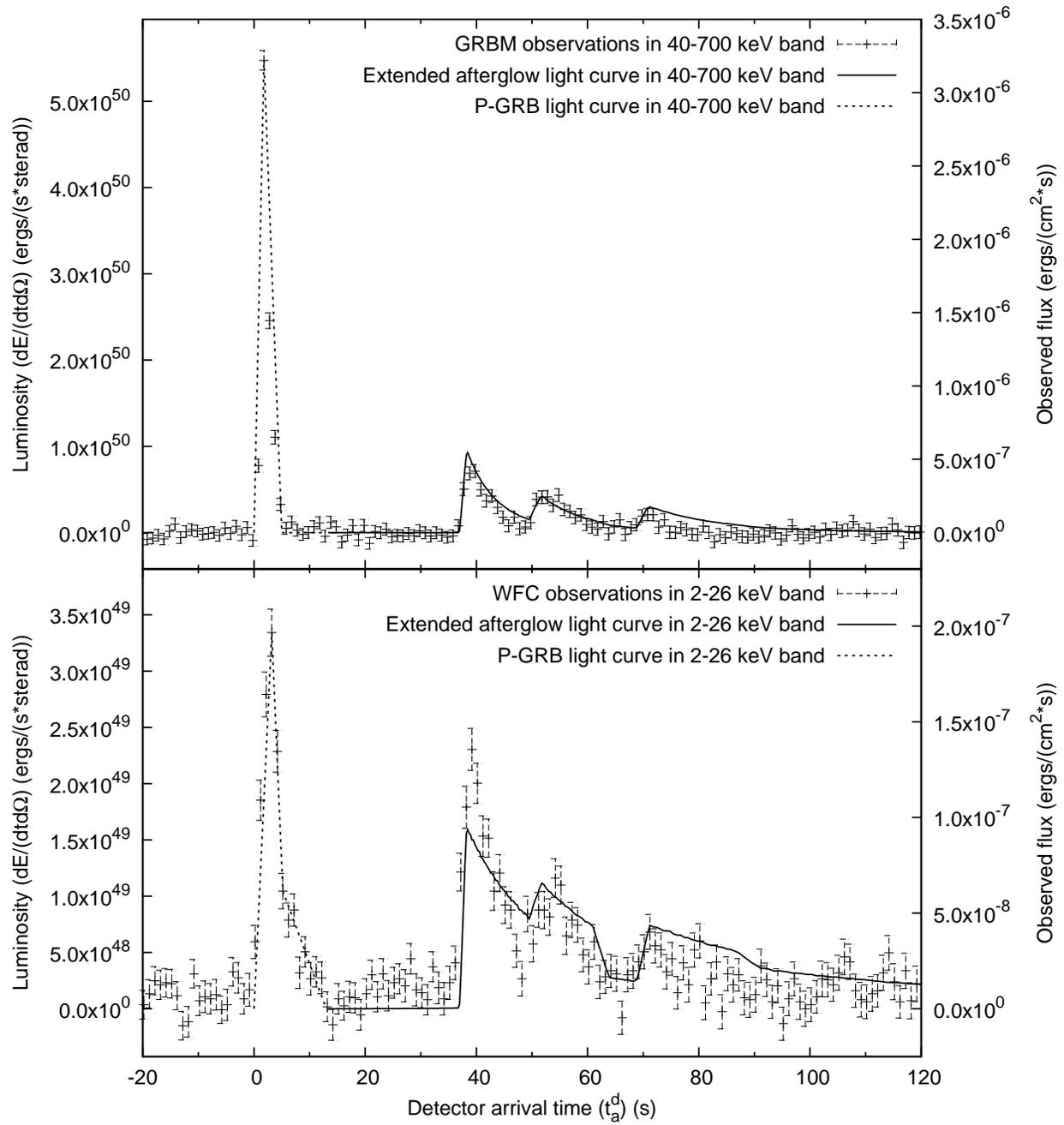}
\caption{The ``canonical GRB'' light curve theoretically computed for the prompt emission of GRB 970228. BeppoSAX GRBM ($40$--$700$ keV, above) and WFC ($2$--$26$ keV, below) light curves (data points) are compared with the extended afterglow peak theoretical ones (solid lines). The onset of the extended afterglow coincides with the end of the P-GRB (represented qualitatively by the dotted lines). For this source we have $B\simeq 5.0\times 10^{-3}$ and $\langle n_{cbm} \rangle \sim 10^{-3}$ particles/cm$^3$. Details in \citet{2007A&A...474L..13B,2008AIPC..966....7B}.}
\label{970228_fit_prompt}
\end{figure}
 
\begin{figure}
\includegraphics[width=\hsize,clip]{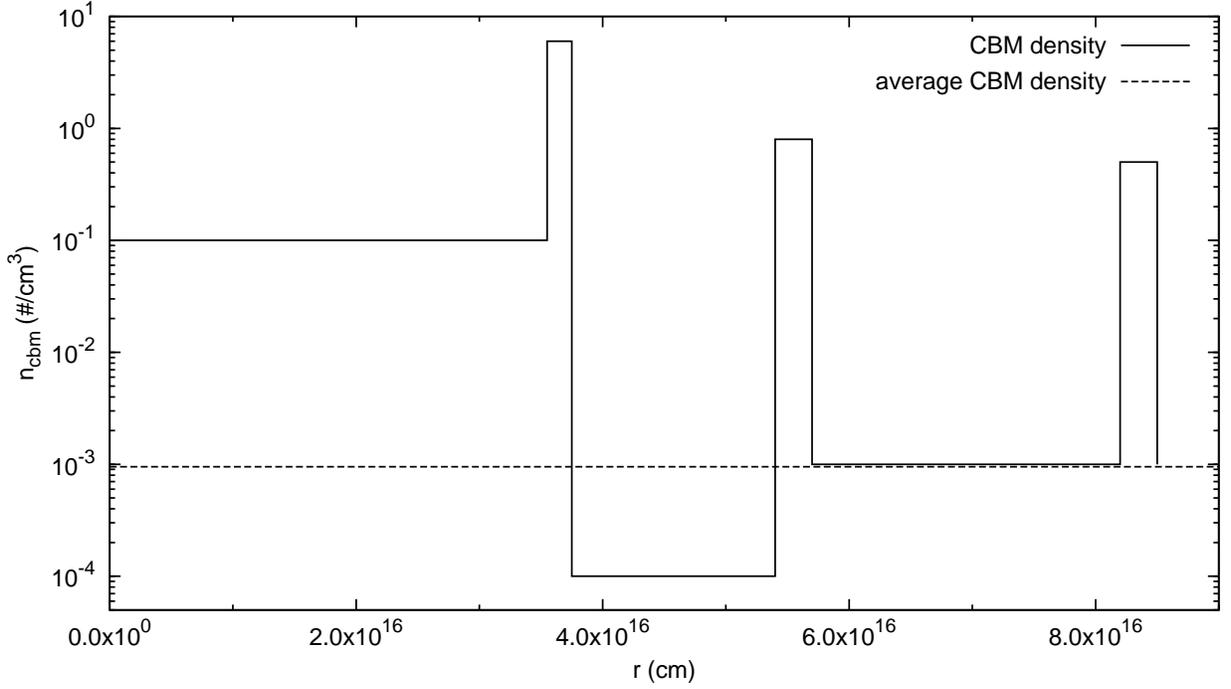}
\caption{The CBM density profile we assumed to reproduce the last three pulses of the GRB 970228 prompt emission (red line), together with its average value $\langle n_{cbm} \rangle = 9.5\times 10^{-4}$ particles/cm$^3$ (green line).}
\label{mask}
\end{figure}

In Fig.~\ref{970228_fit_prompt} we present the theoretical fit of BeppoSAX GRBM ($40$--$700$ keV) and WFC ($2$--$26$ keV) light curves of GRB 970228 prompt emission \citep{1998ApJ...493L..67F}. Within our ``canonical GRB'' scenario we identify the first main pulse with the P-GRB and the three additional pulses with the extended afterglow peak emission, consistently with the above mentioned observations by \citet{1997Natur.387..783C} and \citet{1998ApJ...493L..67F}. Such last three pulses have been reproduced assuming three overdense spherical CBM regions (see Fig.~\ref{mask}) with a very good agreement (see Fig.~\ref{970228_fit_prompt}).

We therefore obtain for the two parameters characterizing the source in our model $E_{e^\pm}^{tot}=1.45\times 10^{54}$ erg and $B = 5.0\times 10^{-3}$. This implies an initial $e^\pm$ plasma created between the radii $r_1 = 3.52\times10^7$ cm and $r_2 = 4.87\times10^8$ cm with a total number of $e^{\pm}$ pairs $N_{e^\pm} = 1.6\times 10^{59}$ and an initial temperature $T = 1.7$ MeV. The theoretically estimated total isotropic energy emitted in the P-GRB is $E_{P-GRB}=1.1\% E_{e^\pm}^{tot}=1.54 \times 10^{52}$ erg, in excellent agreement with the one observed in the first main pulse ($E_{P-GRB}^{obs} \sim 1.5 \times 10^{52}$ erg in $2-700$ keV energy band, see Fig.~\ref{970228_fit_prompt}), as expected due to their identification. After the transparency point at $r_0 = 4.37\times 10^{14}$ cm from the progenitor, the initial Lorentz gamma factor of the fireshell is $\gamma_0 = 199$. On average, during the extended afterglow peak emission phase we have for the CBM $\langle {\cal R} \rangle = 1.5\times 10^{-7}$ and $\langle n_{cbm} \rangle = 9.5\times 10^{-4}$ particles/cm$^3$. This very low average value for the CBM density is compatible with the observed occurrence of GRB 970228 in its host galaxy's halo \citep{1997Natur.387R.476S,1997Natur.386..686V,2006MNRAS.367L..42P} and it is crucial in explaining the light curve behavior.

The values of $E_{e^\pm}^{tot}$ and $B$ we determined are univocally fixed by two tight constraints. The first one is the total energy emitted by the source all the way up to the latest extended afterglow phases (i.e. up to $\sim 10^6$ s). The second one is the ratio between the total time-integrated luminosity of the P-GRB and the corresponding one of the whole extended afterglow (i.e. up to $\sim 10^6$ s). In particular, in GRB 970228 such a ratio results to be $\sim 1.1\%$ (see Fig. \ref{f2}). However, the P-GRB peak luminosity actually results to be much more intense than the extended afterglow one (see Fig.~\ref{970228_fit_prompt}). This is due to the very low average value of the CBM density $\langle n_{cbm} \rangle = 9.5\times 10^{-4}$ particles/cm$^3$, which produces a less intense extended afterglow emission. Since the extended afterglow total time-integrated luminosity is fixed, such a less intense emission lasts longer than what we would expect for an average density $\langle n_{cbm} \rangle \sim 1$ particles/cm$^3$.

\subsubsection{Rescaling the CBM density}

\begin{figure}
\includegraphics[width=\hsize,clip]{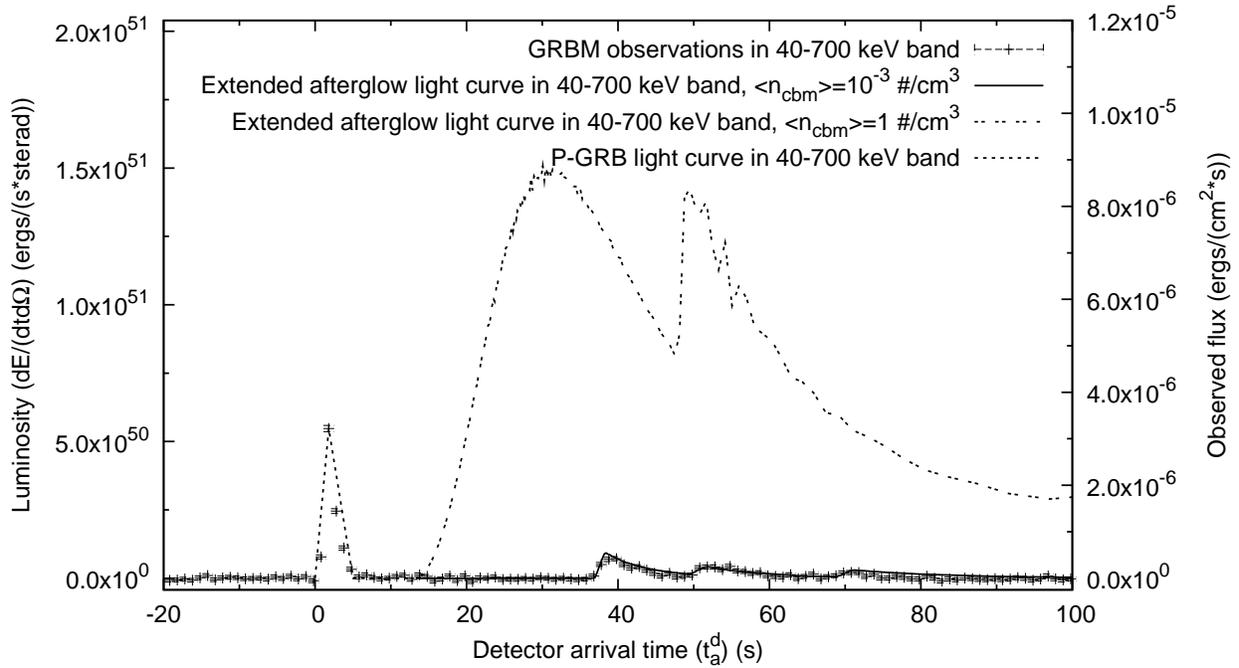}
\caption{The theoretical fit of the BeppoSAX GRBM observations (solid line, see Fig. \ref{970228_fit_prompt}) is compared with the extended afterglow light curve in the $40$--$700$ keV energy band obtained rescaling the CBM density to $\langle n_{cbm} \rangle = 1$ particle/cm$^3$ keeping constant its shape and the values of the fundamental parameters of the theory $E_{e^\pm}^{tot}$ and $B$ (double dotted line). The P-GRB duration and luminosity (dotted line), depending only on $E_{e^\pm}^{tot}$ and $B$, are not affected by this process of rescaling the CBM density.}
\label{picco_n=1}
\end{figure}

We present now an explicit example in order to probe the crucial role of the average CBM density in explaining the relative intensities of the P-GRB and of the extended afterglow peak in GRB 970228. We keep fixed the basic parameters of the source, namely the total energy $E_{e^\pm}^{tot}$ and the baryon loading $B$, therefore keeping fixed the P-GRB and the extended afterglow total time-integrated luminosities. Then we rescale the CBM density profile given in Fig. \ref{mask} by a constant numerical factor in order to raise its average value to the standard one $\langle n_{ism} \rangle = 1$ particle/cm$^3$. We then compute the corresponding light curve, shown in Fig. \ref{picco_n=1}.

We notice a clear enhancement of the extended afterglow peak luminosity with respect to the P-GRB one in comparison with the fit of the observational data presented in Fig. \ref{970228_fit_prompt}. The two light curves actually crosses at $t_a^d \simeq 1.8\times 10^4$ s since their total time-integrated luminosities must be the same. The GRB ``rescaled'' to $\langle n_{ism} \rangle = 1$ particle/cm$^3$ appears to be totally similar to, e.g., GRB 050315 \citep{2006ApJ...645L.109R} and GRB 991216 \citep{2003AIPC..668...16R,2004IJMPD..13..843R,2005AIPC..782...42R}.

It is appropriate to emphasize that, although the two underlying CBM density profiles differ by a constant numerical factor, the two extended afterglow light curves in Fig. \ref{picco_n=1} do not. This is because the absolute value of the CBM density at each point affects in a non-linear way all the following evolution of the fireshell due to the feedback on its dynamics \citep{2005ApJ...633L..13B}. Moreover, the shape of the surfaces of equal arrival time of the photons at the detector (EQTS) is strongly elongated along the line of sight \citep{2005ApJ...620L..23B}. Therefore photons coming from the same CBM density region are observed over a very long arrival time interval.

\subsubsection{GRB 970228 and the Amati relation}

We turn now to the ``Amati relation'' \citep{2002A&A...390...81A,2006MNRAS.372..233A} between the isotropic equivalent energy emitted in the prompt emission $E_{iso}$ and the peak energy of the corresponding time-integrated spectrum $E_{p,i}$ in the source rest frame. It has been shown by \citet{2002A&A...390...81A,2006MNRAS.372..233A} that this correlation holds for almost all the ``long'' GRBs which have a redshift and an $E_{p,i}$ measured, but not for the ones classified as ``short'' \citep{2006MNRAS.372..233A}. If we focus on the ``fake'' or ``disguised'' short GRBs, namely the GRBs belonging to this new class, at least in one case \citep[GRB 050724, see Ref.][]{2006A&A...454..113C} it has been shown that the correlation is recovered if also the extended emission is considered \citep{amatiIK}. 

It clearly follows from our treatment that for the ``canonical GRBs'' with large values of the baryon loading and high $\left\langle n_{cbm}\right\rangle$, which presumably are most of the GRBs for which the correlation holds, the leading contribution to the prompt emission is the extended afterglow peak emission. The case of the ``fake'' or ``disguised'' short GRBs is completely different: it is crucial to consider separately the two components since the P-GRB contribution to the prompt emission in this case is significant.

\begin{figure}
\includegraphics[width=\hsize,clip]{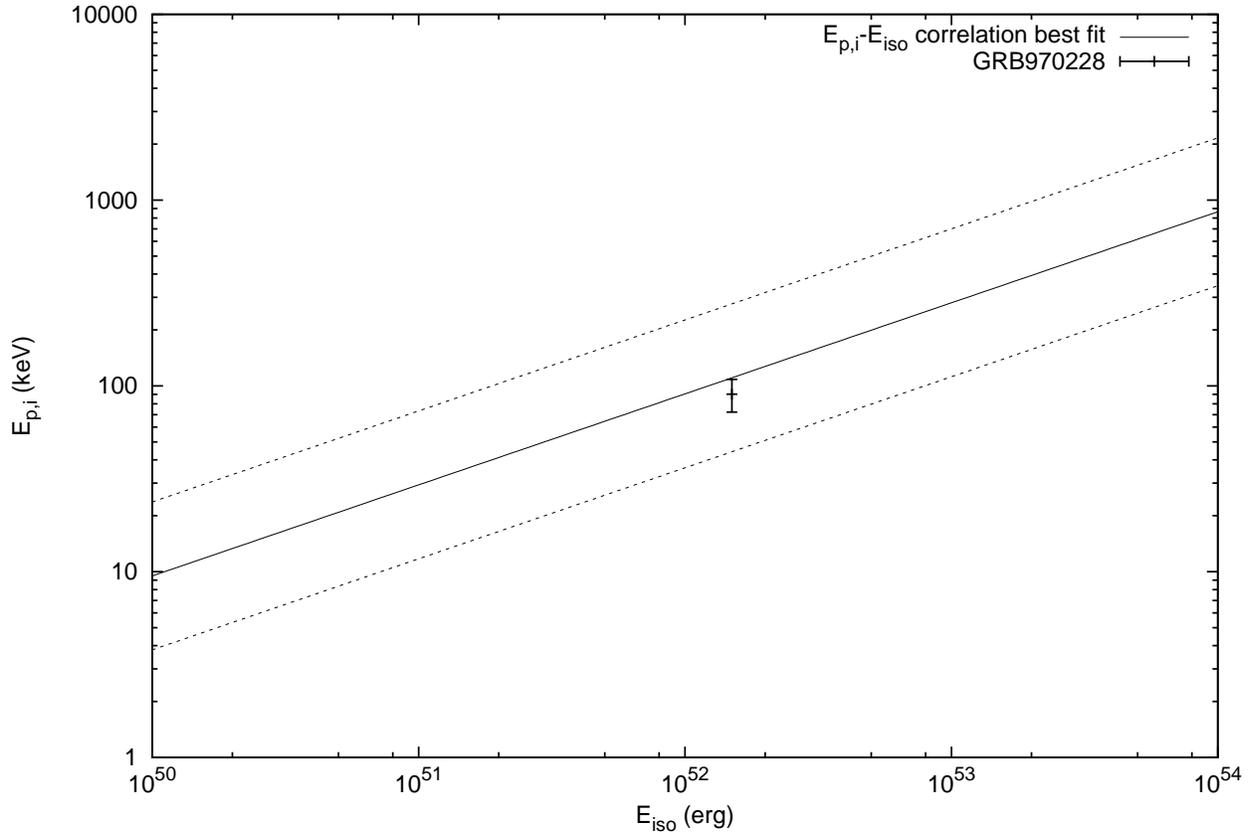}
\caption{The estimated values for $E_{p,i}$ and $E_{iso}$ obtained by our analysis (black dot) compared with the ``Amati relation'' \citep{2002A&A...390...81A}: the solid line is the best fitting power law \citep{2006MNRAS.372..233A} and the dashed lines delimit the region corresponding to a vertical logarithmic deviation of $0.4$ \citep{2006MNRAS.372..233A}. The uncertainty in the theoretical estimated value for $E_{p,i}$ has been assumed conservatively as $20\%$.}
\label{amati}
\end{figure}

To test this scenario, we evaluated from our fit of GRB 970228 $E_{iso}$ and $E_{p,i}$ only for the extended afterglow peak emission component, i.e. from $t_a^d= 37$ s to $t_a^d= 81.6$ s. We found an isotropic energy emitted in the $2$--$400$ keV energy band $E_{iso}=1.5 \times 10^{52}$ erg, and $E_{p,i}=90.3$ keV. As it is clearly shown in Fig. \ref{amati}, the sole extended afterglow component of GRB 970228 prompt emission is in perfect agreement with the Amati relation. If this behavior is confirmed for other GRBs belonging to this new class, this will enforce our identification of the ``fake'' or ``disguised'' short GRBs. This result will also provide a theoretical explanation for the the apparent absence of such correlation for the initial spikelike component in the different nature of the P-GRB.

\subsection{Conclusions}

We conclude that GRB 970228 is a ``canonical GRB'' with a large value of the baryon loading quite near to the maximum $B \sim 10^{-2}$ (see Fig. \ref{f2}). The difference with e.g. GRB 050315 \citep{2006ApJ...645L.109R} or GRB 991216 \citep{2003AIPC..668...16R,2004IJMPD..13..843R, 2005AIPC..782...42R} is the low average value of the CBM density $\langle n_{cbm} \rangle \sim 10^{-3}$ particles/cm$^3$ which deflates the extended afterglow peak luminosity. Hence, the predominance of the P-GRB, coincident with the initial spikelike emission, over the extended afterglow is just apparent: $98.9\%$ of the total time-integrated luminosity is indeed in the extended afterglow component. Such a low average CBM density is consistent with the occurrence of GRB 970228 in the galactic halo of its host galaxy \citep{1997Natur.387R.476S,1997Natur.386..686V}, where lower CBM densities have to be expected \citep{2006MNRAS.367L..42P}.

We propose GRB 970228 as the prototype for the new class of GRBs comprising GRB 060614 and the GRBs analyzed by \citet{2006ApJ...643..266N}. We naturally explain the hardness and the absence of spectral lag in the initial spikelike emission with the physics of the P-GRB originating from the gravitational collapse leading to the black hole formation. The hard-to-soft behavior in the extended afterglow is also naturally explained by the physics of the relativistic fireshell interacting with the CBM, clearly evidenced in GRB 031203 \citep{2005ApJ...634L..29B} and in GRB 050315 \citep{2006ApJ...645L.109R}. Also justified is the applicability of the Amati relation to the sole extended afterglow component \citep[see Refs.][]{2006MNRAS.372..233A,amatiIK}.

This class of GRBs with $z \sim 0.4$ appears to be nearer than the other GRBs detected by \emph{Swift} \citep[$z \sim 2.3$, see Ref.][]{2006NCimB.121.1061G}. This may be explained by the extended afterglow peak luminosity deflation. The absence of a jet break in those afterglows has been pointed out \citep{2006A&A...454..113C,2006A&A...454L.123W}, consistently with our spherically symmetric approach. Their association with non-star-forming host galaxies appears to be consistent with the merging of a compact object binary \citep{2005Natur.438..994B,2005Natur.437..845F}. It is here appropriate, however, to caution on this conclusion, since the association of GRB 060614 and GRB 970228 with the explosion of massive stars is not excluded \citep{2006Natur.444.1050D,2000ApJ...536..185G}.

Most of the sources of this class appear indeed not to be related to bright ``Hypernovae'', to be in the outskirts of their host galaxies \citep[][see above]{2005Natur.437..845F} and a consistent fraction of them are in galaxy clusters with CBM densities $\langle n_{cbm} \rangle \sim 10^{-3}$ particles/cm$^3$ \citep[see e.g. Ref.][]{2003ApJ...586..135L,2007ApJ...660..496B}. This suggests a spiraling out binary nature of their progenitor systems \citep{KMG11} made of neutron stars and/or white dwarfs leading to a black hole formation.

Moreover, we verified the applicability of the Amati relation to the sole extended afterglow component in GRB 970228 prompt emission, in analogy with what happens for some of the GRBs belonging to this new class. In fact it has been shown by \citet{2006MNRAS.372..233A,amatiIK} that the ``fake'' or ``disguised'' short GRBs do not fulfill the $E_{p,i}$--$E_{iso}$ correlation when the sole spiklike emission is considered, while they do if the long soft bump is included. Since the spikelike emission and the soft bump contributions are comparable, it is natural to expect that the soft bump alone will fulfill the correlation as well.

Within our ``canonical GRB'' scenario the sharp distinction between the P-GRB and the extended afterglow provide a natural explanation for the observational features of the two contributions. We naturally explain the hardness and the absence of spectral lag in the initial spikelike emission with the physics of the P-GRB originating from the gravitational collapse leading to the black hole formation. The hard-to-soft behavior in the extended afterglow is also naturally explained by the physics of the relativistic fireshell interacting with the CBM, clearly evidenced in GRB 031203 \citep{2005ApJ...634L..29B} and in GRB 050315 \citep{2006ApJ...645L.109R}. Therefore, we expect naturally that the $E_{p,i}$--$E_{iso}$ correlation holds only for the extended afterglow component and not for the P-GRB. Actually we find that the correlation is recovered for the extended afterglow peak emission of GRB 970228.

In the original work by \citet{2002A&A...390...81A,2006MNRAS.372..233A} only the prompt emission is considered and not the late afterglow one. In our theoretical approach the extended afterglow peak emission contributes to the prompt emission and continues up to the latest GRB emission. Hence, the meaningful procedure within our model to recover the Amati relation is to look at a correlation between the total isotropic energy and the peak of the time-integrated spectrum of the whole extended afterglow. A first attempt to obtain such a correlation has already been performed using GRB 050315 as a template, giving very satisfactory results (see section \emph{Theoretical background for GRBs' empirical correlations}).

	\section{The ``fireshell'' model and GRB progenitors}\label{progenitors}

``Long'' GRBs are traditionally related in the current literature to the idea of a single progenitor, identified as a ``collapsar'' \citep{1993ApJ...405..273W}. Similarly, short GRBs are assumed to originate from binary mergers formed by white dwarfs, neutron stars, and black holes in all possible combinations \citep[see e.g. Refs.][and references therein]{1984SvAL...10..177B,1986ApJ...308L..43P,1986ApJ...308L..47G,1989Natur.340..126E,2005RvMP...76.1143P,2006RPPh...69.2259M}. It has been also suggested that short and long GRBs originate from different galaxy types. In particular, short GRBs are proposed to be associated with galaxies with low specific star forming rate \citep[see e.g. Ref.][]{2009ApJ...690..231B}. Some evidences against such a scenario have been however advanced, due to the small sample size and the different estimates of the star forming rates \citep[see e.g. Ref.][]{2008arXiv0803.2718S}. However, the understanding of GRB structure and of its relation to the CBM distribution, within the fireshell model, leads to a more complex and interesting perspective than the one in the current literature.

\begin{figure}
\includegraphics[width=\hsize,clip]{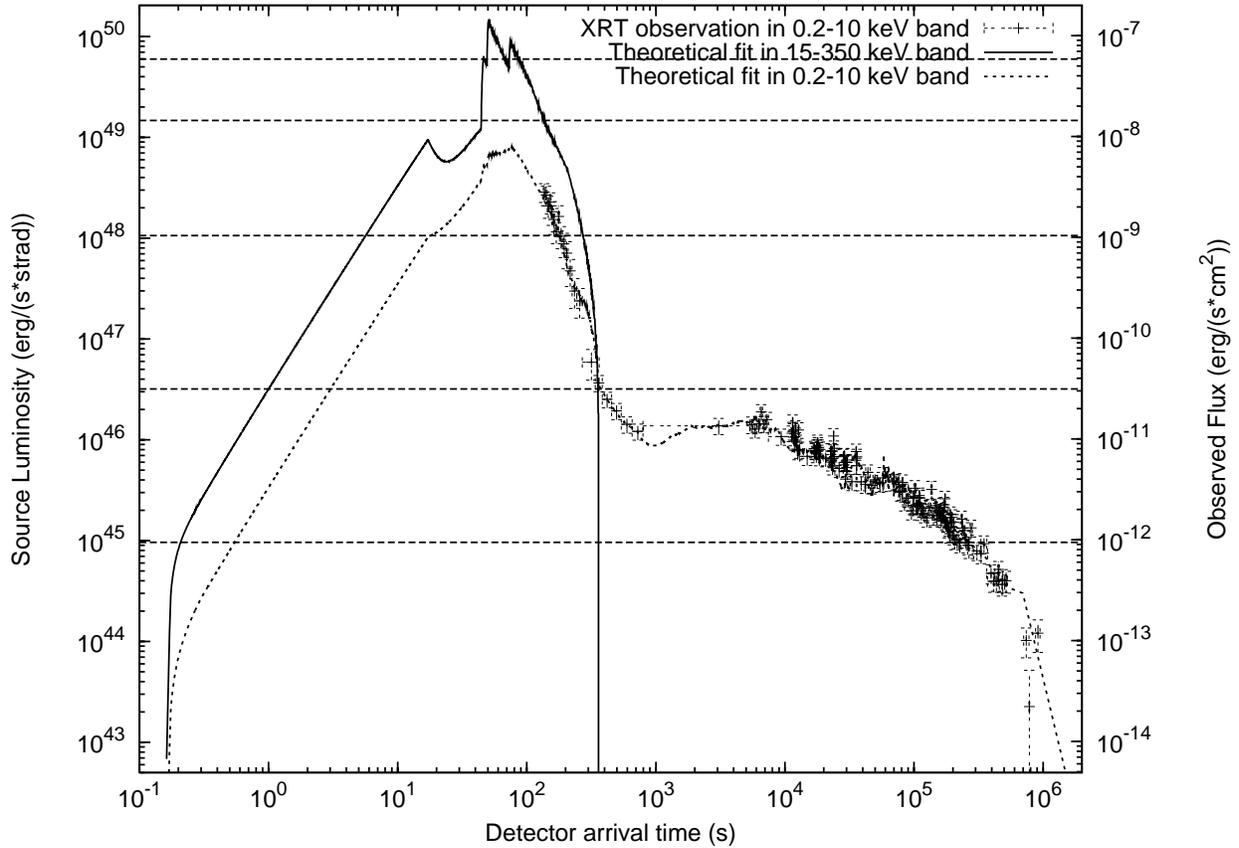}
\caption{The theoretical light curves in the $15-150$ keV (solid line) and $0.2-10$ kev (dotted line) energy bands compared with XRT observations of GRB 050315 \citep{2006ApJ...638..920V}. The horizontal dashed lines correspond to different possible instrumental thresholds. It is clear that long GRB durations are just functions of the observational threshold. Details in \citet{Venezia_Orale}.}
\label{global_th}
\end{figure}

The first general conclusion of the ``fireshell'' model \citep{2001ApJ...555L.113R} is that, while the time scale of ``short'' GRBs is indeed intrinsic to the source, this does not happen for the ``long'' GRBs: their time scale is clearly only a function of the instrumental noise threshold. This has been dramatically confirmed by the observations of the Swift satellite \citep[see Fig. \ref{global_th} and Ref.][]{Venezia_Orale}. Among the traditional classification of ``long'' GRBs we distinguish two different sub-classes of events, none of which originates from collapsars.

The first sub-class contains ``long'' GRBs particularly weak ($E_{iso} \sim 10^{50}$ erg) and associated with Supernovae (SNe) Ib/c. In fact, it has been often proposed that such GRBs, only observed at smaller redshift $0.0085 < z < 0.168$, form a different class, less luminous and possibly much more numerous than the high luminosity GRBs at higher redshift \citep{2006Natur.442.1011P,2004Natur.430..648S,2007ApJ...658L...5M,2006AIPC..836..367D}. Therefore in the current literature they have been proposed to originate from a separate class of progenitors \citep{2007ApJ...662.1111L,2006ApJ...645L.113C}. Within our ``fireshell'' model, they originate in a binary system formed by a neutron star, close to its critical mass, and a companion star, evolved out of the main sequence. They produce GRBs associated with SNe Ib/c, via the ``induced gravitational collapse'' process \citep{2001ApJ...555L.117R}. The low luminosity of these sources is explained by the formation of a black hole with the smallest possible mass: the one formed by the collapse of a just overcritical neutron star \citep{Mosca_Orale,2007A&A...471L..29D}.

A second sub-class of ``long'' GRBs originates from merging binary systems, formed either by two neutron stars or a neutron star and a white dwarf. A prototypical example of such systems is GRB 970228. The binary nature of the source is inferred by its migration from its birth location in a star forming region to a low density region within the galactic halo, where the final merging occurs \citep{2007A&A...474L..13B}. The location of such a merging event in the galactic halo is indeed confirmed by optical observations of the GRB 970228 afterglow  \citep{1997Natur.387R.476S,1997Natur.386..686V}. The crucial point is that, as recalled above, GRB 970228 is a ``canonical'' GRB with $B > 10^{-4}$ ``disguised'' as a short GRB. We are going to see in the following that GRB 060614 also comes from such a progenitor class.

If the binary merging would occur in a region close to its birth place, with an average density of $1$ particle/cm$^3$, the GRB would appear as a traditional high-luminosity ``long'' GRB, of the kind currently observed at higher redshifts (see above, Fig. \ref{picco_n=1}), similar to, e.g., GRB 050315 \citep{2006ApJ...645L.109R}.

Within our approach, therefore, there is the distinct possibility that all GRB progenitors are formed by binary systems, composed by neutron stars, white dwarfs, or stars evolved out of the main sequence, in different combinations.

The case of the ``genuine'' short GRBs is currently being examined within the ``fireshell'' model.

\subsection{GRB 060614: a ``fake'' or ``disguised'' short GRB from a merging binary system}

GRB 060614 \citep{2006Natur.444.1044G,2007A&A...470..105M} has imposed to the general attention of the Gamma-Ray Burst's (GRB's) scientific community because it is the first clear example of a nearby ($z=0.125$), long GRB not associated with a bright Ib/c Supernova (SN) \citep{2006Natur.444.1050D,2006Natur.444.1053G}. It has been estimated that, if present, the SN-component should be about $200$ times fainter than the archetypal SN 1998bw associated to GRB 980425; moreover, it would also be fainter (at least $30$ times) than any stripped-envelope SN ever observed \citep{2006AJ....131.2233R}.

Within the standard scenario, long duration GRBs ($T_{90} > 2$ s) are thought to be produced by SN events during the collapse of massive stars in star forming regions \citep[``collapsar'', see Ref.][]{1993ApJ...405..273W}. The observations of broad-lined and bright type Ib/c SNe associated with GRBs are often reported to favor this scenario \citep[see Ref.][and references therein]{2006ARA&A..44..507W}. The \emph{ansatz} has been advanced that every long GRB should have a SN associated with it \citep{2007ApJ...655L..25Z}. Consequently, in all nearby long GRBs ($z \leq 1$) the SN emission should be observed.

For these reasons the case of GRB 060614 is indeed revolutionary. Some obvious hypothesis have been proposed and ruled out: the chance superposition with a galaxy at low redshift \citep{2006Natur.444.1053G} and the strong dust obscuration and extinction \citep{2006Natur.444.1047F}. Appeal has been made to the possible occurrence of an unusually low luminosity stripped-envelope core-collapse SN \citep{2006Natur.444.1050D}.

The second novelty of GRB 060614 is that it challenges the traditional separation between Long Soft GRBs and Short Hard GRBs. Traditionally \citep{1992grbo.book..161K,1992AIPC..265..304D}, the ``short'' GRBs have $T_{90} < 2$ s, present an harder spectrum and negligible spectral lag, and are assumed to originate from merging of two compact objects, i.e. two neutron stars or a neutron star and a black hole \citep[see e.g. Ref.][and references therein]{1984SvAL...10..177B,1986ApJ...308L..43P,1986ApJ...308L..47G,1989Natur.340..126E,2005RvMP...76.1143P,2006RPPh...69.2259M}. GRB 060614 lasts about one hundred seconds \citep[$T_{90}=(102 \pm 5)$ s; see Ref.][]{2006Natur.444.1044G}, it fulfills the $E_{p}^{rest}-E_{iso}$ correlation \citep{2007A&A...463..913A}, and therefore it should be traditionally classified as a ``long'' GRB. However, its morphology is different from typical long GRBs, similar to the one of GRB 050724, traditionally classified as a short GRB \citep{2007ApJ...655L..25Z,2005Natur.437..822P}. Its optical afterglow luminosity is intermediate between the traditional long and short ones \citep{2008arXiv0804.1959K}. Its host galaxy has a moderate specific star formation rate \citep[$R_{Host}\approx2M_{s}y^{-1}(L^{*})^{-1}$, $M_{vHost}\approx-15.5$; see  Refs.][]{2006Natur.444.1047F,2006Natur.444.1050D}. The spectral lag in its light curves is very small or absent \citep{2006Natur.444.1044G}. All these features are typical of the short GRBs.

A third peculiarity of GRB 060614 is that its $15$--$150$ keV light curve presents a short, hard and multi-peaked episode (about $5$ s). Such an episode is followed by a softer, prolonged emission that manifests a strong hard to soft evolution in the first $400$ s of data \citep{2007A&A...470..105M}. The total fluence in the $15$--$150$ keV energy band is $F=(2.17\pm0.04)\times10^{-5}$ erg/cm$^2$, the 20\% emitted during the initial spikelike emission, where the peak luminosity reaches the value of $300$ keV before decreasing until $8$ keV during the BAT-XRT overlap time (about $80$ s).

These apparent contradictions find a natural explanation in the framework of the ``fireshell'' model. Within the fireshell model, the occurrence of a GRB-SN is not a necessity. The origin of all GRBs is traced back to the formation of a black hole, either occurring in a single process of gravitational collapse, or in a binary system composed by a neutron star and a companion star evolved out of the main sequence, or in a merging binary system composed by neutron stars and/or white dwarfs in all possible combinations. The occurrence of a GRB-SN is indeed only one of the possibilities, linked, for example, to the process of ``induced gravitational collapse'' \citep{2001ApJ...555L.117R,Mosca_Orale,2007A&A...471L..29D}.

We here show how the ``fireshell'' model can explain all the above mentioned GRB 060614 peculiarities and solve the apparent contradictions. In doing so, we also infer constraints on the astrophysical nature of the GRB 060614 progenitors. In turn, these conclusions lead to a new scenario for all GRBs. We can confirm a classification of GRBs in ``genuine'' short, ``fake'' or ``disguised'' short, and, finally, all the remaining ``canonical'' GRBs. The connection between this new classification and the nature of GRB progenitors is quite different from the traditional one in the current literature.

\subsubsection{The fit of the observed luminosity}

\begin{figure}
\includegraphics[width=\hsize]{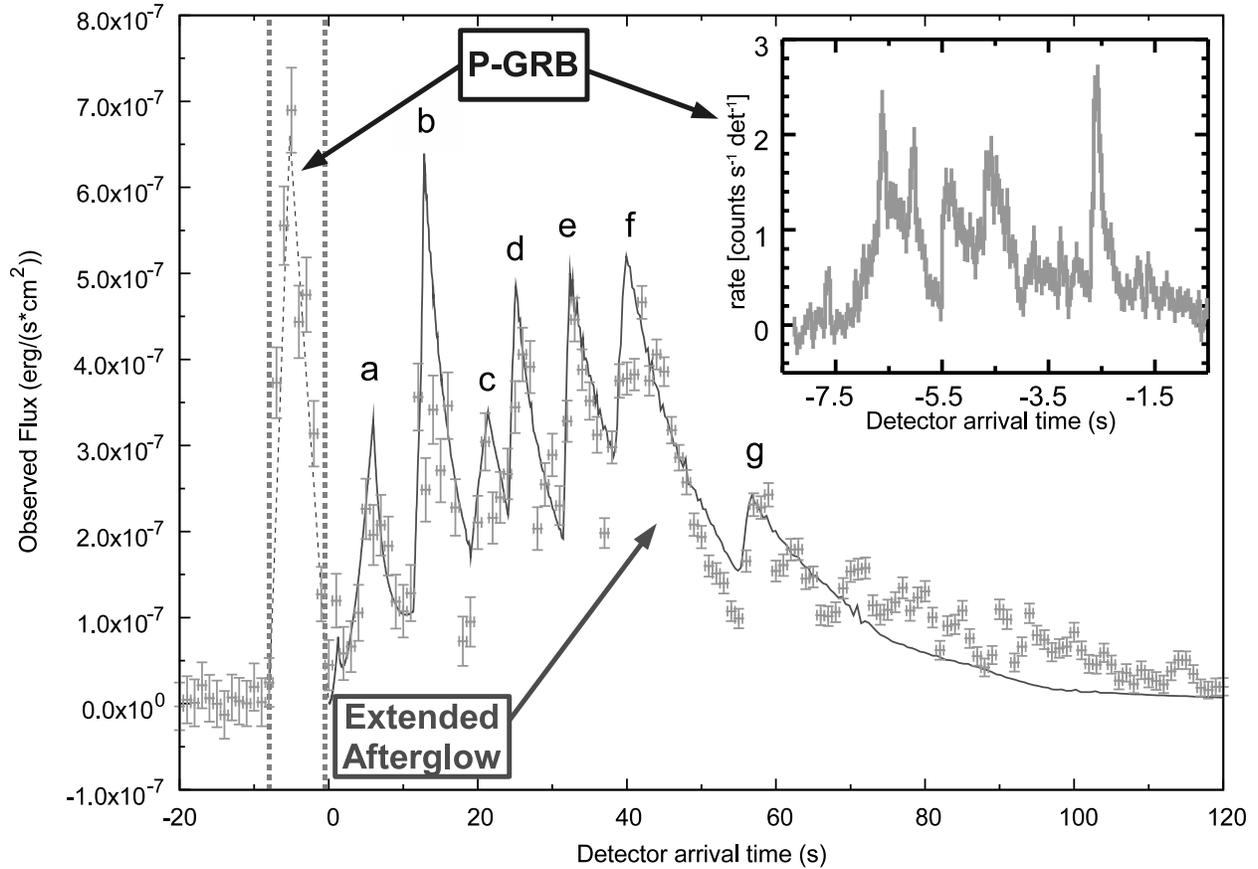}
\caption{The BAT $15$--$150$ keV light curve (points) at $1$ s time resolution is compared with the corresponding theoretical extended afterglow light curve we compute (solid line). The onset of the extended afterglow is at the end of the P-GRB (qualitatively sketched in dashed lines and delimited by dashed tick vertical lines). Therefore the zero of the temporal axis is shifted by $5.5$ s with respect to the BAT trigger time. The peaks of the extended afterglow light curves are labeled to match them with the corresponding CBM density peak in Fig. \ref{f4a}. In the upper right corner there is an enlargement of the P-GRB at $50$ms time resolution \citep[reproduced from Ref.][]{2007A&A...470..105M}, showing its structure.}
\label{f2a}
\end{figure}

\begin{figure}
\includegraphics[width=\hsize]{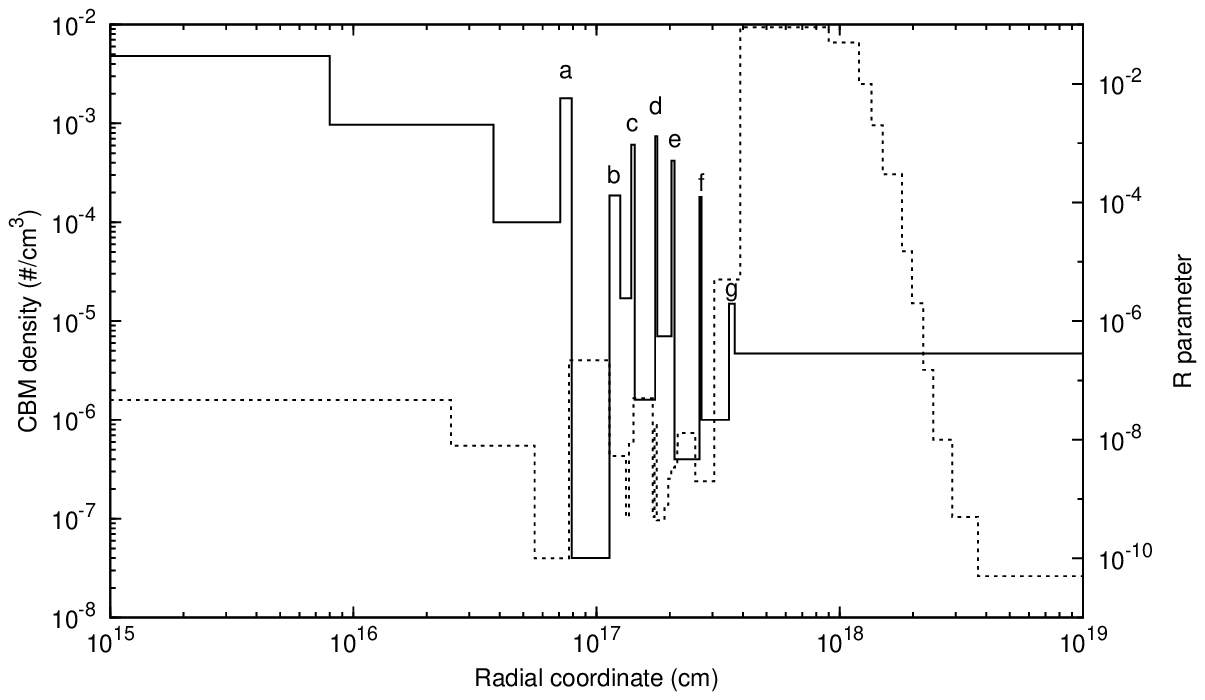}
\caption{Here are the plot of the effective CBM density (solid line) and of the ${\cal R}$ parameter (dotted line) versus the radial coordinate of the shell. The CBM density peaks are labeled to match them with the corresponding extended afterglow light curve peaks in Fig. \ref{f2a}. They corresponds to filaments of characteristic size $\Delta r \sim 10^{15}$ cm and density contrast $\Delta n_{cbm}/\langle n_{cbm} \rangle \sim 20$ particles/cm$^3$.}
\label{f4a}
\end{figure}

\begin{figure}
\includegraphics[width=\hsize]{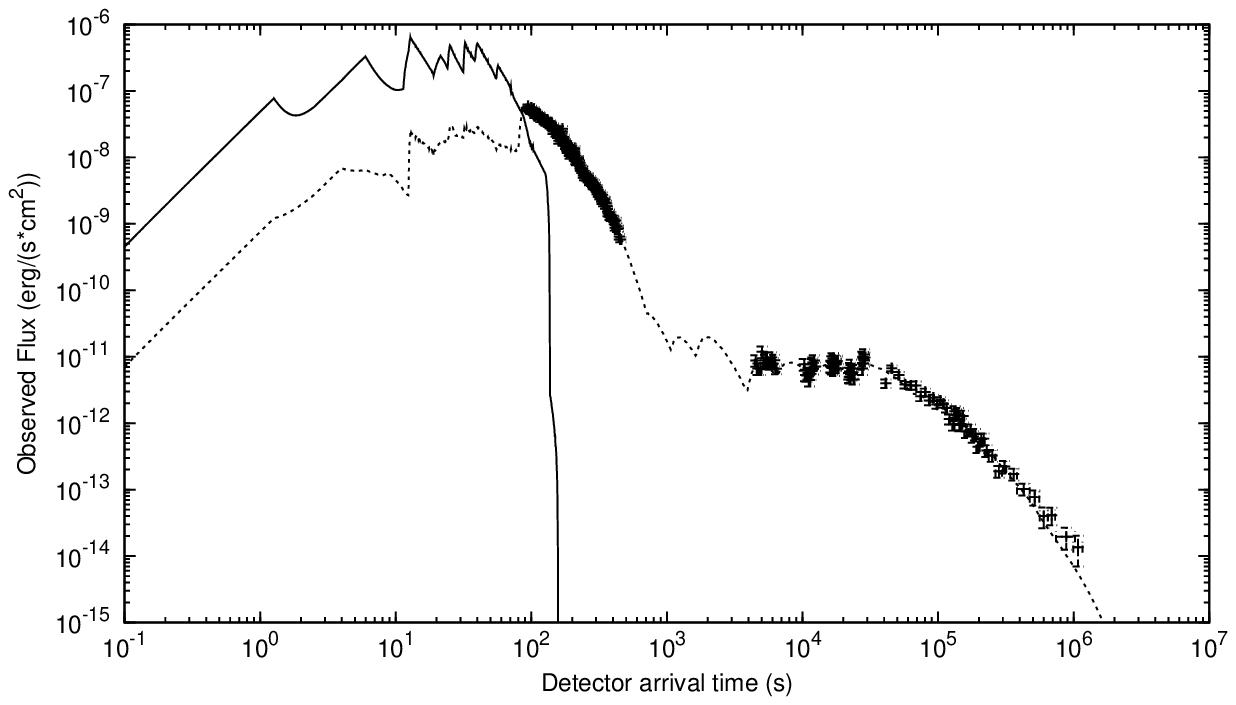}
\caption{The XRT $0.2$--$10$ keV light curve (points) is compared with the corresponding theoretical extended afterglow light curve we compute (dotted line). Also in this case we have a good correspondence between data and theoretical results. For completeness, the solid line shows again the theoretical extended afterglow light curve in the $15$--$150$ keV energy range presented in Fig. \ref{f2a}.}
\label{f4}
\end{figure}

In this scenario, GRB 060614 is naturally interpreted as a ``disguised'' short GRB. We have performed the analysis of the observed light curves in the $15$--$150$ keV energy band, corresponding to the $\gamma$-ray emission observed by the BAT instrument on the Swift satellite, and in the $0.2$--$10$ keV energy band, corresponding to the X-ray component from the XRT instrument on Swift satellite. We do not address in this paper the issue of the optical emission, that represent less than 10\% of the total energy of the GRB. From this fit (see Figs. \ref{f2a}, \ref{f4}) we have derived the total initial energy $E_{tot}^{e^\pm}$, the value of $B$ as well as the effective CBM distribution (see Fig. \ref{f4a}). We find $E_{tot}^{e^\pm}=2.94\times10^{51}$ erg, that accounts for the bolometric emission of both the P-GRB and the extended afterglow. Such a value is compatible with the observed $E_{iso} \simeq 2.5\times10^{51}$ erg \citep{2006Natur.444.1044G}. The value of $B$ is $B=2.8\times10^{-3}$, that corresponds to the lowest one of all the GRBs we have examined (see Fig. \ref{f2}). It corresponds to a canonical GRB with a very clear extended afterglow predominance over the P-GRB. From the model, having determined $E_{tot}^{e^\pm}$ and $B$, we can compute the theoretical expected P-GRB energetics $E_{P-GRB}$ \citep{2001ApJ...555L.113R}. We obtain $E_{P-GRB} \simeq 1.15 \times 10^{50}$ erg, that is in good agreement with the observed one $E_{iso,1p} \simeq 1.18\times10^{50}$ erg \citep{2006Natur.444.1044G}. The Lorentz Gamma Factor at the transparency results to be $\gamma_\circ=346$, one of the highest of all the GRBs we have examined.

In Fig. \ref{f2a} we plot the comparison between the BAT observational data of the GRB 0606014 prompt emission in the $15$--$150$ keV energy range and the P-GRB and extended afterglow light curves computed within our model. The temporal variability of the extended afterglow peak emission is due to the inhomogeneities in the effective CBM density (see Figs. \ref{f2a}, \ref{f4a}). Toward the end of the BAT light curve, the good agreement between the observations and the fit is affected by the Lorentz gamma factor decrease and the corresponding increase of the maximum viewing angle. The source visible area becomes larger than the typical size of the filaments. This invalidates the radial approximation we use for the CBM description. To overcome this problem it is necessary to introduce a more detailed three-dimensional CBM description, in order to avoid an over-estimated area of emission and, correspondingly, to describe the sharpness of some observed light curves. We are still working on this issue \citep{2002ApJ...581L..19R,C07,Venezia_Flares,G07}.

We turn now to the crucial determination of the CBM density, which is derived from the fit. At the transparency point it resulted to be $n_{cbm} = 4.8 \times 10^{-3}$ particles/cm$^3$ (see Fig. \ref{f4a}). This density is compatible with the typical values of the galactic halos. During the peak of the extended afterglow emission the effective average CBM density decreases reaching $\left\langle n_{cbm}\right\rangle = 2.25 \times 10^{-5}$ particles/cm$^3$, possibly due to an occurring fragmentation of the shell \citep{2007A&A...471L..29D} or due to a fractal structure in the CBM. The ${\cal R}$ value resulted to be on average $\left\langle {\cal R}\right\rangle = 1.72 \times 10^{-8}$. It is interesting to emphasize the striking analogy of the numerical value and the overall radial dependence of the CBM density in the present case of GRB 060614 when compared and contrasted with the ones of GRB 970228 \citep{2007A&A...474L..13B}.

Concerning the $0.2$--$10$ keV light curve of the decaying phase of the afterglow, observed by the XRT instrument, we have also reproduced very satisfactorily both the hard decrease in the slope and the plateau of the light curve keeping constant the effective CBM density and changing only ${\cal R}$. The result of this analysis is reported in Fig. \ref{f4}. We assume in this phase $n_{cbm} = 4.70 \times 10^{-6}$ particles/cm$^{-3}$. The average value of the ${\cal R}$ parameter is $\left\langle {\cal R}\right\rangle=1.27\times10^{-2}$. The drastic enhancement in the ${\cal R}$ parameter with respect to the values at the peak of the extended afterglow is consistent with similar features encountered in other sources we have studied: GRB 060218 presents a bump of five orders of magnitude \citep{2007A&A...471L..29D}, in GRB 060710 the bump is of about four orders of magnitude (see Izzo et al., in preparation) while in GRB 050315 there is a three orders of magnitude bump \citep{2006ApJ...645L.109R}. In these last two cases, we find the occurrence of the enhancement of ${\cal R}$ between $r$=$2\times10^{17}$ cm and $r$=$3\times10^{17}$ cm, just like for GRB 060614, for which we have the bump at $r$=$3.5\times10^{17}$ cm. The time of the bump approximately corresponds to the appearance of the optical emission observed in GRB 060614 and, more in general, to the onset of the second component of the \citet{2007ApJ...662.1093W} scheme for GRBs.
 
\subsection{Conclusions}

GRB 060614 presents three major novelties, which challenges the most widespread theoretical models and which are strongly debated in the current literature. The first one is that it challenges the traditional separation between Long Soft GRBs and Short Hard GRBs \citep{2006Natur.444.1044G}. The second one is that it presents a short, hard and multi-peaked episode, followed by a softer, prolonged emission with a strong hard to soft evolution \citep{2006Natur.444.1044G,2007A&A...470..105M}. The third one is that it is the first clear example of a nearby, long GRB not associated with a bright SN Ib/c \citep{2006Natur.444.1050D,2006Natur.444.1053G}. All these three issues are naturally explained within our ``fireshell'' model, which allows a detailed analysis of the temporal behavior of the signal originating up to a distance $r \sim 10^{17}$--$10^{18}$ cm from the black hole, and relates, with all the relativistic transformations, the arrival time to the CBM structure and the relativistic parameters of the fireshell.

One of the major outcome of the Swift observation of, e.g., GRB 050315 \citep{2006ApJ...638..920V,2006ApJ...645L.109R} has been the confirmation that long GRB duration is not intrinsic to the source but it is merely a function of the instrumental noise threshold \citep{Venezia_Orale}. GRB 060614 represents an additional fundamental progress in clarifying the role of the CBM density in determining the GRB morphology. It confirms the results presented in GRB 970228 \citep{2007A&A...474L..13B}, that is the prototype of the new class of ``fake'' or ``disguised'' short GRBs. They correspond to canonical GRBs with an extended afterglow emission energetically predominant with respect to the P-GRB one and a baryon loading $B > 10^{-4}$. The sharp spiky emission corresponds to the P-GRB. As recalled above, a comparison of the luminosities of the P-GRB and of the extended afterglow is indeed misleading: it follows from the low average CBM density inferred from the fit of the fireshell model, which leads to $n_{cbm} \sim 10^{-3}$ particles/cm$^3$. Therefore such a feature is neither intrinsic to the progenitor nor to the black hole, but it is only indicative of the CBM density at the location where the final merging occurs. GRB 060614 is a canonical GRB and it is what would be traditionally called a ``long'' GRB if it had not exploded in a specially low CBM density environment. GRB 060614 must necessarily fulfill, and indeed it does, the Amati relation. This happens even taking into account the entire prompt emission mixing together the P-GRB and the extended afterglow \citep{2007A&A...463..913A}, due to the above recalled energetic predominance of the extended afterglow \citep[see also Ref.][]{2008A&A...487L..37G}. These results justify the occurrence of the above mentioned first two novelties.

The low value of the CBM density is compatible with a galactic halo environment. This result points to an old binary system as the progenitor of GRB 060614 and it justifies the above mentioned third novelty: the absence of an associated SN Ib/c \citep[see also Ref.][]{2007AIPC..906...69D}. Such a binary system departed from its original location in a star forming region and spiraled out in a low density region of the galactic halo \citep[see e.g. Ref.][]{KMG11}. The energetic of this GRB is about two orders of magnitude smaller than the one of GRB 970228 \citep{2007A&A...474L..13B}. A natural possible explanation is that instead of a neutron star - neutron star merging binary system we are in presence of a white dwarf - neutron star binary. We therefore agree, for different reasons, with the identification proposed by \citet{2007AIPC..906...69D} for the GRB 060614 progenitor. In principle, the nature of the white dwarf, with typical radius on the order of $10^3$ km, as opposed to the one of the neutron star, typically on the order of $10$ km, may manifest itself in characteristic signatures in the structure of the P-GRB (see Fig. \ref{f2a}).

It is interesting that these results lead also to three major new possibilities:
\begin{itemize}
\item The majority of GRBs declared as shorts \citep[see e.g. Ref.][]{2005Natur.437..822P} are likely ``disguised'' short GRBs, in which the extended afterglow is below the instrumental threshold.
\item The observations of GRB 060614 offer the opportunity, for the first time, to analyze in detail the structure of a P-GRB lasting $5$ s. This feature is directly linked to the physics of the gravitational collapse originating the GRB. Recently, there has been a crucial theoretical physics result, showing that the characteristic time constant for the thermalization for an $e^\pm$ plasma is on the order of $10^{-13}$ s \citep{2007PhRvL..99l5003A}. Such a time scale still applies for an $e^\pm$ plasma with a baryon loading on the order of the one observed in GRBs \citep{2009PhRvD..79d3008A}. The shortness of such a time scale, as well as the knowledge of the dynamical equations of the optically thick phase preceding the P-GRB emission \citep{brvx06}, implies that the structure of the P-GRB is a faithful representation of the gravitational collapse process leading to the formation of the black hole \citep{2005IJMPD..14..131R}. In this respect, it is indeed crucial that the Swift data on the P-GRB observed in GRB 060614 (see Fig. \ref{f2a}) appear to be highly structured all the way to time scale of $0.1$ s. This opens a new field of research: the study of the P-GRB structure in relation to the process of gravitational collapse leading to the GRB.
\item If indeed the binary nature of the progenitor system and the peculiarly low CBM density $n_{cbm} \sim 10^{-3}$ particles/cm$^3$ will be confirmed for all ``fake'' or ``disguised'' GRBs, then it is very likely that the traditionally ``long'' high luminosity GRBs at higher redshift also originates from the merging of binary systems formed by neutron stars and/or white dwarfs occurring close to their birth location in star forming regions with $n_{cbm} \sim 1$ particle/cm$^3$ (see Fig. \ref{picco_n=1}).
\end{itemize}

\section{Open issues in current theoretical models}

The ``fireshell'' model addresses mainly the $\gamma$ and X-ray emission, which are energetically the most relevant part of the GRB phenomenon. The model allows a detailed identification of the fundamental three parameters of the GRB source: the total energy, the baryon loading, as well as the CBM density, filamentary structure and porosity. The knowledge of these phenomena characterizes the region surrounding the black hole up to a distance which in this source reaches $\sim 10^{17}$--$10^{18}$ cm. When applied, however, to larger distances, which corresponds to the latest phases of the X-ray afterglow, since the beginning of the ``plateau'' phase, the model reveals a different regime which has not yet been fully interpreted in its astrophysical implications. To fit the light curve in the soft X-ray regime for $r \gtrsim 4 \times 10^{17}$ cm, we must appeal to an enhancement of about six orders of magnitude in the ${\cal R}$ factor (see above, and Fig. \ref{f4a}). This would correspond to a more diffuse CBM structure, with a smaller porosity, interacting with the fireshell. This points to a different leading physical process during the latest X-ray afterglow phases. When we turn to the optical, IR and radio emission, the fireshell model leads to a much smaller flux than the observed one, especially for $r \sim 10^{17}$--$10^{18}$ cm. Although the optical, IR, and radio luminosities have a minority energetic role, they may lead to the identification of crucial parameters and new phenomena occurring in the source, and they deserve maximum attention. 

In these latest phases for $r \geq 10^{17}$ cm it is currently applied the treatment based on synchrotron emission pioneered by \citet{1997ApJ...476..232M} even before the discovery of the afterglow \citep{1997Natur.387..783C}. Such a model has been further developed \citep[see Refs.][and references therein]{1998ApJ...497L..17S,2005RvMP...76.1143P,2006RPPh...69.2259M}. Also in this case, however, some difficulties remain since it is necessary to invoke the presence of an unidentified energy injection mechanism \citep{2006ApJ...642..354Z}. Such a model appears to be quite successful in explaining the late phases of the X-ray emission of GRB 060614, as well as the corresponding optical emission, in terms of different power-law indexes for the different parts of the afterglow light curves \citep{2007A&A...470..105M,2008arXiv0812.0979X}. However, also in this case an unidentified energy injection mechanism between $\sim 0.01$ days and $\sim 0.26$ days appears to be necessary \citep{2008arXiv0812.0979X}. 

The attempt to describe the prompt emission via the synchrotron process by the internal shock scenario \citep[see e.g. Refs.][and references therein]{1994ApJ...430L..93R,2005RvMP...76.1143P,2006RPPh...69.2259M} also encounters difficulties: \citet{2008MNRAS.384...33K} have shown that the traditional synchrotron model can be applied to the prompt emission only if it occurs at $r > 10^{17}$ cm. A proposed way-out of this problem, via the inverse Compton process, suffers of an ``energy crisis'' \citep[see e.g. Ref.][]{2008arXiv0807.3954P}. 

Interestingly, the declared region of validity of the traditional synchrotron model ($r > 10^{17}$ cm) is complementary to the one successfully described by our model ($r < 10^{17}$--$10^{18}$ cm). Astrophysically, \citet{2008arXiv0812.0979X} have reached, within the framework of the traditional synchrotron model, two conclusions which are consistent with the results of our analysis of GRB 060614. First, they also infer from their numerical fit a very low density environment, namely $n_{cbm} \sim 0.04$ particles/cm$^3$. Second, they also mention the possibility that the progenitor of GRB 0606014 is a merging binary system formed by two compact objects.

	\section{The search for a ``genuine short'' GRB: the case of GRB 050509B}

\begin{figure}
\centering
\includegraphics[width=10.0 cm]{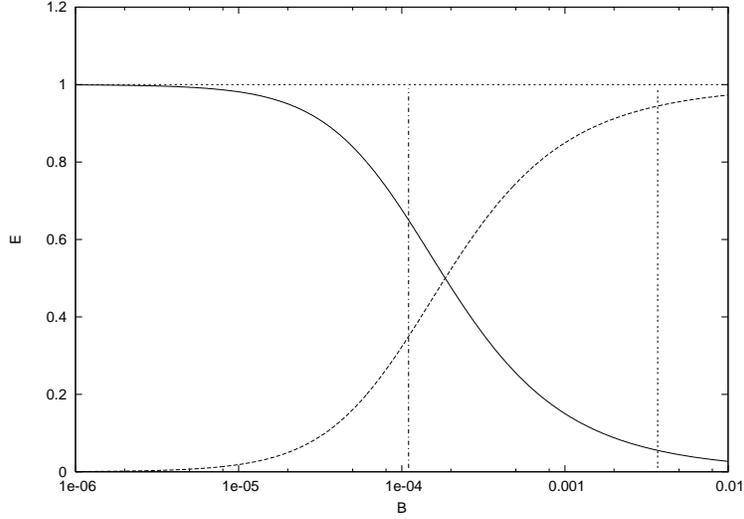}
\caption{Energy emitted in: the P-GRB (solid line) and  extended afterglow (dashed line), in function of the $B$ parameter. The dotted line is the sum of the two lines. The two vertical lines represents the values of the $B$ parameter of our two analyses, respectively. In our second analysis GRB 050509B results to be a ``genuine'' short.}\label{fig:ratio}
\end{figure}

As we already discussed above, within the fireshell model the baryon loading is the key parameter to classify GRBs: if $B \lesssim 10^{-5}$ we have what we call ``genuine'' short GRBs. In order to investigate if this is indeed the case of GRB 050509B \citep{2005Natur.437..851G} we performed two different analyses, respectively with $B= 3.7\times 10^{-3}$ and with $B =1.1\times 10^{-4}$ (see Fig. \ref{fig:ratio}).

\paragraph{Analysis 1}

\begin{figure}
\centering
\includegraphics[width=10.0 cm]{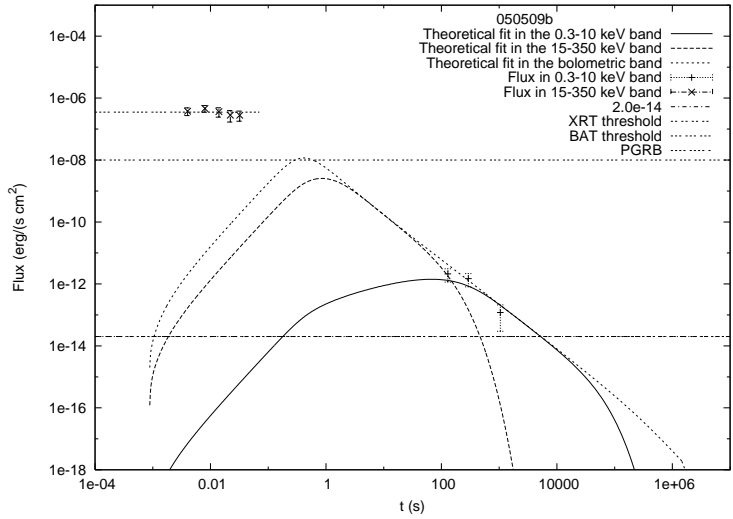}
\caption{Analysis 1: the P-GRB corresponds to the BAT observations and the extended afterglow, that has a total energy that is much greater than the P-GRB one, to the XRT observations.}\label{fig:P-GRB}
\end{figure}

We identify the prompt emission of this GRB \citep[see Ref.][]{2005Natur.437..851G} with our P-GRB. Consequently, the extended afterglow corresponds to the observed X-ray afterglow (see Fig. \ref{fig:P-GRB}). In this case, we have the total energy of the GRB estimated in $E_{tot}^{e\pm}=2.11\times 10^{48}$ erg (which is a low energy for GRBs) and the baryon loading is $B=3.7\times 10^{-3}$. With this choice of the fireshell parameters, we obtain that the P-GRB energy is $E_{P-GRB}=1.6\times 10^{47}$ erg. More than $90\%$ of the total energy is released in the extended afterglow, hence GRB 050509B cannot be classified as ``genuine'' short GRB. The reason for the non observability of the peak of this extended afterglow is that it results under the BAT threshold in the gamma-ray energy band, and before the beginning of the XRT observations (which sets around $100$ s).

\paragraph{Analysis 2}

\begin{figure}
\centering
\includegraphics[width=10.0 cm]{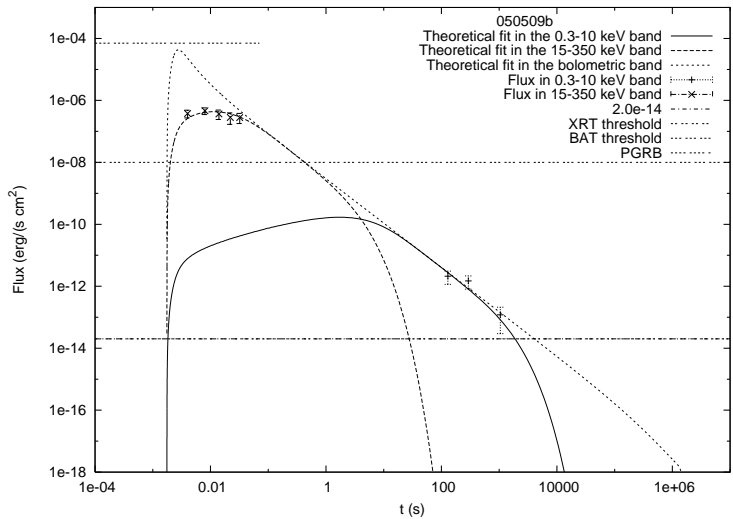}
\caption{Analysis 2: both the BAT observations and the XRT ones are identified with the extended afterglow emission. The P-GRB flux is more then twice the extended afterglow one, but it is too hard to be observed by BAT.}\label{fig:After}
\end{figure}

We performed an alternative analysis interpreting GRB 050509B as a ``genuine'' short GRB. We fit the BAT observations as the peak of the extended afterglow (see Fig. \ref{fig:After}). In this case the total energy is $E_{tot}^{e\pm}=5.07\times 10^{49}$ erg, the baryon loading is $B=1.1\times 10^{-4}$, and this implies that the energy emitted on the P-GRB is almost $60\%$ of the total one, $E_{P-GRB}=3.30\times 10^{49}$ erg. Differently with the previous case, it was not observable by BAT since its peak energy would be about $817$ keV. According this second interpretation, GRB 050509B is a ``genuine'' short GRB.

\paragraph{GRB 050509B within the Amati relation}

In order to discriminate between the above analyses we checked if their results are compatible with the Amati relation \citep{2002A&A...390...81A,2006MNRAS.372..233A}. According to the fireshell model, only the extended afterglow component should satisfy the Amati relation, while the P-GRB component should not. This accounts for the fact that the ``short'' GRBs are outliers of the correlation. Therefore, according to the first interpretation the P-GRB which coincides with the BAT observation should be out from the correlation: this is indeed true. 

On the other hand, the second analysis reveals that the BAT observation should satisfy the correlation. Hence, this possibility have to be ruled out. We continue investigating the first possibility in order to obtain further information that from the astrophysical setting of this GRB constraint better the fit of the extended afterglow.

Apart from this result, the new generation of high energy satellites are very important for the observation and study of the P-GRB, and for the identification of `genuine'' short GRBs.

\begin{figure}
\centering
\includegraphics[width=10.0 cm]{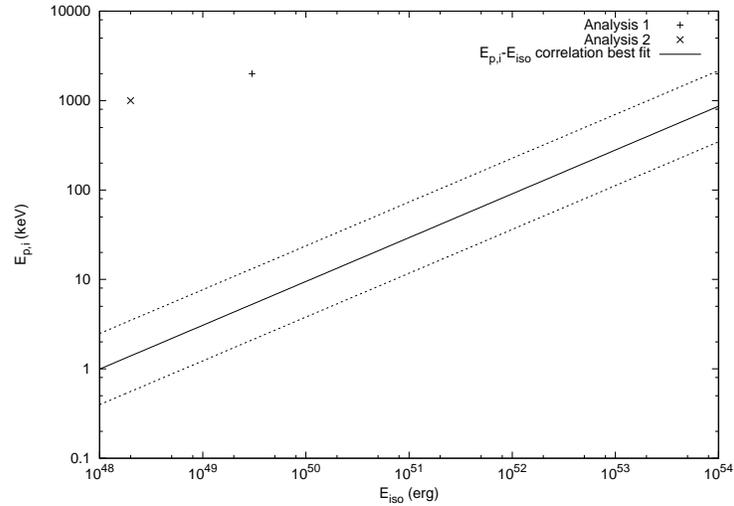}
\caption{The Amati relation with our predictions about GRB 050509B in the two analyses. In the first one, the P-GRB results out from the correlation, as it should be. In the second one, the peak of the extended afterglow emission should satisfy the correlation, but it does not happen. Hence, the first analysis turns to be more correct.}\label{fig:Amati}
\end{figure}

	\section{GRBs and SNe: the induced gravitational collapse}

The \textit{Collapsar model} \citep{1993ApJ...405..273W,1998ApJ...494L..45P,1999ApJ...524..262M,2003ApJ...586..356Z} proposes that GRBs arise from the collapse of a single Wolf-Rayet star endowed with fast rotation. This idea is purported by the evidence that many GRBs are close to star-forming regions and that this suggests that GRBs are linked to cataclysmic deaths of massive stars ($M>30M_\odot$). In such a model very massive stars are able to fuse material in their centers all the way to iron, at which point a star cannot continue to generate energy by fusion and collapses, in this case, immediately forming a black hole. Matter from the star around the core rains down toward the center and (for rapidly rotating stars) swirls into a high-density accretion disk. The mass of the accretion disk is around $0.1\, M_\odot$. The infall of this material into the black hole is assumed to drive a pair of jets (with opening angles $<10$ degrees) out along the rotational axis, where the matter density is much lower than in the accretion disk, toward the poles of the star at velocities approaching the speed of light, creating a relativistic shock wave at the front \citep{1976PhFl...19.1130B}. The processes of core collapse and of accretion along the polar column and the jet propagation through the stellar envelope all together last $\sim 10$ sec \citep{1999ApJ...524..262M}. The jet, as it passes through the star, is modulated by its interaction with the surrounding medium. In this way the Collapsar model attempts to explain the time structure of GRB prompt emission and to produce the variable Lorentz factor necessary for the internal shocks occurrence \citep{2006ARA&A..44..507W}. Moreover it is a prediction of this model that the central engine remains active for a long time after the principal burst is over, potentially contributing to the GRB afterglow \citep{2005Sci...309.1833B}.

Three very special conditions are required for a star to evolve all the way to a gamma-ray burst according to this theory: the star should be very massive ($25M_\odot$ \citet{1993ApJ...405..273W}, $35-40M_\odot$ on the main sequence \citet{1999ApJ...520..650F}) to form a central black hole, the star rapidly rotates to develop an accretion torus capable of launching jets, and the star should have low metallicity in order to strip off its hydrogen envelope so the jets can reach the surface. As a result, gamma-ray bursts are far rarer than ordinary core-collapse supernovae, which only require the star to be massive enough to fuse all the way to iron. 

The consensus and the difficulties for the Collapsar model can be simply summarized:
\begin{itemize}
\item \citet{2006ASPC..353...15L} asserts that long gamma-ray bursts are found in systems with abundant recent star formation, low metallicity environment.
\item The second evidence in favor of the Collapsar model is that there are several observed cases where a supernova is practical coeval with GRBs.
\item However, strong evidence against the Collapsar model comes from the fact that there were recently discovered two nearby long gamma-ray bursts which lacked a signature of any type Ib/c supernova: both GRB 060614 \citep[][see also above]{2006Natur.444.1050D,2006Natur.444.1053G} and GRB 060505 \citep{2006Natur.444.1047F} defied predictions that a supernova would emerge despite intense scrutiny from ground-based telescopes. 
\end{itemize}

\begin{figure}
\includegraphics[width=\hsize,clip]{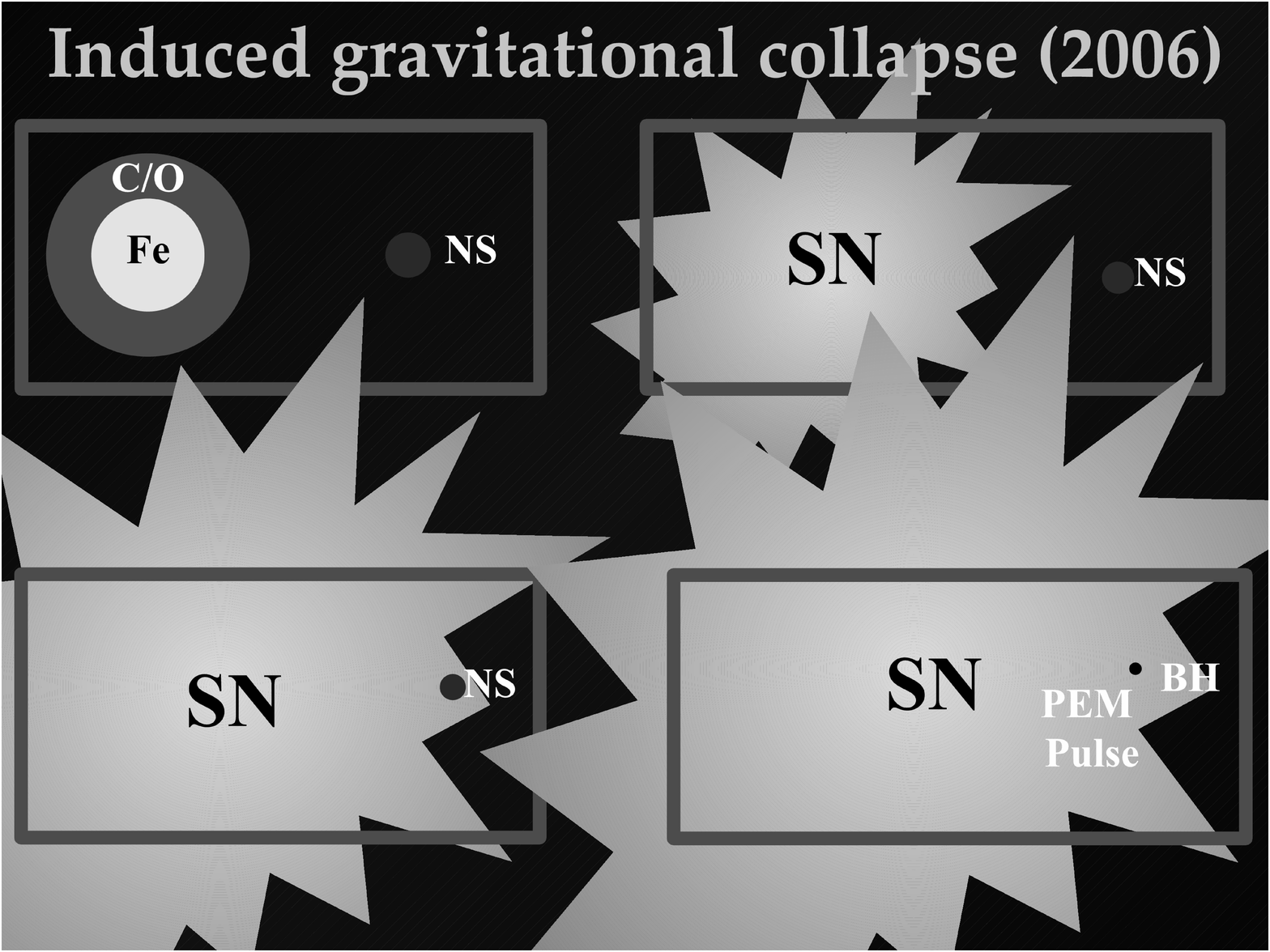}
\caption{A sketch summarizing the induced gravitational collapse scenario.}
\label{fig_indcoll}
\end{figure}

Within our fireshell model, we recall, the approach is drastically different, as already introduced in \citet{2001ApJ...555L.117R}. In fact in this framework, the SN which is often observed in temporal and spatial coincidence with the GRB cannot be interpreted as its progenitor because of the high quantity of ejected matter from the supernova explosion would prevent the GRB occurrence. Moreover:
\begin{itemize}
\item It is very unlikely that a core collapse SN produces directly a black hole.
\item GRBs originate from gravitational collapses to black holes (see above). The possible explanation for the GRB-SN connection proposed in \citet{2001ApJ...555L.117R} is that both the GRB and the supernova progenitors belong to a binary system. Under special conditions it is possible that the GRB emission triggers the supernova explosion of the companion star. Alternatively, it is possible that the process of gravitational collapse to a black hole producing the GRB is ``induced'' by the supernova Ib/c on a companion neutron star \citep[see Fig.~\ref{fig_indcoll} and Refs.][]{Mosca_Orale,2007A&A...471L..29D}. The faintness of this GRB class could be in this case naturally explained by the formation of the smallest possible black hole, just over the critical mass of the neutron star \citep{mg11}. Moreover these systems occur in a low density CBM ($10^{-2}$--$1$ particle/cm$^3$).
\item Also the observation of the occurrence of ``long'' GRBs in star forming regions is explained by identifying the progenitor with a binary system formed by a neutron star and a star evolved out of the main sequence. 
\end{itemize}

	\subsection{Application to GRB 060218}

GRB 060218 triggered the BAT instrument of {\em Swift} on 18 February 2006 at 03:36:02 UT and has a $T_{90} = (2100 \pm 100)$ s \citep{2006GCN..4775....1C}. The XRT instrument \citep{2006GCN..4776....1K,2006GCN..4775....1C} began observations $\sim 153$ s after the BAT trigger and continued for $\sim 12.3$ days \citep{2006GCN..4822....1S}. The source is characterized by a flat $\gamma$-ray light curve and a soft spectrum \citep{2006GCN..4780....1B}. It has an X-ray light curve with a long, slow rise and gradual decline and it is considered an X-Ray Flash (XRF) since its peak energy occurs at $E_p=4.9^{+0.4}_{-0.3}$ keV \citep{2006Natur.442.1008C}. It has been observed by the \emph{Chandra} satellite on February 26.78 and March 7.55 UT ($t\simeq 8.8$ and $17.4$ days) for $20$ and $30$ ks respectively \citep{2006Natur.442.1014S}. The spectroscopic redshift has been found to be $z=0.033$ \citep{2006A&A...454..503S,2006ApJ...643L..99M}. The corresponding isotropic equivalent energy is $E_{iso}=(1.9\pm 0.1)\times 10^{49}$ erg \citep{2006GCN..4822....1S} which sets this GRB as a low luminous one, consistent with most of the GRBs associated with SNe \citep{2007ApJ...662.1111L,2006ApJ...645L.113C,2007ApJ...657L..73G}.

GRB 060218 is associated with SN2006aj whose expansion velocity is $v\sim 0.1c$ \citep{2006Natur.442.1011P,2006GCN..4809....1F,2006GCN..4804....1S,2006ApJ...645L.113C}. The host galaxy of SN2006aj is a low luminosity, metal poor star forming dwarf galaxy \citep{2007AIPC..924..120F} with an irregular morphology \citep{2007A&A...464..529W}, similar to the ones of other GRBs associated with SNe \citep{2006ApJ...645L..21M,2006A&A...454..503S}.

\subsubsection{The fit of the observed data}

\begin{figure}
\includegraphics[width=\hsize,clip]{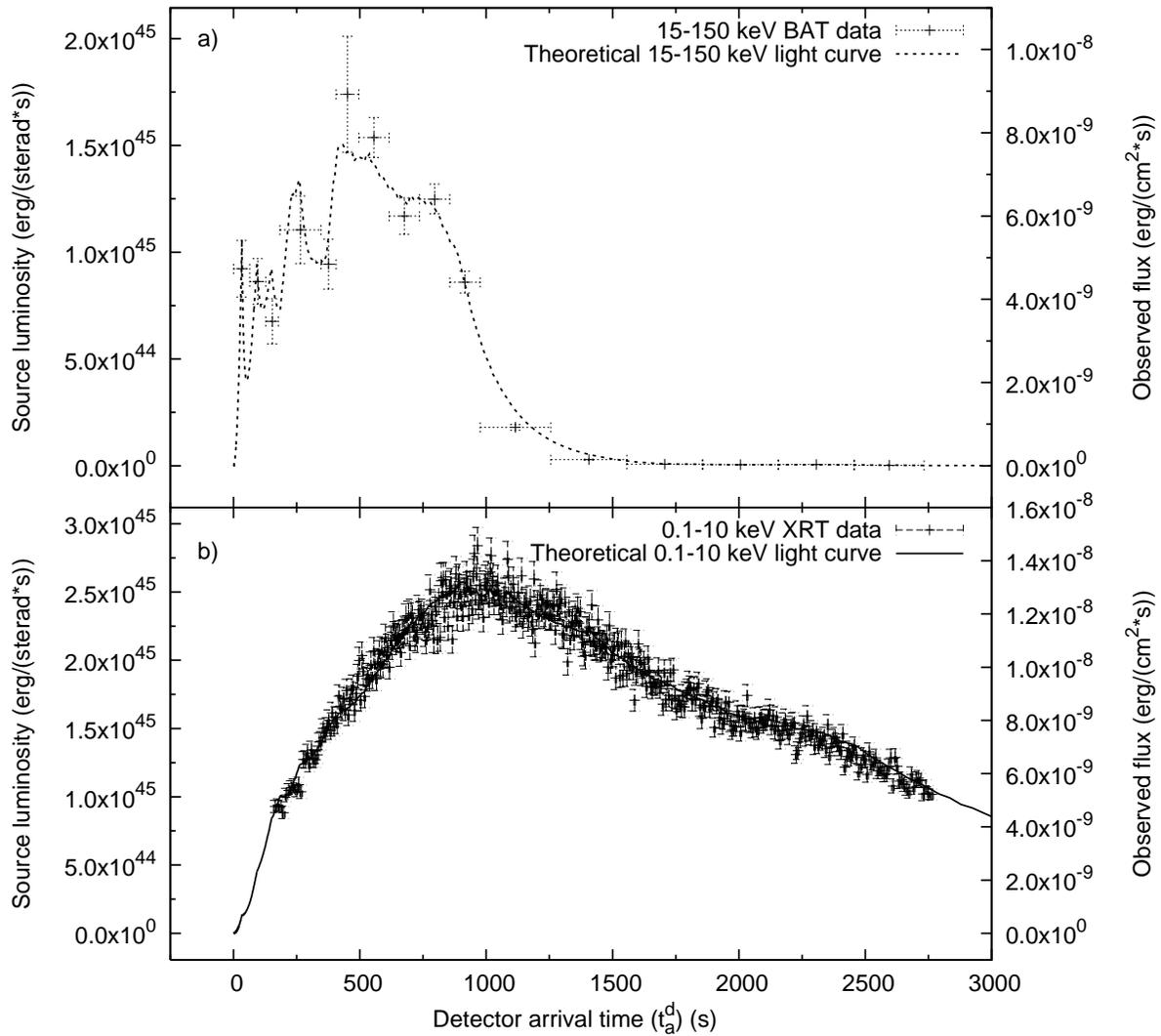}
\caption{GRB 060218 prompt emission: a) our theoretical fit (dotted line) of the BAT observations in the $15$--$150$ keV energy band (dotted points); b) our theoretical fit (solid line) of the XRT observations in the $0.3$--$10$ keV energy band (solid points) \citep[Data from Ref.][]{2006Natur.442.1008C}.}
\label{060218_tot}
\end{figure}

In this section we present the fit of our fireshell model to the observed data (see Figs. \ref{060218_tot}, \ref{global2}). The fit leads to a total energy of the $e^\pm$ plasma $E_{e^\pm}^{tot}= 2.32\times 10^{50}$ erg, with an initial temperature $T = 1.86$ MeV and a total number of pairs $N_{e^\pm} = 1.79\times 10^{55}$. The second parameter of the theory, $B = 1.0 \times 10^{-2}$, is the highest value ever observed and is close to the limit for the stability of the adiabatic optically thick acceleration phase of the fireshell \citep[for further details see Ref.][]{2000A&A...359..855R}. The Lorentz gamma factor obtained solving the fireshell equations of motion \citep{2005ApJ...620L..23B,2005ApJ...633L..13B} is $\gamma_\circ=99.2$ at the beginning of the extended afterglow phase at a distance from the progenitor $r_\circ=7.82\times 10^{12}$ cm. It is much larger than $\gamma \sim 5$ estimated by \citet{2007ApJ...654..385K} and \citet{2007ApJ...659.1420T}.

In Fig. \ref{060218_tot} we show the extended afterglow light curves fitting the prompt emission both in the BAT ($15$--$150$ keV) and in the XRT ($0.3$--$10$ keV) energy ranges, as expected in our ``canonical GRB'' scenario \citep{2007A&A...471L..29D}. Initially the two luminosities are comparable to each other, but for a detector arrival time $t_a^d > 1000$ s the XRT curves becomes dominant. The displacement between the peaks of these two light curves leads to a theoretically estimated spectral lag greater than $500$ s in perfect agreement with the observations \citep{2006ApJ...653L..81L}. We obtain that the bolometric luminosity in this early part coincides with the sum of the BAT and XRT light curves (see Fig. \ref{global2}) and the luminosity in the other energy ranges is negligible.

We recall that at $t_a^d \sim 10^4$ s there is a sudden enhancement in the radio luminosity and there is an optical luminosity dominated by the SN2006aj emission \citep{2006Natur.442.1008C,2006Natur.442.1014S,2006JCAP...09..013F}. Although our analysis addresses only the BAT and XRT observations, for $r > 10^{18}$ cm corresponding to $t_a^d > 10^4$ s the fit of the XRT data implies two new features: \textbf{1)} a sudden increase of the ${\cal R}$ factor from ${\cal R} = 1.0\times 10^{-11}$ to ${\cal R} = 1.6\times 10^{-6}$, corresponding to a significantly more homogeneous effective CBM distribution (see Fig.\ref{060218_global}b); \textbf{2)} an XRT luminosity much smaller than the bolometric one (see Fig. \ref{global2}). These theoretical predictions may account for the energetics of the enhancement of the radio and possibly optical and UV luminosities. Therefore, we identify two different regimes in the extended afterglow, one for $t_a^d < 10^4$ s and the other for $t_a^d > 10^4$ s. Nevertheless, there is a unifying feature: the determined effective CBM density decreases with the distance $r$ monotonically and continuously through both these two regimes from $n_{cbm} = 1$ particle/cm$^3$ at $r = r_\circ$ to $n_{cbm} = 10^{-6}$ particle/cm$^3$ at $r = 6.0 \times 10^{18}$ cm: $n_{cbm} \propto r^{-\alpha}$, with $1.0 \lesssim \alpha \lesssim 1.7$ (see Fig. \ref{060218_global}a).

Our assumption of spherical symmetry is supported by the observations which set for GRB 060218 an opening beaming angle larger than $\sim 37^\circ$ \citep{2007ApJ...662.1111L,2006Natur.442.1008C,2006Natur.442.1014S,2007ApJ...657L..73G}.

\subsubsection{The procedure of the fit}

The arrival time of each photon at the detector depends on the entire previous history of the fireshell \citep{2001ApJ...555L.107R}. Moreover, all the observables depends on the EQTS \citep{2004ApJ...605L...1B,2005ApJ...620L..23B} which, in turn, depend crucially on the equations of motion of the fireshell. The CBM engulfment has to be computed self-consistently through the entire dynamical evolution of the fireshell and not separately at each point. Any change in the CBM distribution strongly influences the entire dynamical evolution of the fireshell and, due to the EQTS structure, produces observable effects up to a much later time. For example if we change the density mask at a certain distance from the black hole we modify the shape of the lightcurve and consequently the evolution changes at larger radii corresponding to later times. 
Anyway the change of the density is not the only problem to face in the fitting of the source, in fact first of all we have to choose the energy in order to have Lorentz gamma factor sufficiently high to fit the entire GRB.  
In order to show the sensitivity of the fitting procedure I also present two examples of fits with the same value of $B$ and different value of $E_{e^\pm}^{tot}$.

\begin{figure}
  \includegraphics[width=\hsize,clip]{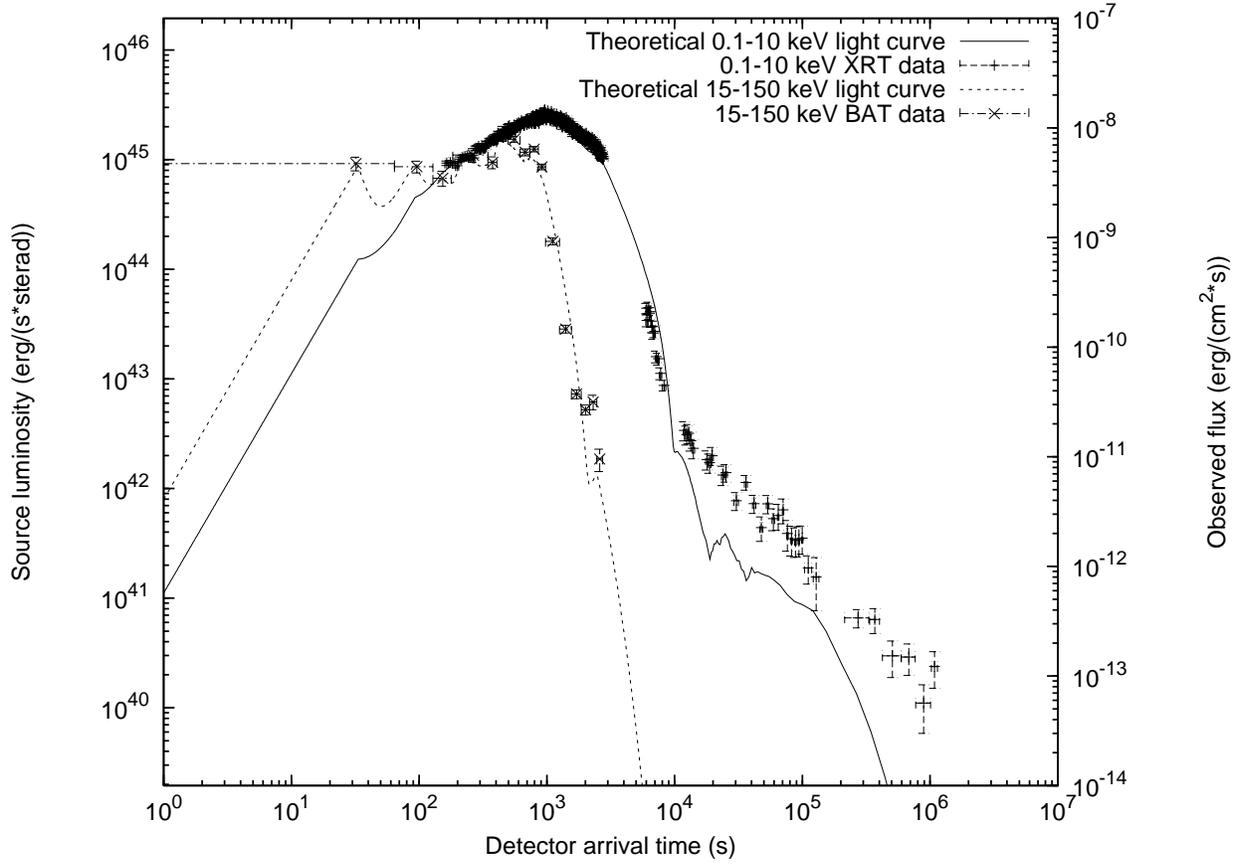}
  \caption{GRB 060218 light curves with $E_{e^\pm}^{tot}= 1.36\times 10^{50}$ erg: our theoretical fit (dotted line) of the $15$--$150$ keV BAT observations and our theoretical fit (solid line) of the $0.3$--$10$ keV XRT observations are represented (Data from: \citet{2006Natur.442.1008C,2006Natur.442.1014S}).}
  \label{060218bolometricasotto}
\end{figure}

The first example has an $E_{e^\pm}^{tot}$ = $1.36\times 10^{50}$ erg . This fit resulted unsuccessfully as we see from the Fig.\ref {060218bolometricasotto}, because the bolometric lightcurve is under the XRT peak of the extended afterglow. This means that the value of the energy chosen is too small to fit any data points after the peak of the extended afterglow. So we have to increase the value of the Energy to a have a better fit. In fact the parameters values have been found with various attempt in order to obtain the best fit. 

\begin{figure}
  \includegraphics[width=\hsize,clip]{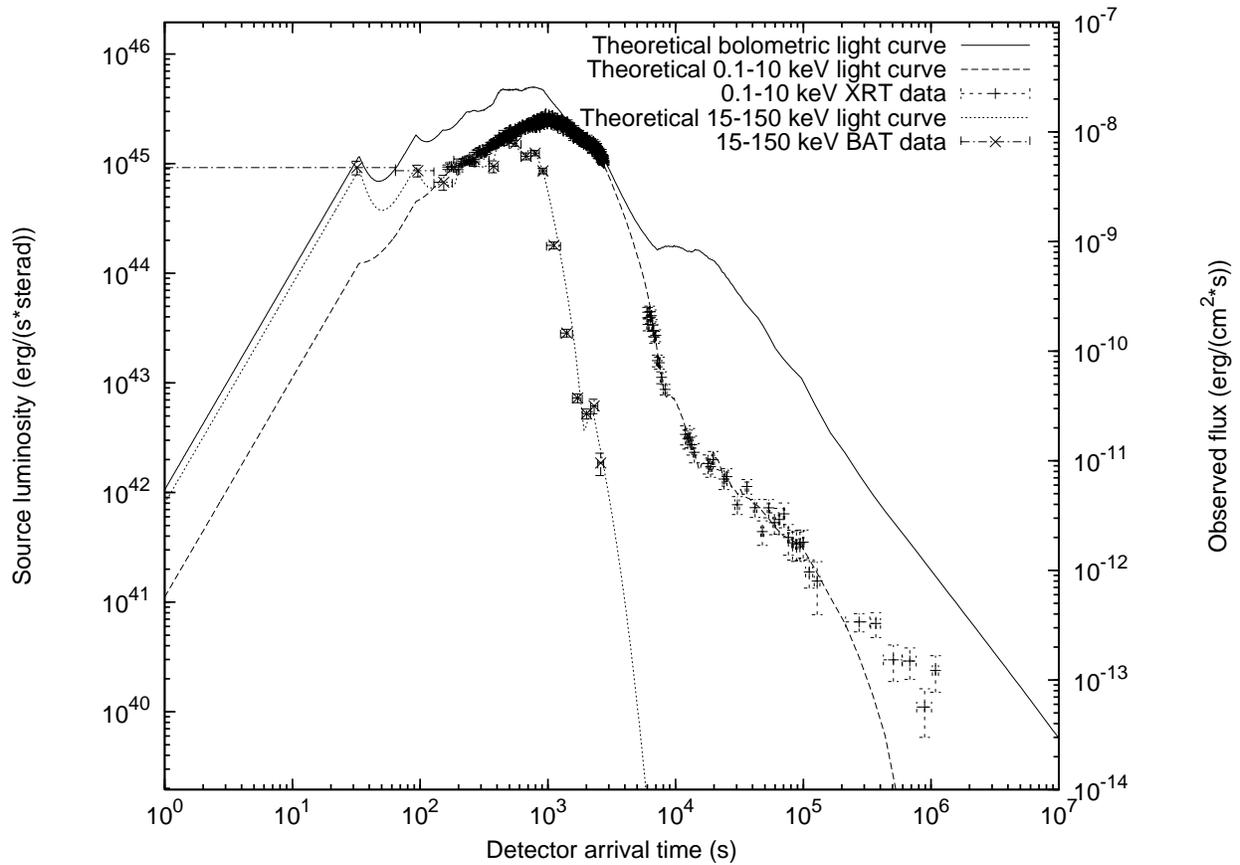}
  \caption{GRB 060218 light curves with $E_{e^\pm}^{tot}= 1.61\times 10^{50}$ erg: our theoretical fit (dotted line) of the $15$--$150$ keV BAT observations and our theoretical fit (dashed line) of the $0.3$--$10$ keV XRT observations are represented together with our theoretically computed bolometric luminosity (solid line). Data from: \citet{2006Natur.442.1008C,2006Natur.442.1014S}.}
  \label{060218senzaultimipunti}
\end{figure}

\begin{figure}
\includegraphics[width=\hsize,clip]{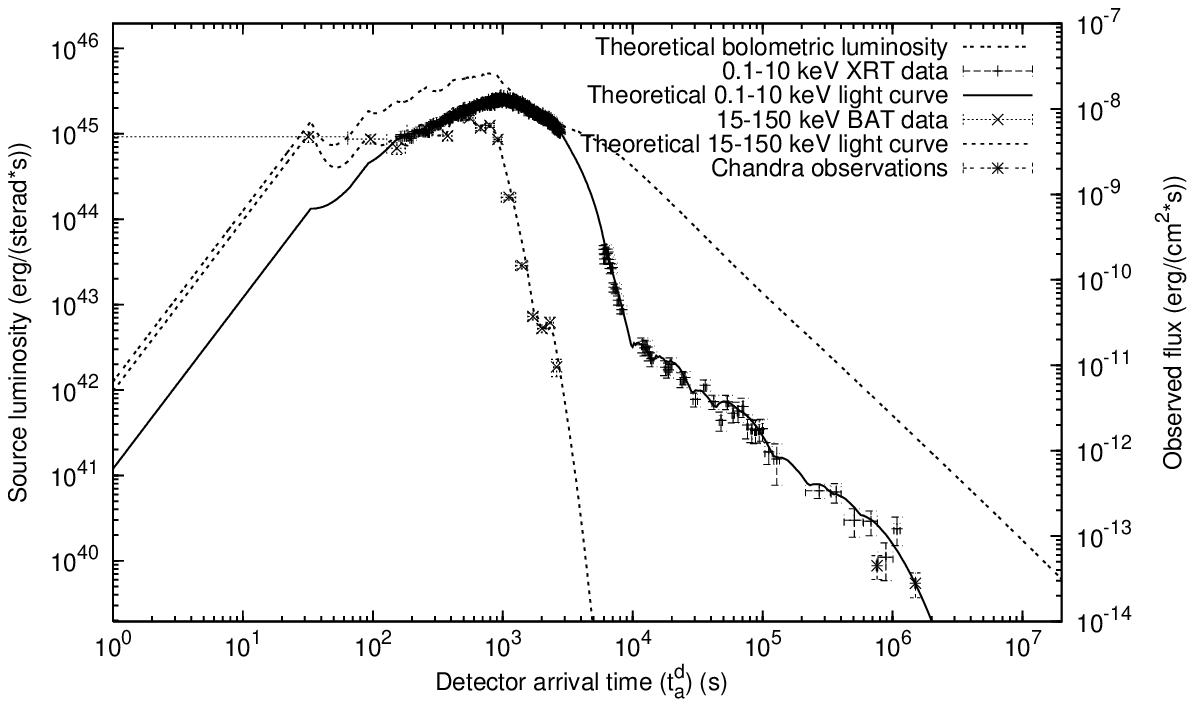}
\caption{GRB 060218 complete light curves: our theoretical fit (dotted line) of the $15$--$150$ keV BAT observations, our theoretical fit (solid line) of the $0.3$--$10$ keV XRT observations and the $0.3$--$10$ keV \textit{Chandra} observations are represented together with our theoretically computed bolometric luminosity (double dotted line) \citep[Data from Refs.][]{2006Natur.442.1008C,2006Natur.442.1014S}.}
\label{global2}
\end{figure}

The second example is characterized by $E_{e^\pm}^{tot}= 1.61\times 10^{50}$ erg and the all the data are fitted except for the last point from $2.0\times 10^{2}$s to the end (see Fig. \ref{060218senzaultimipunti}). I attempt to fit these last points trying to diminuishes the $R$ values in order to enhance the energy emission, but again the low value of the Lorentz gamma factor, that in this case is $3$ prevent the fireshell to expand. So again in this case the value of the Energy chosen is too small, but it is better than the previous attempt. In this case we increased the energy value of the 24\%, but it is not enough so we decide to increase 16\%.

So the final fit is characterized by the $B=1.0\times10^{-2}$ and by the $E_{e^\pm}^{tot}= 2.32\times 10^{50}$ erg. With this value of the energy we are able to fit all the experimental points.

\subsubsection{The fireshell fragmentation}

\begin{figure}
\includegraphics[width=\hsize,clip]{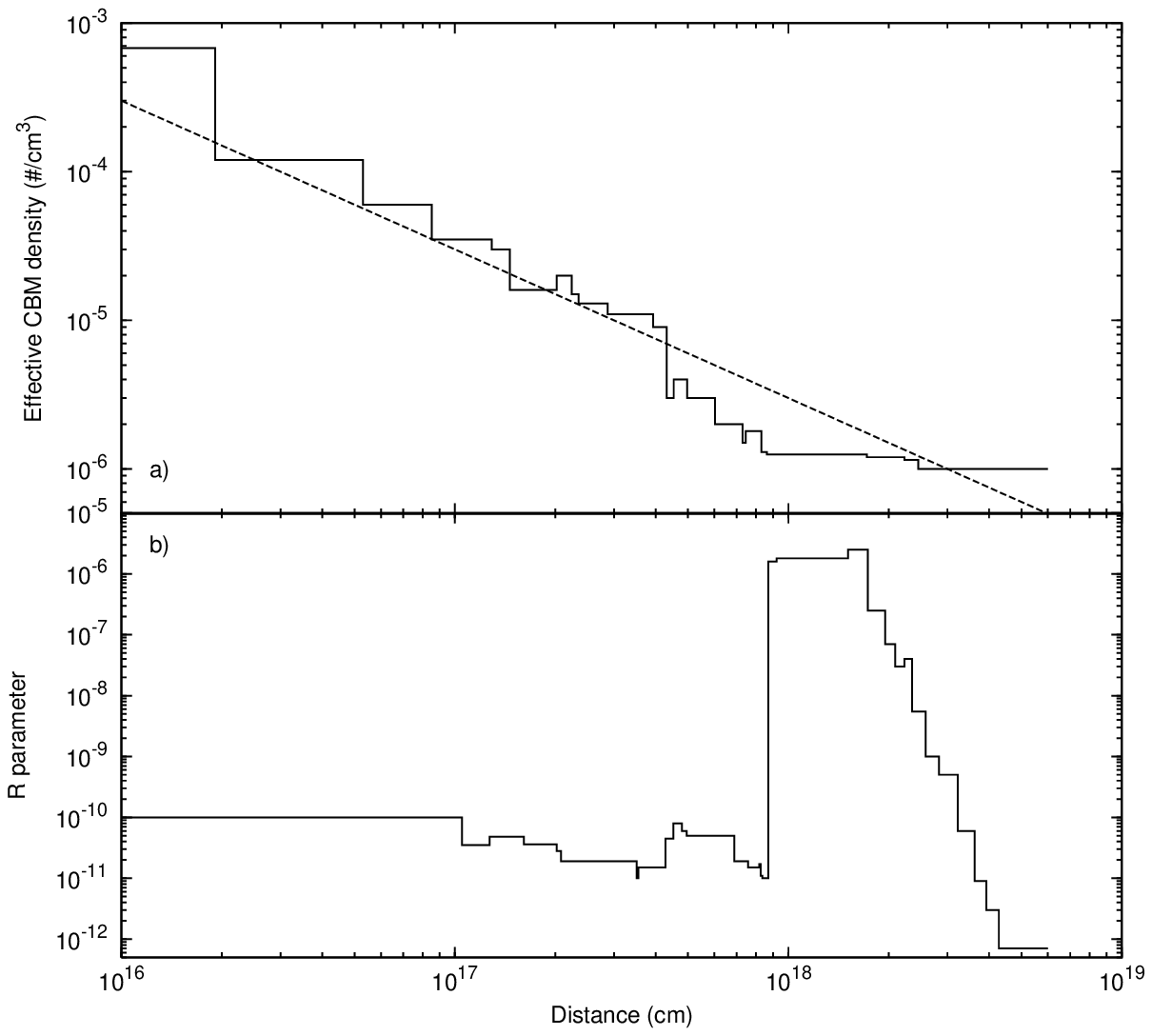}
\caption{The CBM distribution parameters: a) the effective CBM number density (solid line) monotonically decreases with the distance $r$ following Eq.(\ref{nismact}) (dashed line); b) the ${\cal R}$ parameter vs. distance.}
\label{060218_global}
\end{figure}

GRB 060218 presents different peculiarities: the extremely long $T_{90}$, the very low effective CBM density decreasing with the distance and the largest possible value of $B=10^{-2}$. These peculiarities appear to be correlated. Following \citet{Mosca_Orale}, we propose that in the present case the fireshell is fragmented. This implies that the surface of the fireshell does not increase any longer as $r^2$ but as $r^\beta$ with $\beta < 2$. Consequently, the effective CBM density $n_{cbm}$ is linked to the actual one $n_{cbm}^{act}$ by:
\begin{equation}
n_{cbm} = {\cal R}_{shell} n_{cbm}^{act}\, , \quad \mathrm{with} \quad {\cal R}_{shell} \equiv \left(r^\star/r\right)^\alpha\, ,
\label{nismact}
\end{equation}
where $r^\star$ is the starting radius at which the fragmentation occurs and $\alpha = 2 - \beta$ (see Fig. \ref{060218_global}a). For $r^\star = r_\circ$ we have $n_{cbm}^{act}=1$ particles/cm$^3$, as expected for a ``canonical GRB'' \citep{2007AIPC..910...55R} and in agreement with the apparent absence of a massive stellar wind in the CBM \citep{2006Natur.442.1014S,2006JCAP...09..013F,2007MNRAS.375..240L}.

The ${\cal R}$ parameter defined in Eq.(\ref{060218_Rdef}) has to take into account both the effect of the fireshell fragmentation (${\cal R}_{shell}$) and of the effective CBM porosity (${\cal R}_{cbm}$):
\begin{equation}
{\cal R} \equiv {\cal R}_{shell} \times {\cal R}_{cbm}\, .
\label{060218_Rdef}
\end{equation}

The phenomenon of the clumpiness of the ejecta, whose measure is the filling factor, is an aspect well known in astrophysics.  For example, in the case of Novae the filling factor has been measured to be in the range $10^{-2}$--$10^{-5}$ \citep{2006A&A...459..875E}. Such a filling factor coincides, in our case, with ${\cal R}_{shell}$.

\subsubsection{Binaries as progenitors of GRB-SN systems}

The majority of the existing models in the literature appeal to a single astrophysical phenomenon to explain both the GRB and the SN \citep[``collapsar'', see e.g. Ref.][]{2006ARA&A..44..507W}. On the contrary, a distinguishing feature of our theoretical approach is to differentiate between the SN and the GRB process (see above). The GRB is assumed to occur during the formation process of a black hole. The SN is assumed to lead to the formation of a neutron star (NS) or to a complete disruptive explosion without remnants and, in no way, to the formation of a black hole. In the case of SN2006aj the formation of such a NS has been actually inferred by \citet{2007ApJ...658L...5M} because of the large amount of $^{58}$Ni ($0.05 M_\odot$). Moreover the significantly small initial mass of the SN progenitor star $M \approx 20 M_\odot$ is expected to form a NS rather than a black hole when its core collapses \citep{2007ApJ...658L...5M,2007AIPC..924..120F,2006Natur.442.1018M,2006NCimB.121.1207N}. In order to fulfill both the above requirement, we assume that the progenitor of the GRB and the SN consists of a binary system formed by a NS close to its critical mass collapsing to a black hole, and a companion star evolved out of the main sequence originating the SN. The temporal coincidence between the GRB and the SN phenomenon is explained in term of the concept of ``induced'' gravitational collapse \citep{2001ApJ...555L.117R,Mosca_Orale}. There is also the distinct possibility of observing the young born NS out of the SN \citep[see e.g. Ref.][and references therein]{Mosca_Orale}.

It has been often proposed that GRBs associated with SNe Ib/c, at smaller redshift $0.0085 < z < 0.168$ \citep[see e.g. Ref.][and references therein]{2006AIPC..836..367D}, form a different class, less luminous and possibly much more numerous than the high luminosity GRBs at higher redshift \citep{2006Natur.442.1011P,2004Natur.430..648S,2007ApJ...658L...5M,2006AIPC..836..367D}. Therefore they have been proposed to originate from a separate class of progenitors \citep{2007ApJ...662.1111L,2006ApJ...645L.113C}. In our model this is explained by the nature of the progenitor system leading to the formation of the black hole with the smallest possible mass: the one formed by the collapse of a just overcritical NS \citep{RuffiniTF1,Mosca_Orale}.

\subsection{Conclusions}

GRB 060218 presents a variety of peculiarities, including its extremely large $T_{90}$ and its classification as an XRF. Nevertheless, a crucial point of our analysis is that we have successfully applied to this source our ``canonical GRB'' scenario.

Within our model there is no need for inserting GRB 060218 in a new class of GRBs, such as the XRFs, alternative to the ``canonical'' ones. This same point recently received strong observational support in the case of GRB 060218 \citep{2006ApJ...653L..81L} and a consensus by other models in the literature \citep{2007ApJ...654..385K}.

The anomalously long $T_{90}$ led us to infer a monotonic decrease in the CBM effective density giving the first clear evidence for a fragmentation in the fireshell. This phenomenon appears to be essential in understanding the features of also other GRBs \citep[see e.g. GRB 050315 in Refs.][]{Mosca_Orale,2007A&A...474L..13B}.

Our ``canonical GRB'' scenario originates from the gravitational collapse to a black hole and is now confirmed over a $10^6$ range in energy \citep[see e.g. Ref.][and references therein]{2007AIPC..910...55R}. It is clear that, although the process of gravitational collapse is unique, there is a large variety of progenitors which may lead to the formation of black holes, each one with precise signatures in the energetics. The low energetics of the class of GRBs associated with SNe, and the necessity of the occurrence of the SN, naturally leads in our model to identify their progenitors with the formation of the smallest possible black hole originating from a NS overcoming his critical mass in a binary system. For GRB 060218 there is no need within our model for a new or unidentified source such as a magnetar or a collapsar.

GRB 060218 is the first GRB associated with SN with complete coverage of data from the onset all the way up to $\sim 10^6$ s. This fact offers an unprecedented opportunity to verify theoretical models on such a GRB class. For example, GRB 060218 fulfills the \citet{2002A&A...390...81A} relation unlike other sources in its same class. This is particularly significant, since GRB 060218 is the only source in such a class to have an excellent data coverage without gaps. We are currently examining if the missing data in the other sources of such a class may have a prominent role in their non-fulfillment of the \citet{2002A&A...390...81A} relation \citep[Dainotti et al., in preparation; see also Ref.][]{2006MNRAS.372.1699G}.

	\section{Theoretical background for GRBs' empirical correlations}

The detection of GRBs up to very high redshifts \citep[up to $z = 6.7$, see Ref.][]{2008arXiv0810.2314G}, their high observed rate of one every few days, and the progress in the theoretical understanding of these sources all make them useful as cosmological tools, complementary to supernovae Ia, which are observed only up to $z = 1.7$ \citep{2001ARA&A..39...67L,2001ApJ...560...49R}. One of the hottest topics on GRBs is the possible existence of empirical relations between GRB observables \citep{2002A&A...390...81A,2004ApJ...616..331G,2004ApJ...609..935Y,2005ApJ...633..611L,2006MNRAS.370..185F,2008MNRAS.391..577A}, which may lead, if confirmed, to using GRBs as tracers of models of universe. The first empirical relation, discovered when analyzing the \emph{BeppoSAX} so-called ``long'' bursts with known redshift, was the ``Amati relation'' \citep{2002A&A...390...81A}. It was found that the isotropic-equivalent radiated energy of the prompt emission $E_{iso}$ is correlated with the cosmological rest-frame $\nu F_{\nu}$ spectrum peak energy $E_{p,i}$: $E_{p,i}\propto (E_{iso})^{a}$, with $a = 0.52 \pm 0.06$ \citep{2002A&A...390...81A}. The existence of the Amati relation has been confirmed by studying a sample of GRBs discovered by Swift, with $a = 0.49^{+0.06}_{-0.05}$ \citep{2006ApJ...636L..73S,2006MNRAS.372..233A}.

Swift has for the first time made it possible to obtain high quality data in selected energy bands from the GRB trigger time all the way to the latest extended afterglow phases \citep{2004ApJ...611.1005G}. This has given us the opportunity to apply our theoretical ``fireshell'' model, thereby obtaining detailed values for its two free parameters, namely for the total energy $E^{e^\pm}_{tot}$ and the baryon loading $B$ of the fireshell, as well as for the effective density and filamentary structure of the CBM. From this we were able to compute multi-band light curves and spectra, both instantaneous and time-integrated, compared with selected GRB sources, such as GRB 050315.

In the ``fireshell'' model, $E^{e^\pm}_{tot}$ comprises two different components: (i) the P-GRB with energy $E_{P-GRB}$, emitted at the moment when the $e^+e^-$-driven accelerating baryonic matter reaches transparency, and (ii) the following extended afterglow phase with energy $E_{aft}$, with the decelerating baryons interacting with the CBM \citep{2001ApJ...555L.113R}. These two phases are clearly distinguishable by their relative intensity and temporal separation in arrival time. We have
\begin{equation}
E^{e^\pm}_{tot} = E_{P-GRB} + E_{aft}\, .
\label{Etot}
\end{equation}
What is usually called the ``prompt emission'' corresponds within the fireshell model to the P-GRB together with the peak of the extended afterglow \citep[see below, e.g. Ref.][and references therein]{2001ApJ...555L.113R,2006ApJ...645L.109R,2007AIPC..910...55R,2007A&A...471L..29D,2007A&A...474L..13B,2008AIPC..966...16C,2008AIPC..966...12B}.

Among the crucial issues raised by the Amati relation, there are its theoretical explanation and its possible dependence on the assumed cosmological model. We examined a set of ``gedanken'' GRBs, all at the same cosmological redshift of GRB 050315. Such a set assumes the same fireshell baryon loading and effective CBM distribution as GRB 050315 \citep{2006ApJ...645L.109R} and each ``gedanken'' GRB differs from the others uniquely by the value of its total energy $E^{e^\pm}_{tot}$. We then considered a second set of ``gedanken'' GRBs, differing from the previous one by assuming a constant effective CBM density instead of the one inferred for GRB 050315. In both these sets, we looked for a relation between the isotropic-equivalent radiated energy of the \emph{entire} extended afterglow $E_{aft}$ and the corresponding time-integrated $\nu F_{\nu}$ spectrum peak energy $E_p$:
\begin{equation}
E_p\propto (E_{aft})^a\, .
\label{al}
\end{equation}

In this chapter, after briefly recalling the various spectral-energy correlations mentioned above, we present the derivation of the $E_p$ -- $E_{aft}$ relation for the two sets of ``gedanken'' GRBs.

\subsection{Spectral-energy correlations}

Many empirical spectral-energy correlations exist, some are purely phenomenological and assumption free while others are based on assumptions and are dependent on model, basically the standard fireball model \citep{1999PhR...314..575P}. Some correlations assume spherical symmetry while others assume collimated (jet) emission. This last case was triggered by the observation by \citet{2001ApJ...562L..55F} that the collimation corrected energetics of those GRBs of know jet aperture angles clustered into a narrow distribution, $E_\gamma = (1-\cos\theta_j) E_{iso}\sim 10^{51}$ erg. The opening angle of the jet is estimated within the standard model as
\begin{eqnarray}
\theta_{\rm j} &=& 0.161 \,
\left(\frac{ t_{\rm jet,d} }{ 1+z}\right)^{3/8} 
\left(\frac{n \, \eta_{\gamma}}{ E_{\rm iso,52}}\right)^{1/8};
\,\,\,  \quad {\rm H} \nonumber \\
\theta_{\rm j} &=& 0.2016 \,
\left( \frac{t_{\rm jet, d} }{ 1+z}\right)^{1/4} 
\left( \frac{ \eta_\gamma\ A_* }{ E_{\rm iso,52}}\right)^{1/4}; \quad {\rm W } 
\label{theta} 
\end{eqnarray}

where $t_{\rm jet,d}$ is the break time measured in days and $z$ is the redshift. The efficiency $\eta_\gamma$ relates the isotropic kinetic energy of the fireball $E_{\rm k, iso}$ to the prompt emitted energy $E_{\rm iso}$: $E_{\rm k, iso}= E_{\rm iso}/\eta_\gamma$. Usually, it is assumed a constant value for all bursts, i.e. $\eta_\gamma=0.2$ (after its first use by \citet{2001ApJ...562L..55F}, following the estimate of this parameter in GRB 970508 \cite{2000ApJ...537..191F}). In the homogeneous (H) case, $n$ is the CircumBurst density, independent from the radial coordinate; for the wind (W) case, the density is a function of the radial coordinate, $n(r)=A\,r^{-2}$ and $A_*$ is the value of A ($A=\dot{M}_w/(4\pi v_w)=5\times 10^{11}\, A_*$ g cm$^{-1}$) when setting the wind mass loss rate to $\dot{M}_w=10^{-5}M_\odot yr^{-1}$ and the wind velocity to $v_w=10^3$ km s$^{-1}$. Usually, a constant value (i.e. $A_*=1$) is adopted for all bursts.

The most important spectral-energy correlations are:
\begin{itemize}
\item \textbf{The Amati relation:} It was historically the first correlation discovered, considering \emph{BeppoSAX} bursts \citep{2002A&A...390...81A}. It was found that the isotropic-equivalent radiated energy of the prompt emission $E_{iso}$ is correlated with the cosmological rest-frame $\nu F_{\nu}$ spectrum peak energy $E_{p,i}$: $E_{p,i}\propto E_{iso}^{0.5}$. This correlation, recently updated \citep{2006MNRAS.372..233A} to a larger sample, holds for all but two long bursts, while no short burst satisfies it. The long burst outliers are GRB 980425 and the debated GRB 031203 \citep{2006ApJ...636..967W}. As far as short bursts are concerned, there are two cases: the burst with an initial spike-like emission followed by a soft bump (short burst with afterglow) and the short burst with no afterglow. The burst belonging to the first class are what we named \citep{2007A&A...474L..13B} ``fake'' or ``disguised'' short GRBs, while the ones belonging to the second case are the ``genuine'' short GRBs. Both classes, as already said above, does not follow the Amati relation, but, if one excludes the initial spike-like emission and considers only the soft later part of the bursts in the first class, then the Amati relation is recovered \citep{amatiIK,2008AIPC..966....7B}.
\item \textbf{The Yonetoku correlation:} \citet{2004ApJ...609..935Y} showed that also the peak luminosity $L_{p,iso}$ of the prompt emission correlates with $E_p$, in the same way as $E_{iso}$: $E_p\propto L_{p,iso}^{1/2}$. The scatter is similar to the scatter of the Amati correlation, and the outliers are the same as well.
\item \textbf{The Ghirlanda correlation:} Assuming a collimated emission, \citet{2004ApJ...616..331G} found that the collimation corrected (by a factor $(1-\cos\theta_{j})$) energy, $E_\gamma$, is tightly correlated with $E_p$. The correlation is $E_{p,i}\propto E_{\gamma}^{0.7}$. As outlined above, this relation is based on a theoretical model needed to calculate $\theta_j$, that in turns relies on the assumptions adopted for the efficiency and the CircumBurst density and profile.
\item \textbf{The Liang \& Zhang correlation:} To find the jet angle $\theta_j$, as explained above, one needs a model and some assumptions; the \citet{2005ApJ...633..611L} correlation instead is entirely phenomenological, so model independent and assumptions free. It involves three observables (plus the redshift) and it is of the form $E_{iso}\propto E_p^2 t_{jet}^{-1}$. It is consistent \citep{2006A&A...450..471N} with the Ghirlanda correlation, and has similar spread.
\item \textbf{The Firmani correlation:} The \citet{2006MNRAS.370..185F} correlation links three quantities of the prompt emission: the bolometric isotropic peak luminosity $L_{\rm p}$, the peak energy $E_{\rm p, iso}$ of the time integrated spectrum, and a characteristic time: $T_{0.45}$, which is the time interval spanned by the brightest $45\%$ of the total light curve counts above the background. This time is used to characterize the variability properties of the prompt emission \citep{2001ApJ...552...57R}. The correlation is of the form: $L_{\rm p, iso} \propto E_{\rm p}^{3/2} T_{0.45}^{-1/2}$. Also this relation is model independent and assumption free.
\end{itemize}

\subsection{The $E_p$ -- $E_{aft}$ relation}

In our approach, only the \emph{entire} extended afterglow emission is considered in establishing our $E_p$ -- $E_{aft}$ relation. From this assumption one derives, in a natural way, that the Amati relation holds only for long GRBs, where the P-GRB is negligible, and not for short GRBs \citep{2007A&A...463..913A}.

We can compute the ``instantaneous'' spectrum of GRB 050315 at each value of the detector arrival time during the entire extended afterglow emission. Such a spectrum sharply evolves in the arrival time, presenting a typical hard-to-soft behavior \citep{2006ApJ...645L.109R}. We then computed the $\nu F_\nu$ time-integrated spectrum over the total duration of our extended afterglow phase, that is, from the end of the P-GRB up to when the fireshell reaches a Lorentz gamma factor close to unity. We can then define the energy $E_p$ as the energy of the peak of this $\nu F_\nu$ time-integrated spectrum, and we look at its relation with the total energy $E_{aft}$ of the extended afterglow.

\begin{figure}
\includegraphics[width=\hsize,clip]{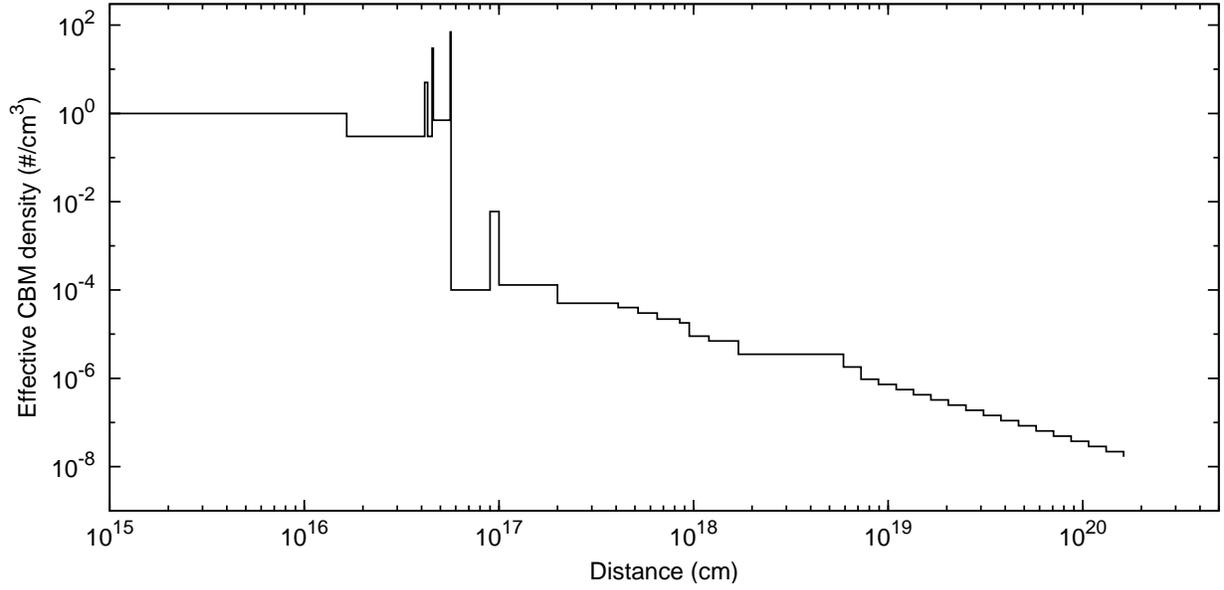}
\caption{The effective CBM number density inferred from the theoretical analysis of GRB 050315. Details in \citet{2006ApJ...645L.109R}.}
\label{mask050315}
\end{figure}

We construct two sets of ``gedanken'' GRBs at a fixed cosmological redshift, therefore independently of the cosmological model. The first set assumes the same fireshell baryon loading and effective CBM distribution as GRB 050315 (see Fig. \ref{mask050315}) and each ``gedanken'' GRB differs from the others uniquely by the value of its total energy $E^{e^\pm}_{tot}$. The second set assumes a constant effective CBM density $\sim 1$ particle/cm$^3$ instead of the one inferred for GRB 050315.

\begin{figure}
\includegraphics[width=\hsize,clip]{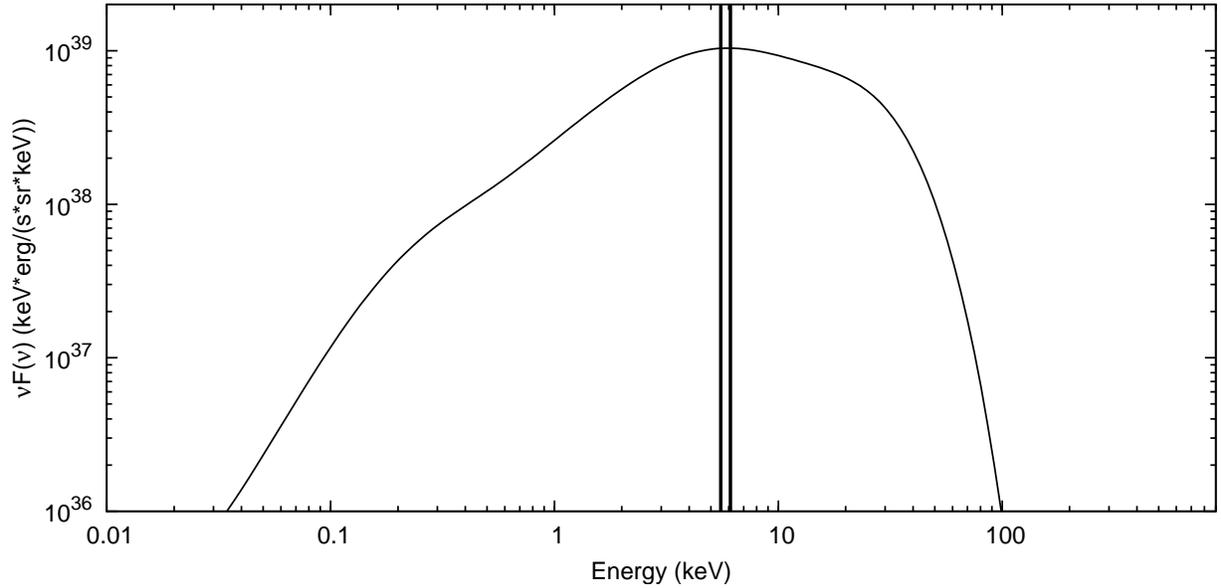}
\caption{The $\nu F_\nu$ time-integrated spectrum over the total duration of our extended afterglow phase for the ``gedanken'' GRB of the first set with total energy $E^{e^\pm}_{tot} = 3.40\times 10^{51}$. The two vertical lines constrain the $5\%$ error region around the peak. We determine $E_p = 5.82$ keV $\pm 5\%$.}
\label{maximum_spectrum}
\end{figure}

In our model, $E_{aft}$ is a fixed value determined by $E^{e^\pm}_{tot}$ and $B$, so clearly there are no errors associated to it. Instead, $E_p$ is evaluated from the numerically calculated spectrum, and its determination is therefore affected by the numerical resolution. Choosing a $5\%$ error on $E_p$, which is consistent with our numerical resolution, we checked that this value is reasonable looking at each spectrum. Figure \ref{maximum_spectrum} shows the time-integrated spectrum corresponding to $E^{e^\pm}_{tot} = 3.40\times 10^{51}$ erg with the error around $E_p$.

\subsection{Results and discussion}

\begin{figure}
\includegraphics[width=\hsize,clip]{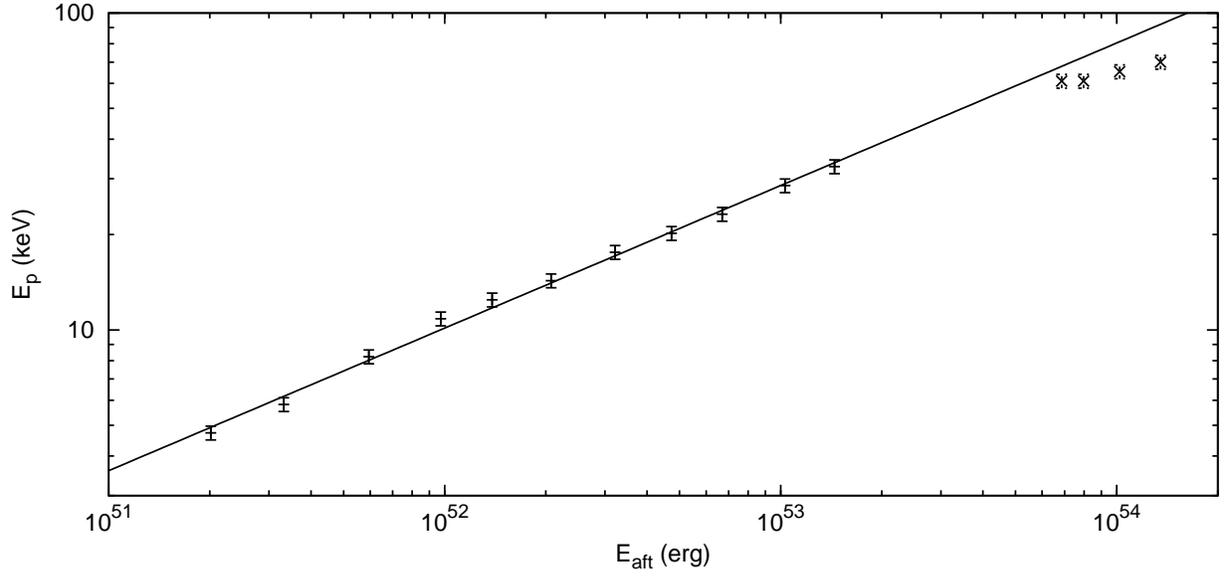}
\caption{The $E_p$ -- $E_{aft}$ relation: the results of the simulations of the first set of ``gedanken'' GRBs (points marked as crosses) are well-fitted by a power law (solid line) $E_p\propto (E_{aft})^a$ with $a = 0.45 \pm 0.01$. The points marked as ``X'' are the results of the extension of the first set above $10^{53}$ erg.}
\label{fig-correlation_extended}
\end{figure}

Figure \ref{fig-correlation_extended} shows the $E_p$ -- $E_{aft}$ relation of the ``gedanken'' GRBs belonging to the first set (red points). It extends over two orders of magnitude in energy, from $10^{51}$ to $10^{53}$ erg, and is well-fitted by a power law $E_p\propto (E_{aft})^a$ with $a = 0.45 \pm 0.01$. We emphasize that such a power-law slope strictly agrees with the Amati relation, namely $E_{p,i}\propto (E_{iso})^{a}$, with $a = 0.49 ^{+0.06}_{-0.05}$ \citep{2006MNRAS.372..233A}. We recall that $E_p$ is the observed peak energy; i.e., it is not rescaled for the cosmological redshift, because all the ``gedanken'' GRBs of the set are at the same redshift of GRB 050315, namely $z=1.949$ \citep{2006ApJ...638..920V}. The normalization is clearly different from the Amati one.

\begin{figure}
\includegraphics[width=\hsize,clip]{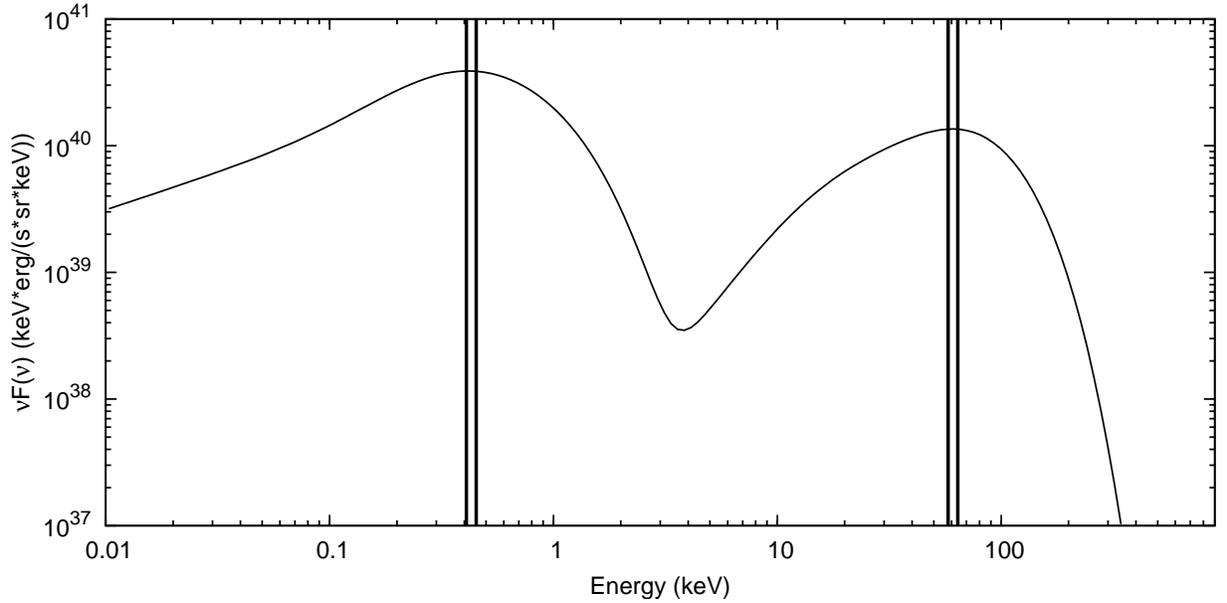}
\caption{The $\nu F_\nu$ time-integrated spectrum over the total duration of our extended afterglow phase for the ``gedanken'' GRB of the extended first set with total energy $E^{e^\pm}_{tot} = 6.95\times 10^{53}$ erg. The vertical lines constrain the $5\%$ error region around each peak.}
\label{spectrum_G_2_peaks}
\end{figure}

If we try to extend the first sample of ``gedanken'' GRBs below $10^{51}$ erg, the relevant CBM distribution would be for $r \lesssim 10^{16}$ cm, where no data are available from the GRB 050315 observations. If we try to extend the first set of ``gedanken'' GRBs above $10^{53}$ erg, we notice that for $E^{e^\pm}_{tot} \gtrsim 10^{54}$ erg the small ``bump'', which can be noticed between $0.2$ and $1.0$ keV in the spectrum of Fig. \ref{maximum_spectrum}, evolves into a low-energy second spectral peak that is even higher than the high-energy one (see Fig. \ref{spectrum_G_2_peaks}). We are currently investigating whether this low-energy second peak is a real, theoretically predicted spectral feature that may be observed in the future in highly energetic sources. There is also the other possibility that the low-energy and late part of our GRB 050315 fit is not enough constrained by the XRT observational data so that this effect is magnified by the $E^{e^\pm}_{tot}$ rescaling.

The high-energy spectral peak is due to the emission at the peak of the extended afterglow, and therefore due to the so-called ``prompt emission''. The low-energy one is due to late-time soft X-ray emission. Therefore, the high-energy spectral peak is the relevant one for the Amati relation. We find indeed that such a high-energy spectral peak still fulfills the $E_p$ -- $E_{aft}$ relation for $E^{e^\pm}_{tot} \sim 10^{54}$ erg, with a possible saturation for $E^{e^\pm}_{tot} > 10^{54}$ erg (see Fig. \ref{fig-correlation_extended}).

\begin{figure}
\includegraphics[width=\hsize,clip]{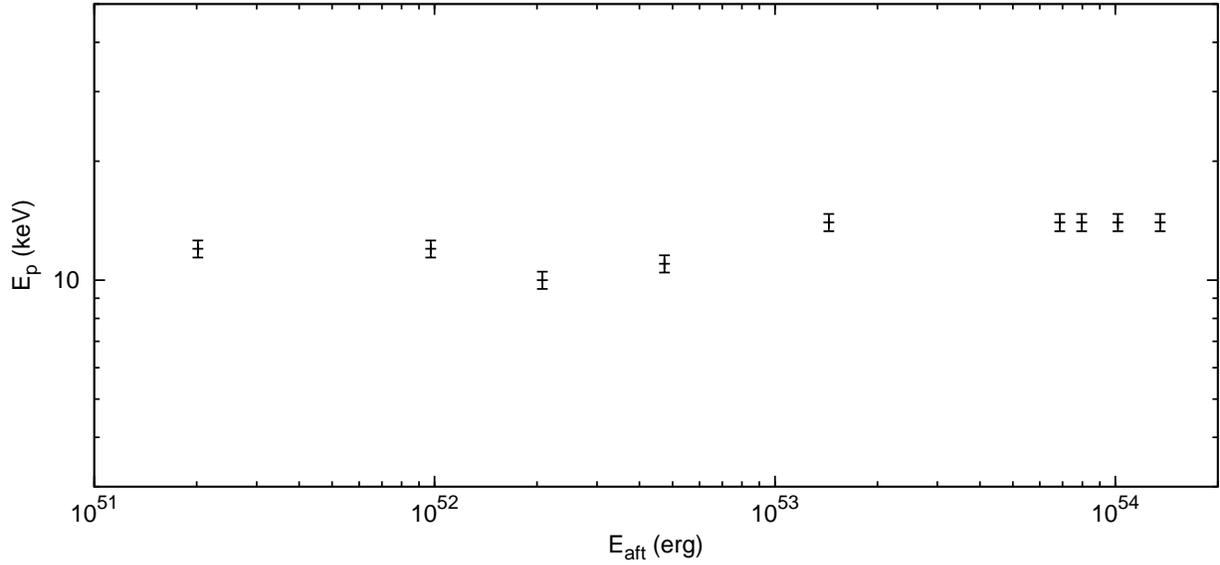}
\caption{The second set of ``gedanken'' GRBs. Clearly, in this case there in no relation between $E_p$ and $E_{aft}$.}
\label{correlation-no-mask}
\end{figure}

Figure \ref{correlation-no-mask} clearly shows that in the second set of ``gedanken'' GRBs, built assuming a constant effective CBM density $\sim 1$ particle/cm$^3$, instead of the one specifically inferred for GRB 050315, there in no relation between $E_p$ and $E_{aft}$.

\subsection{Conclusions}

The high-quality Swift data, for the first time giving gapless and multiwavelength coverage from the GRB trigger all the way to the latest extended afterglow phases, have led to a complete fit of the GRB 050315 multiband light curves based on our fireshell model. We fixed the free parameters describing the source and determined the instantaneous and time-integrated spectra during the entire extended afterglow.

Starting from this, we examined two sets of ``gedanken'' GRBs, constructed at a fixed cosmological redshift. The first set assumes the same fireshell baryon loading and effective CBM distribution as GRB 050315, and each ``gedanken'' GRB differs from the others uniquely by the value of its total energy $E^{e^\pm}_{tot}$. The second set assumes a constant effective CBM density $\sim 1$ particle/cm$^3$ instead of the one inferred for GRB 050315.

Recalling that the ``canonical'' GRB light curve in the fireshell model is composed of two well-separated components, the P-GRB, and the entire extended afterglow, we looked for a relation in both sets between the isotropic-equivalent radiated energy of the \emph{entire} extended afterglow $E_{aft}$ and the corresponding time-integrated $\nu F_{\nu}$ spectrum peak energy $E_p$: $E_p\propto (E_{aft})^a$. In doing so, we assumed that the Amati relation is directly linked to the interaction between the accelerated baryons and the CBM. The P-GRBs, which originate from the fireshell transparency, do not fulfill the Amati relation in our approach. Consequently, the short GRBs, which have a vanishing extended afterglow with respect to the P-GRB, should also not fulfill the Amati relation. This last point is supported by the observational evidence \citep{2006MNRAS.372..233A}.

We notice that the first set of ``gedanken'' GRBs fulfills the $E_p\propto (E_{aft})^a$ relation very well with $a = 0.45 \pm 0.01$. This slope strongly agrees with the Amati relation. In contrast, no relation between $E_p$ and $E_{aft}$ seems to hold for the second set. We conclude that the Amati relation originates from the detailed structure of the effective CBM.

Turning now to the analogies and differences between our $E_p$ -- $E_{aft}$ relation and the Amati one, our analysis excludes the P-GRB from the prompt emission, extends all the way to the latest extended afterglow phases, and is independent of the assumed cosmological model, since all ``gedanken'' GRBs are at the same redshift. The Amati relation, on the other hand, includes the P-GRB, focuses only on the prompt emission, being therefore influenced by the instrumental threshold that fixes the end of the prompt emission, and depends on the assumed cosmology. This might explain the intrinsic scatter observed in the Amati relation \citep{2006MNRAS.372..233A}. Our theoretical work is a first unavoidable step toward supporting the use of the empirical Amati relation for measuring the cosmological parameters.

	\section{Thermalization process of electron-positron plasma with baryon loading}

Initial evolution of electron-positron-photon plasma in the source of a GRB has a key role in the subsequent dynamics of the fireshell. In particular, particle spectra, temperatures, chemical potentials all need to be known in order to describe acceleration of the fireshell and in general its expansion in terms of hydrodynamics. Since quite different theoretical arguments existed on the initial state of optically thick electron-positron-photon plasma in GRBs \citep[see e.g. Refs.][]{1978MNRAS.183..359C,1986ApJ...308L..47G} we turned to analysis of kinetic properties of nonequilibrium electron-positron pairs.

Having this goal in mind \citet{2007PhRvL..99l5003A} solved numerically relativistic Boltzmann equations for distribution functions of electrons, positrons and photons, assuming their uniform spatial distribution. Considering energy density in the range, typical for GRBs, the relevant thermalization timescales were determined. It turns out that particles reach kinetic equilibrium on a timescale $t<10^{-14}$ sec, when distribution functions of electrons/positrons (photons) acquire Fermi-Dirac (Bose-Einstein) form, all particles have a common temperature but nonzero chemical potentials. Further, on a timescale $t<10^{-12}$ sec chemical potentials vanish and particles reach thermal distribution.

Since in many bursts baryon loading is dynamically significant \citet{2009PhRvD..79d3008A} considered proton admixture parametrized by
the parameter $B=n_{p}m_{p}c^{2}/\rho_{r}$, where $n_{p}$ is the number density of protons, $m_{p}$ is their mass, $\rho_{r}$ is the radiative energy density (including the energy density of electron-positron pairs). Independent on the baryon loading the thermalization timescale was found to be $t<10^{-11}$ sec for such a plasma, despite thermalization process is more
complicated.

\begin{figure}
\centering
\includegraphics[width=\hsize]{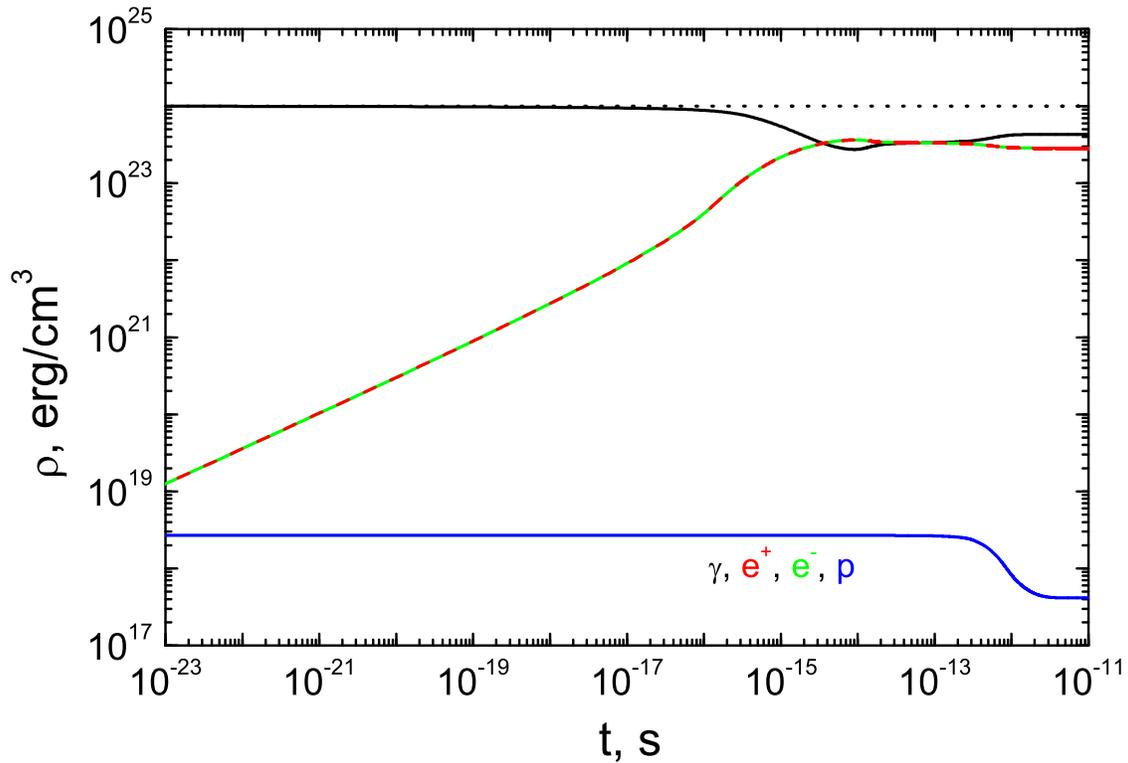}
\caption{Depencence on time of energy densities of electrons (green), positrons (red), photons (black) and protons (blue). Total energy density is shown by dotted black line. Interaction between pairs and photons operates on very short timescales up to $10^{-23}$ sec. Quasi-equilibrium state is established at $t_{\mathrm{k}}\simeq10^{-14}$ sec which corresponds to kinetic
equilibrium for pairs and photons. Protons start to interact with them as late as at $t_{\mathrm{th}}\simeq10^{-13}$ sec.}
\label{rho1}
\end{figure}

\begin{figure}
\centering
\includegraphics[width=\hsize]{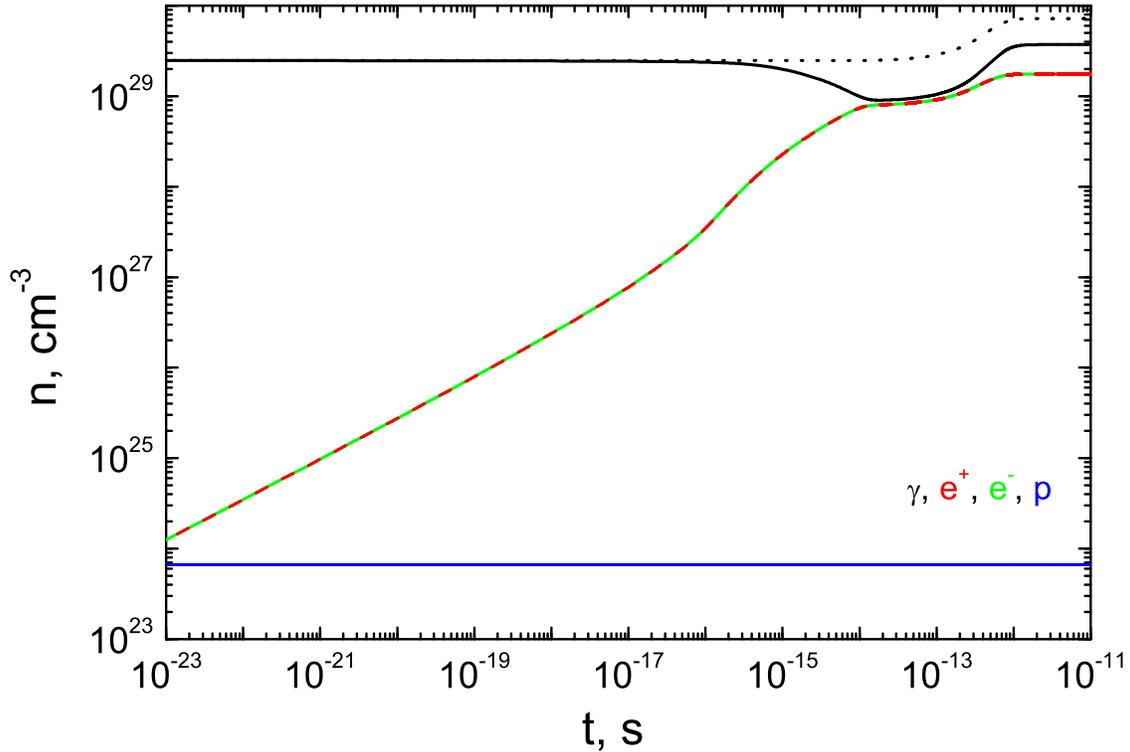}
\caption{Depencence on time of concentrations of electrons (green), positrons(red), photons (black) and protons (blue). Total number density is shown by dotted black line. In this case kinetic equilibrium between electrons, positrons and photons is reached at $t_{\mathrm{k}}\simeq10^{-14}$ sec. Protons join thermal equilibrium with other particles at $t_{\mathrm{th}}\simeq4\times10^{-12}$ sec.}
\label{n1}
\end{figure}

\begin{figure}
\centering
\includegraphics[width=\hsize]{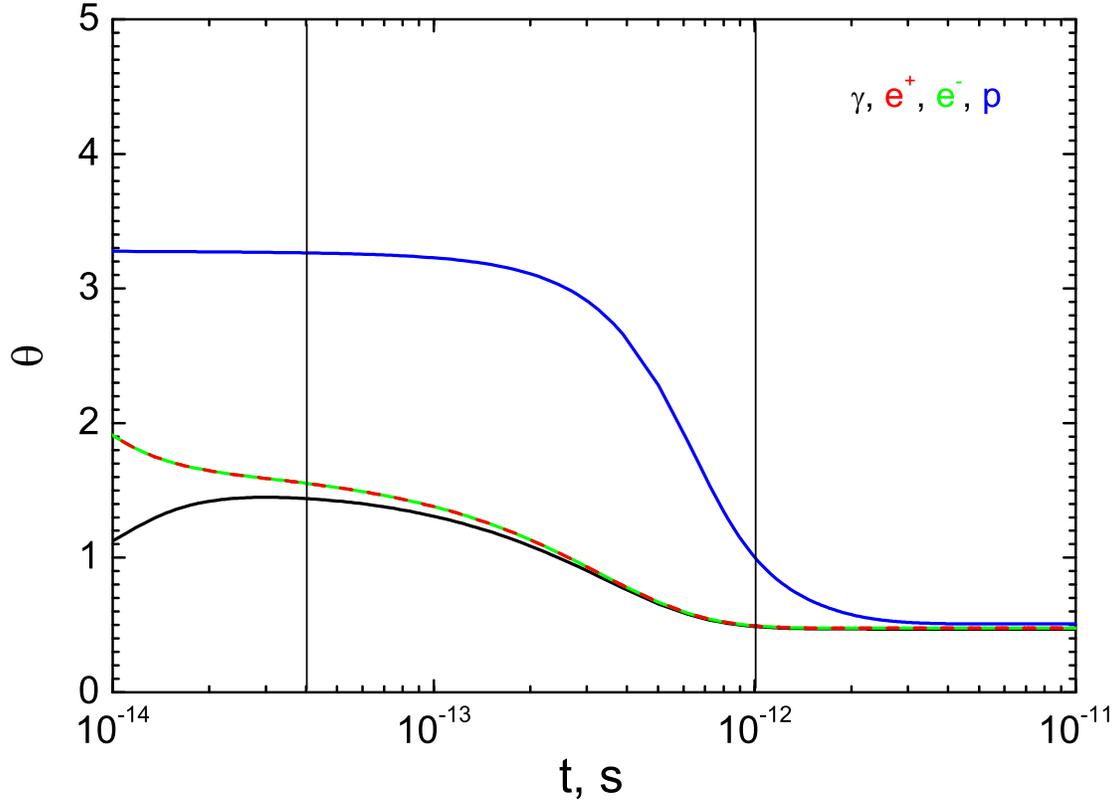}
\caption{Depencence on time of dimensionless temperature of electrons (green), positrons (red), photons (black) and protons (blue). The temperature for pairs and photons acquires physical meaning only in kinetic equilibrium at $t_{\mathrm{k}}\simeq10^{-14}$ sec. Protons are cooled by the pair-photon plasma and acquire common temperature with it as late as at $t_{\mathrm{th}}\simeq4\times10^{-12}$ sec.}
\label{theta1}
\end{figure}

\begin{figure}
\centering
\includegraphics[width=\hsize]{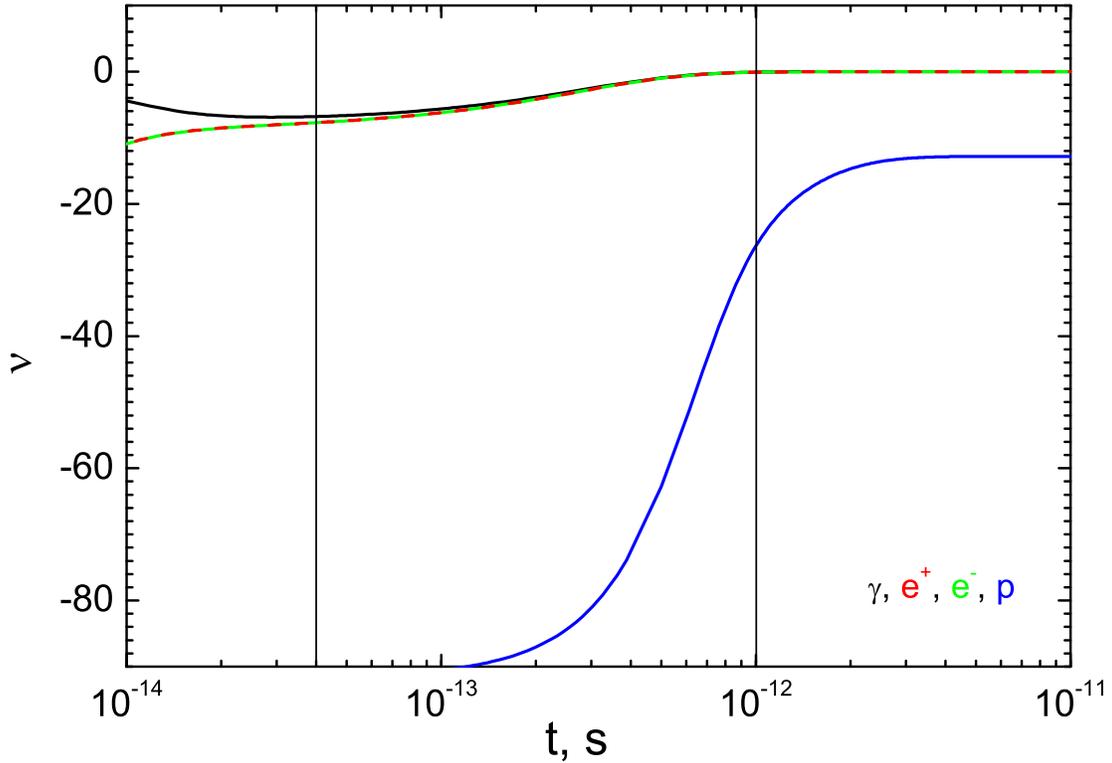}
\caption{Depencence on time of dimensionless chemical potential of electrons (green), positrons (red), photons (black) and protons (blue). The chemical potential for pairs and photons acquires physical meaning only in kinetic equilibrium at $t_{\mathrm{k}}\simeq10^{-14}$ sec, while for protons this happens at $t_{\mathrm{th}}\simeq4\times10^{-12}$ sec. At this time chemical potential of photons has evolved to zero and thermal equilibrium has been already reached.}
\label{phi1}
\end{figure}

\begin{figure}
\centering
\includegraphics[width=\hsize]{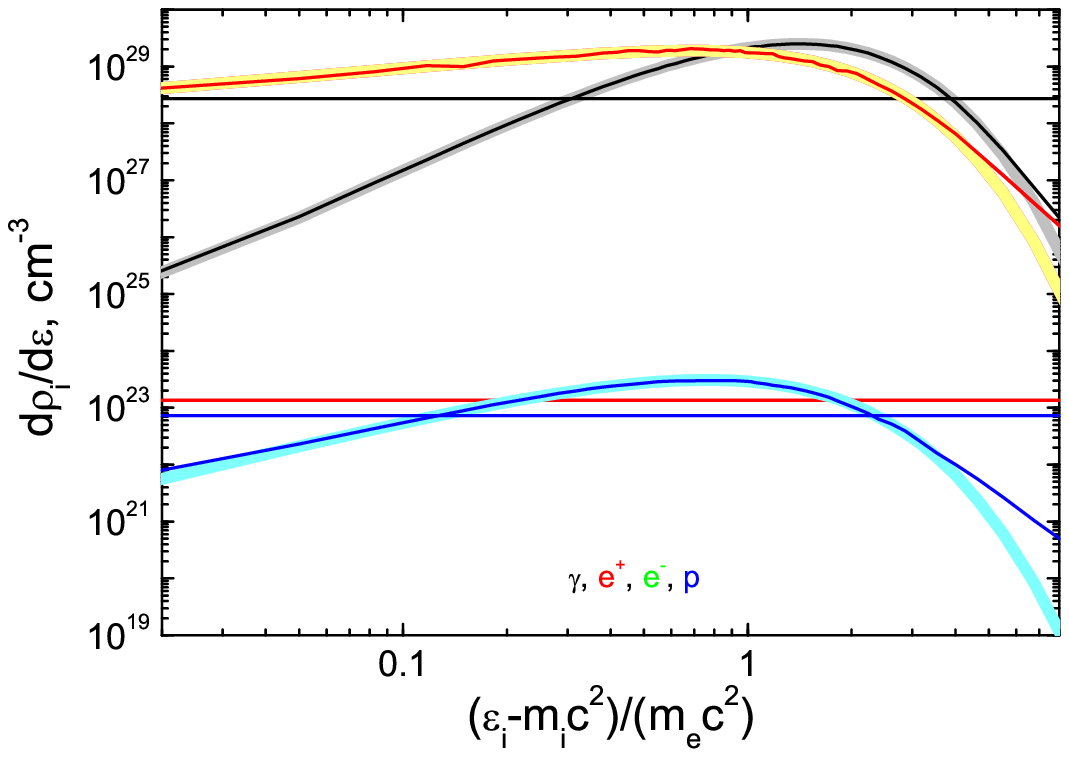}
\caption{Spectral density as function of particle energy for electrons (green), positrons (red), photons (black) and protons (blue) in initial and final time moments of the computation. Fits of the spectra with chemical potentials and temperatures corresponding to thermal equilibrium state are also shown by yellow (electrons and positrons), gray (photons) and light blue (protons) thick lines. The final photon spectrum is black body one.} 
\label{spectra1}
\end{figure}

As example, we show energy densities (Fig. \ref{rho1}), number densities (Fig. \ref{n1}), temperatures and chemical potentials (Fig.\ref{theta1}, \ref{phi1} respectively) depending on time. Initial and final spectra of particles are shown in Fig. \ref{spectra1}. Initial conditions were chosen with flat spectral densities and total energy density $\rho=10^{24}$ erg/cm$^{3}$. This initial state is clearly far from equilibrium. Interactions between particles change distribution functions such that at the moment $t_{1}=4\times10^{-14}$ sec, shown by the vertical line on the left in Fig. \ref{theta1} and \ref{phi1}, distribution functions acquire an equilibrium form with temperature of photons and pairs is $\theta_{\mathrm{k}}=k_{B}T_{\mathrm{k}}/(m_{e}c^{2})\simeq1.5$, while the chemical potentials of these particles are $\nu_{\mathrm{k}}=\varphi_{\mathrm{k}}/(m_{e}c^{2})\simeq-7$, where $k_{B}$ is Boltzmann's constant, $m_{e}$ is electron mass, and $c$ is the speed of light. These changes are essentially due to binary interactions, which do not change number of particles. Further evolution of the system is due to triple interactions. Only triple interactions, which do not conserve the number of particles, are able to change chemical potentials of particles. The chemical potential of photons reaches zero at $t_{2}=10^{-12}$ sec, shown by the vertical line on the right in Fig. \ref{theta1} and \ref{phi1}, which means that electrons, positrons and photons come to thermal equilibrium. Protons join thermal equilibrium the last. Simulations with different initial conditions show that thermalization timescale depends essentially on total energy density and the baryonic loading parameter $B$ \citep[for details see Ref.][]{2008AIPC..966..191A}.

Such short timescales, compared to a typical expansion timescale $t_{ex}\sim R_{0}/c\sim10^{-3}$ sec where $R_{0}$ is initial size of the plasma allow to speak of completely thermalized plasma long before expansion starts \citep[for details see Ref.][]{2008AIPC..966..191A}.

	\section{Critical electric Fields on the surface of massive cores and Dyadotorus of the Kerr-Newman Geometry}

\subsection{Critical electric Fields on the surface of massive cores}

\begin{figure}
\centering
\includegraphics[width=\hsize]{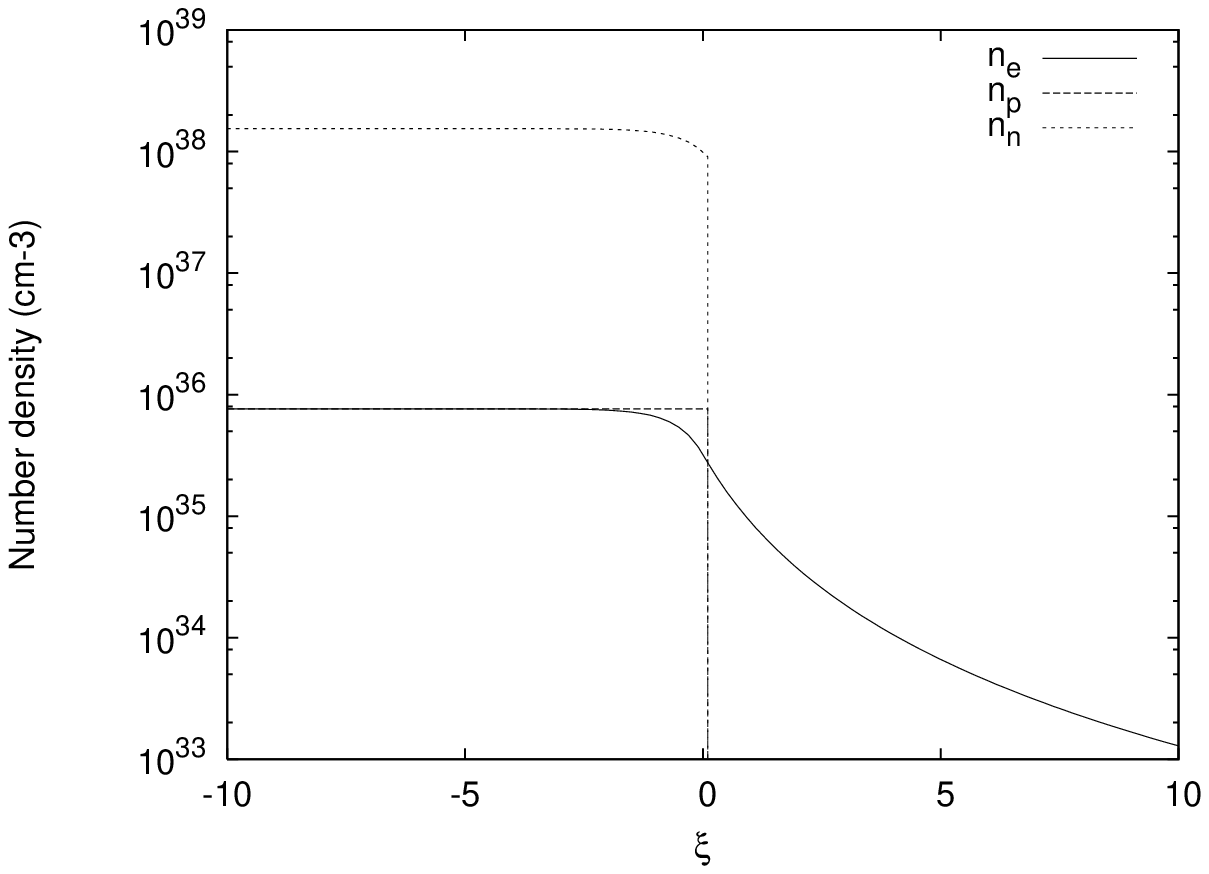}
\caption{Number density of electrons, protons and neutrons.}
\label{introa} 
\end{figure}

\begin{figure}
\centering 
\includegraphics[width=\hsize]{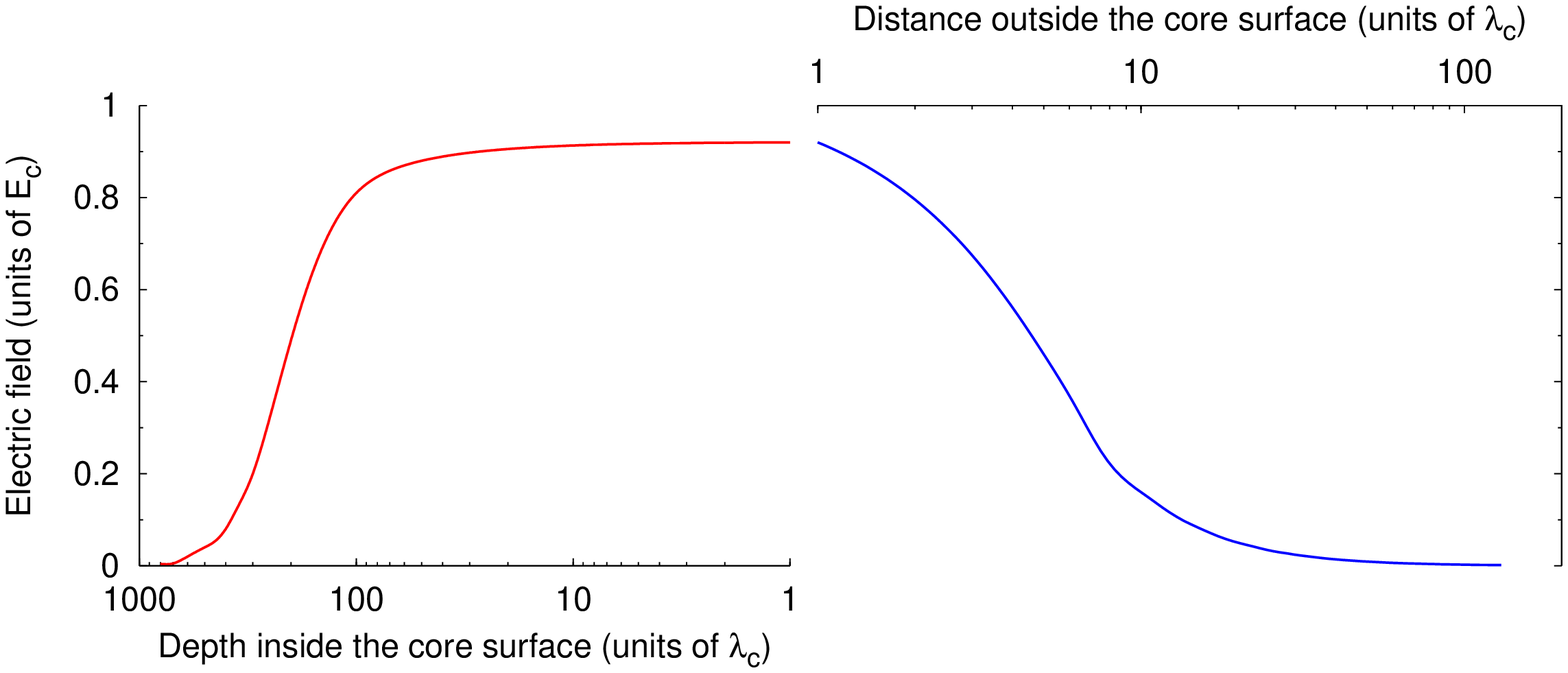}
\caption{Electric Field in units of the critical field.}
\label{introb}
\end{figure}

One of the most active field of research has been to analyse a general approach to Neutron Stars based on the Thomas-Fermi ultrarelativistic equations amply adopted in the study of superheavy nuclei. The aim is to have a unified approach both to superheavy nuclei, up to atomic numbers of the order of $10^5$--$10^6$, and to what we have called ``Massive Nuclear Cores'', which are
\begin{itemize}
\item characterized by atomic number of the order of $10^{57}$;
\item composed by neutrons, protons and electrons in $\beta$--equilibrium;
\item expected to be kept at nuclear density by self gravity.
\end{itemize}
The analysis of superheavy nuclei has historically represented a major field of research \citep{1969ZPhy..218..327P,1971JETP...32..526P,1972SvPhU..14..673Z,1972PhRvL..28.1235M,1982PhT....35h..24G}, guided by Prof. V. Popov and Prof. W. Greiner and their schools. This same problem was studied in the context of the relativistic Thomas-Fermi equation also by R. Ruffini and L. Stella \citep{1980PhLB...91..314F,1981PhLB..102..442R}, already in the 80s. The recent numerical approach has shown the possibility to extrapolate this treatment of superheavy nuclei to the case of Massive Nuclear Cores \citep{2007IJMPD..16....1R}. The very unexpected result has been that also around these massive cores there is the distinct possibility of having an electromagnetic field close to the critical value $E_c = \frac{m_e^2 c^3}{e \hbar}$, although localized in a very narrow shell of the order of the electron Compton wavelength (see Fig. \ref{introa}, \ref{introb}).

The welcome result has been that all the analytic work \citep{1976ZhPmR..24..186M} developed by Prof. Popov and his  Russian collaborators can be straightforwardly applied to the case of massive cores, and the $\beta$--equilibrium condition is properly taken into account. In Ref. \citep{prrx2009}, we show that globally neutral massive cores can be gravitationally bound and the value of the charge-to-mass ratios predicted at the surface of massive cores coincides with the range of values expected in astrophysical scenarios for Kerr-Newman black holes, in addition to a further verification of the over critical electric field on the massive core surface, numerically obtained by Ruffini, Xue and Rotondo already in 2007 \citep{2007IJMPD..16....1R}. A large variety of problems has emerged, and working progress in solving these problems have been going on in the direct discussion with or participation by Prof. Greiner, Prof. Popov, and Prof. 't Hooft at ICRANet center located at Pescara. The crucial issue to be debated is the stability of such cores under the competing effects of self gravity and Coulomb repulsion. In order to probe this stability, we have started a new approach to the problem within the framework of general relativity. The object of the work by Patricelli and Rueda is the generalization of the Tolman-Oppenheimer-Volkoff equation duly taking into account the elecrodynamical contribution. The major scientific issue here is to have a unified approach solving the coupled system of the general relativistic self gravitating electrodynamical problem with the corresponding formulation of the Thomas-Fermi equation in the framework of general relativity. Prof. 't Hooft, in a series of lectures at Pescara, has forcefully expressed the opinion that necessarily, during the process of gravitational collapse, it should occur a more extended distribution of the electromagnetic field to the entire core of the star and not only confined to a thin shell. This is a necessary condition in order to transmit the gravitational energy of the collapse to the electrodynamical component of the field giving possibly rise to large pair creation processes. This crucial idea is currently being pursued by the application to this system of a classical work of Feynmann-Metropolis and Teller, who considered in relativistic Thomas-Fermi the crucial role of non-degeneracy.

\subsection{On the Dyadotorus of the Kerr-Newman Geometry}

In the merging process of two neutron stars and in the final process of gravitational collapse of a black hole it is possible that very large electromagnetic field strength larger than the critical value of vacuum polarization $E_c$ do occur \citep{2007AIPC..910...55R}. The description of the time evolution of the gravitational collapse and the associated electrodynamical process (occurring on characteristic time scales $\tau=GM/c^3\simeq 5\times 10^{-5} M/M_\odot$ s) are too complex for a direct description.  A more confined problem is  the case of an already formed Kerr-Newman black hole.

This deserves analysis in itself as a theoretical problem and may represent a physical condition asymptotically reached in the process of gravitational collapse. Such an asymptotic configuration will be reached when all the multipoles departing from the Kerr-Newman geometry have been radiated away either by process of vacuum polarization or electromagnetic and gravitational waves. This simplified problem may lead to a direct evaluation of the energetics as well as of the created $e^-e^+$ pairs occurring on time scales $\Delta t=\hbar/(m_e c^2) \simeq 10^{-21}$ s.

Therefore we explore the initial condition for such a process by the definition of the spatial extent of a ``dyadotorus'' which generalizes to the Kerr-Newman geometry the concept of the ``dyadosphere'' previously introduced in the case of the spherically symmetric Reissner-Nordstr\"om geometry \citep{1998bhhe.conf..167R,1998A&A...338L..87P}.

\citet{1975PhRvL..35..463D} showed that vacuum polarization processes {\it \`{a} la} Sauter-Heisenberg-Euler-Schwinger \citep{1936ZPhy...98..714H} can occur in the field of a Kerr-Newman black hole endowed with a mass ranging from the maximum critical mass for neutron stars $(3.2M_{\odot})$ all the way up to $7.2\times10^6M_{\odot}$. It is an almost perfectly reversible process in the sense defined by \citet{1971PhRvD...4.3552C}, leading to a very efficient mechanism of extracting energy from the black hole.

In the case of absence of rotation in spacetime, we have a Reissner-Nordstr\"{o}m black hole as the background geometry. The region where vacuum polarization processes take place is a sphere centered about the hole, and has been called dyadosphere \citep{1998bhhe.conf..167R,1998A&A...338L..87P}. We investigate how the presence of rotation in spacetime modifies the shape of the surface containing the region where electron-positron pairs are created.

\begin{figure}
\centering
\includegraphics[width=0.5\hsize]{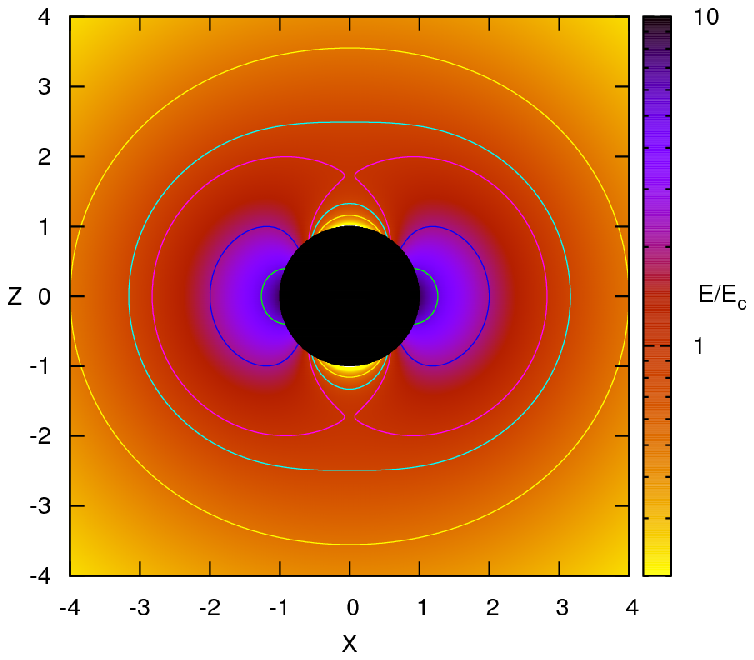}
\includegraphics[width=0.5\hsize]{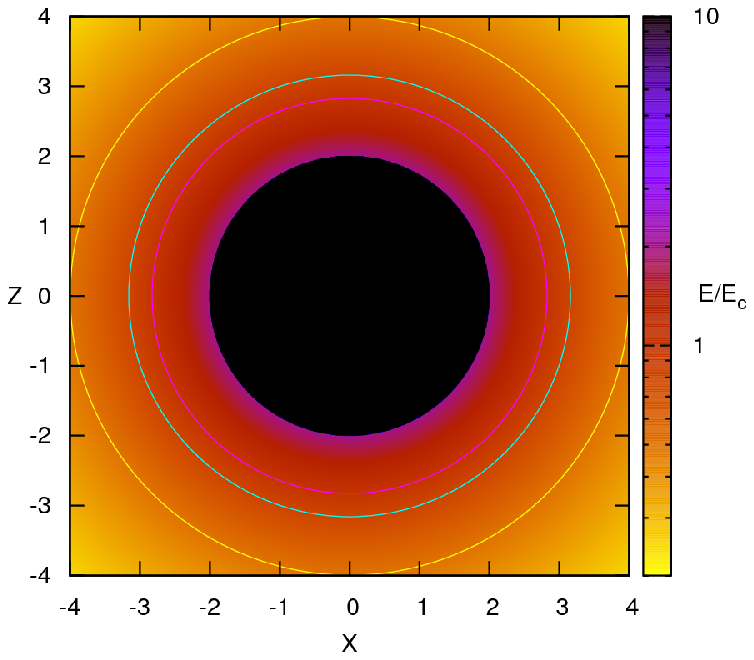}
\caption{The projections of the dyadotorus on the $X-Z$ plane corresponding to different values of the ratio $|{\bf E}|/E_c\equiv k$ are shown in the left pane for $\mu=10$ and $\xi=1.49\times 10^{-4}$.  The corresponding plot for the dyadosphere with the same mass energy  $M$ and charge to mass ratio $\xi$ is shown in the right pane for comparison.}
\label{fig:2bis}
\end{figure}

Due to the axial symmetry we call that region as dyadotorus and we give the conditions for its existence.  We have defined the dyadotorus as the locus of points where $E=k E_c$ with $k$ some positive constant which can be less than one \citep[see Ref.][for details]{dyadotorus}.  We have found that the geometry of the dyadotorus is indeed torus-like when 
\begin{equation}
k \geq \frac{\xi}{8 E_c M_\odot \mu \alpha^2}\approx 6.6\times 10^4 \frac{\xi}{\mu \alpha^2}\, ,
\end{equation}
where $\mu = M/M\odot$, $\xi=Q/M$ and $\alpha=J/M^2$ being $M$, $Q$ and $J$ the mass, the charge and angular momentum of the black hole.  Otherwise, it becomes ellipsoid-like.  This can be seen from Fig.\ref{fig:2bis}.

An estimate of the electromagnetic energy contained in the dyadotorus can be calculated by using, for example, three different definitions of it commonly adopted in the literature, i.e. the standard definition in terms of the timelike Killing vector \citep[see e.g. Ref.][]{2002PhLB..545..233R}, the one recently suggested by Katz, Lynden-Bell and Bi{\v c}\'ak \citep{2006CQGra..23.7111K,2007PhRvD..75b4040L} for axially symmetric asymptotically flat spacetimes, which is an observer dependent definition of energy, and the last one involving the theory of pseudotensors \citep[see e.g.][]{1996GReGr..28.1393A}. All these approaches are shown to give the same results.  

From this, we find that in addition to the topological differences between the dyadotorus and the dyadosphere, larger field strengths and electromagnetic energy are allowed in the case of a Kerr-Newman geometry close to the horizon, when compared with a Reissner-Nordstr\"om black hole of the same mass energy and charge to mass ratio.

	\section{Selected processes originating high-energy emission}

The knowledge of the radiation mechanisms is crucial for the correct
understanding of many astronomical observations. In particular in gamma
ray astronomy the observational data can be often explained by two or
more production mechanisms. It is therefore important to model correctly
the different interactions producing gamma rays. Among all mechanisms
producing gamma rays hadronic interactions between nucleons which produce
pions, which in turn decay into photons and neutrinos, are one of the most
studied models. High energy collisions of nucleons cannot be treated
perturbatively because of the large value of the interaction constant in nuclear
interactions. Many authors have occupied themselves with this
problem and already in $1950$ Fermi developed an elegant statistical method for
computing the multiple production of particles in collisions of high
energetic protons. In the meantime a very large set of data has been acquired from high
energy accelerators. Our aim here is 1. to rederive the Fermi theoretical
equations, 2. to compare them to the experimental data and 3. to explore possibilities of observing such phenomena.

\subsection{Fermi's approach to the study of hadronic interactions}

In treating high energy collisions of nucleons, Fermi made the
assumption that the possible final configurations of the system are
determined by the statistical weights of the various possible final
configurations and accordingly developed a statistical
method to determine the final particles produced \citep{1950PThPh...5..570F}. First of
all one might think of many different final configurations for the system
after the collision, but conservation laws of charge and of momentum, as
well as the feasibly of the processes, have to be taken into account.
Transitions in Yukawa's theory, in which charged and neutral pions are
created, are therefore the most probable processes taking place. So
during the collisions of high energetic hadrons a large amount of energy is released
in a small volume around the hadrons and used to form pions. In view of the strong
interactions between these pions, one can imagine that the energy available in
the small volume will be rapidly distributed to the different pions
having different energies. In other words the energy will be statistically distributed among
all degrees of freedom of the system. Fermi himself said: ``When two nucleons
collide with very great energy in their center of mass
system this energy will be suddenly released in a small volume surrounding
the two nucleons. The event is a collision
in which the nucleons with their surrounding retinue of pions hit against
each other so that all the portion of space occupied by the nucleons and
by their surrounding pion field will be suddenly loaded with a very great
amount of energy. Since the interactions of the pion field are strong we may
expect that rapidly this energy will be distributed among the various
degrees of freedom present in this volume according to statistical laws.
One can then compute statistically the probability that in this tiny volume
a certain number of pions will be created with a given energy distribution.
It is then assumed that the concentration of energy will rapidly dissolve
and that the particles into which the energy has been converted will fly
out in all directions.''

Fermi's method resembles somehow Heisenberg's approach
\citep{1949ZPhy..126..569H} to treat high energy collisions of nucleons, with the
difference that Heisenberg used qualitative ideas of turbulence whereas,
Fermi believed that in high energetic processes statistical equilibrium is
reached.

According to Fermi the process proceeds as follows: in the
laboratory frame a very energetic proton scatters off a proton
target. For convenience the process is examined in the center of
mass of the system. The only parameter which has to be tuned in
Fermi's theory is the volume in which the energy is dumped. The
value of this parameter can be modified to improve the agreement
with the experiments. Fermi defined $\Omega$ being the
volume at rest in the laboratory frame that contains the energy of
the colliding particles. Since the particles mediating the Yukawa interactions
are the pions, the volume is taken as a sphere with radius of order
of the pion Compton wavelength $\lambda_{\pi}^{c} = \hbar/m_\pi c = 1.4\times 10^{-13}$ cm.

We will use the subscript ``$0$'' to indicate quantities in the c.m. frame, and no subscript to indicate the same quantities in the laboratory frame. Analogously, we use the superscript $'$ to indicate quantities after the collision, and no superscript to indicate the same quantities before the collision.

Let's focus for the moment on the target proton, which is at rest in the laboratory frame. Its associated volume $\Omega$ is:
\begin{equation}
\Omega = \frac{4}{3}\pi R^3\, ,
\label{Fermi-f2}
\end{equation}
with
\begin{equation}
R = a\lambda_\pi^c\, ,
\label{Fermi-f3}
\end{equation}
where $a$ is a free parameter of the order of unity which Fermi leaves free to better fit the experimental data. The same volume measured in the c.m. frame, where the target proton is moving with Lorentz factor $\gamma$, is given by:
\begin{equation}
\Omega_0 = \frac{1}{\gamma}\Omega\, .\label{Fermi-f1}
\end{equation}

In order to determine the Lorentz factor $\gamma$, we consider the case in which, in the initial reference frame, the laboratory frame, a particle with mass $m_1$ and energy $E_1$ collides with a particle with mass $m_2$ which is at rest. The total energy of the two particles is:
\begin{equation}
E = E_1 + E_2 = E_1 + m_2c^2\, .
\label{Fermi-fx1}
\end{equation}
Their total momentum is given by:
\begin{equation}
\vec{P} = \vec{p_1}\, .
\label{Fermi-fx2}
\end{equation}
We can observe the same process in the c.m. frame, where the two colliding particles have zero total momentum. The square $s$ of the 4-momentum in the c.m. frame is given by:
\begin{equation}
s = P_0^\alpha {P_0}_\alpha\, .
\label{Fermi-fx3}
\end{equation}
Since $s$ is Lorentz invariant, it must be:
\begin{equation}
s = P^\alpha P_\alpha = \left(E_1 + m_2c^2\right)^2 - p_1^2c^2 = \left(E_1 + m_2c^2\right)^2 - \left(E_1^2 - m_1^2c^4\right) = m_1^2c^4 + m_2^2c^4 + 2E_1m_2c^2\, .
\label{Fermi-fx4}
\end{equation}
If the two particles are protons, $m_1 = m_2 = m_p$. We then have
\begin{equation}
s = 2m_pc^2\left(E_1 + m_pc^2\right)\, .
\label{Fermi-fx5}
\end{equation}
The energy of each proton in the c.m. frame can be written as:
\begin{equation}
E_0 = \gamma m_p c^2\, ,
\label{Fermi-fx6}
\end{equation}
in fact, in the c.m. frame both of them moves with the same Lorentz factor $\gamma$. Since
\begin{equation}
s = \left(2 E_0 \right)^2\, ,
\label{Fermi-fx7}
\end{equation}
we have:
\begin{equation}
\gamma = \frac{\sqrt{s}}{2 m_p c^2} = \frac{\sqrt{2m_pc^2\left(E_1 + m_pc^2\right)}}{2 m_p c^2} = \sqrt{\frac{E_1 + m_pc^2}{2 m_p c^2}}\, .
\label{Fermi-fx8}
\end{equation}

An alternative derivation of Eq.(\ref{Fermi-fx8}) starts from the fact that $\gamma$ is also the Lorentz factor of the motion of the c.m. in the laboratory frame, since, we recall, the target proton is at rest in the laboratory frame. The speed $v$ of the c.m. in the laboratory frame is given by (see e.g. Eq.(11.4) in \citet{1975ctf..book.....L}):
\begin{equation}
v = \frac{\left|\vec{P}\right|c^2}{E} = \frac{\left|\vec{p_1}\right|c^2}{E_1 + m_pc^2}\, ,
\label{Fermi-fx8-1}
\end{equation}
where we used Eqs.(\ref{Fermi-fx1})-(\ref{Fermi-fx2}) together with the fact that for two protons $m_1 = m_2 = m_p$. From Eq.(\ref{Fermi-fx8-1}), by definition of $\gamma$, we have:
\begin{equation}
\gamma = \sqrt{\frac{1}{1-(v/c)^2}} = \sqrt{\frac{\left(E_1 + m_pc^2\right)^2}{\left(E_1 + m_pc^2\right)^2 - p_1^2c^2}} = \sqrt{\frac{\left(E_1 + m_pc^2\right)^2}{E_1^2 + m_p^2c^4 + 2E_1m_pc^2 - p_1^2c^2}} = \sqrt{\frac{E_1 + m_pc^2}{2 m_p c^2}}\, ,
\label{Fermi-fx8-2}
\end{equation}
where we used the fact that $E_1^2 = p_1^2c^2 + m_p^2c^4$. As we expected, Eq.(\ref{Fermi-fx8-2}) is identical to Eq.(\ref{Fermi-fx8}).

Substituting Eq. (\ref{Fermi-fx8}) in Eq. (\ref{Fermi-f1}), we get the expression of Fermi for the Lorentz contraction of the volume $\Omega_0$ in the c.m. frame with respect to $\Omega$:
\begin{equation}
\Omega_0 = \sqrt{\frac{2m_pc^2}{E_1 + m_pc^2}}\Omega\, ,
\label{Fermi-f15}
\end{equation}
Note that if the energy $E_1$ increase, the volume $\Omega_0$ decreases, as is predicted in special relativity. The parameter volume will be therefore energy dependent. Fermi calculated also the cross-section as the area available for collisions around the pion cloud
\begin{equation}
\sigma_{tot}=\pi R^2.\label{Fermi-f16}
\end{equation}
Substituting $\lambda_\pi^c$ of Eq. (\ref{Fermi-f3}) in Eq. (\ref{Fermi-f16}), we get $\sigma_{tot}=6\times 10^{-26}\,\rm{cm}^2$, where $\hbar = 1.054\times 10^{-27}\,$erg.s and $c=2.9979\times 10^{10}\,$cm/s a value close to the modern experimental value.

In treating the collisions of extremely high energy nucleons Fermi make use of thermodynamic laws, instead of considering a detailed statistical treatment. The energy density around the colliding nucleons is so high that multiple pions as well as antiprotons will be produced.

From Planck law, the spectral intensity (dimension $I_\nu (\nu,T)\rightarrow dE/dt\,dA\,d\Omega\,d\nu$) of the black body is given as
\begin{equation}
I_\nu (\nu,T) = \frac{2h\nu^3}{c^2}\frac{1}{e^{\scriptscriptstyle h\nu/kT}-1},\label{Fermi-f26a}
\end{equation}
where $k$ is the Boltzmann constant, $\nu$ the frequency and $T$ the temperature.

From Stefan-Boltzmann law of the black-body (radiation flux $R(T)=\sigma T^4$, dimension $dE/dt\,dA$), the energy density $\rho (T)$ (dimension $dE/dV$) is
\begin{gather}
I(T) = \int_0^\infty I_\nu (\nu,T)d\nu = \frac{c}{4\pi}\rho (T) = \frac{1}{\pi}R(T)\Rightarrow\notag\\
\rho(T) = \frac{4\pi}{c}\int_0^\infty I_\nu (\nu,T)d\nu = \frac{4}{c}\sigma T^4,\label{Fermi-f27a}\\
\rho (T) = \left(\frac{\pi^2}{15c^3\hbar^3}\right)(kT)^4 =
\left(\frac{6.494}{\pi^2c^3\hbar^3}\right)(kT)^4,\label{Fermi-f27}
\end{gather}
where $\pi^4/15 = 6 \overset{\scriptscriptstyle\infty}{\underset{{\scriptscriptstyle
n=1}}{\Sigma}} 1/n^4 = 6.494$ (from Gamma and Riemann Zeta
functions, respectively, $\Gamma(z)$ and $\zeta(s)$), $\sigma$ is
the Stefan-Boltzmann constant and $I(T)$ the spectral intensity
integer in all frequency of the black-body. According to Fermi:
``Consequently the Stefan's law for the pions will be quite similar
to the ordinary Stefan's law of the black-body radiation. The
difference is only in a statistical weight factor. For the photons
the statistical wight is the factor, $2$, because of the two
polarization directions. If we assume that the pions have spin zero
and differ only by their charge $\pm e$ or $0$, their statistical
weight will be $3$. Consequently, the energy density of the pions
will be obtained by multiplying the energy density of the ordinary
Stefan's law by the factor $3/2$.'' Then, multiplying
the energy density (\ref{Fermi-f27}) by $3/2$, the energy density via
pions is
\begin{equation}
\rho_\pi (T) = \frac{3}{2}\rho (T) = \frac{3\times
6.494}{2\pi^2\hbar^3c^3}(kT)^4.\label{Fermi-f28}
\end{equation}

The total energy of the system (Eq. \ref{Fermi-fx5}) is divided among pions,
protons and anti-protons. Then it is necessary to get the energy density
via protons and anti-protons. In this case is used Planck law modified
(for fermions) as
\begin{equation}
I_{\scriptscriptstyle fermion} (E,T) = \frac{2}{h^2 c^2}\frac{E^3}{e^{\scriptscriptstyle E/kT} + 1},
\label{Fermi-f28a}
\end{equation}
and
\begin{equation}
\rho_{\scriptscriptstyle fermion}(T) = \frac{4\pi}{c}\int_0^\infty I_{\scriptscriptstyle fermion}
(E,T)\frac{dE}{h}.\label{Fermi-f28b}
\end{equation}
It is necessary to use the Planck law modified because the protons and
anti-protons are fermions, where they obey Fermi-Dirac statistical.
According to Fermi: ``The contribution of the nucleons and anti-nucleons
to the energy density is given by a similar formula. The differences are
that the statistical weight of the nucleons is eight since we have four
different types of nucleons and anti-nucleons and for each, two spin
orientations. A further difference is due to the fact that these
particles obey the Pauli principle.'' Then it is necessary to multiply
the Stefan-Boltzmann law by fermions to $8/2$. The process is
$pp\,\rightarrow\pi + X$, where ``$X$'' represent the protons and
anti-protons, then the energy density via protons and anti-protons is
\begin{equation}
\rho_{\scriptscriptstyle X}(T) = \frac{8}{2}\rho_{\scriptscriptstyle fermion}(T) = \frac{4\times
5.682}{\pi^2\hbar^3 c^3}(kT)^4\label{Fermi-f29}
\end{equation}
where $6 \overset{\scriptscriptstyle\infty}{\underset{{\scriptscriptstyle n=1}}{\Sigma}} (-1)^{n+1}/n^4 = 5.682$.

The total energy density $\rho_{tot}$ of the system during the collision
is given by sum $\rho_{tot} = \rho_\pi +\rho_{\scriptscriptstyle X}$, but also it is
the energy of c.m. divided per volume, $\rho_{tot} = \sqrt{s}/\Omega_0$.
Then
\begin{equation}
\rho_{tot} = \frac{\sqrt{s}}{\Omega_0} = \rho_\pi + \rho_{\scriptscriptstyle X}.\label{Fermi-f30}
\end{equation}

Substituting Eqs. (\ref{Fermi-f28}) and (\ref{Fermi-f29}) in Eq. (\ref{Fermi-f30})
\begin{gather}
\rho_{tot} = \frac{\sqrt{s}}{\frac{2m_pc^2}{\sqrt{s}}\Omega} =
\frac{3\times 6.494}{2\pi^2\hbar^3c^3}(kT)^4 +\frac{4\times 5.682}
{\pi^2\hbar^3 c^3}(kT)^4,\notag\\
(kT)^4 = 0.152\frac{\hbar^3c^3s}{m_pc^2\Omega}.\label{Fermi-f31}
\end{gather}
Note that the energy density is frame invariant,
$\rho_{tot} = E_{0tot}/\Omega_0 = E_{tot}/\Omega$, because cancel the
Lorentz factors.

Analogous to the Eqs. (\ref{Fermi-f26a}) and (\ref{Fermi-f27a}) it has the definition of numerical density of the black-body radiation is \citep[see][]{1980stph.book.....L}
\begin{gather}
n(T) = \frac{4\pi}{c}\int_0^\infty \frac{I_\nu (\nu,T)}{h\nu}d\nu =
\frac{8\pi}{c^3}\int_0^\infty \frac{\nu^2 d\nu}{e^{h\nu/kT}-1}\Rightarrow\notag\\
n(T) = \frac{1}{\pi^2}\left(\frac{kT}{c\hbar}\right)^3\int_0^\infty
\frac{x^2dx}{e^x-1} = \frac{\Gamma(3)\zeta(3)}{\pi^2}\left(\frac{kT}
{c\hbar}\right)^3,\notag\\
n(T) = 0.243576\frac{(kT)^3}{\hbar^3 c^3},\label{Fermi-fp6}
\end{gather}
where $x=h\nu/kT$. In the case of pions,
(${\scriptscriptstyle\frac{\#\,\rm{pions}}{\rm{volume}}}$), it is necessary
multiply the last equation by $3/2$ (analogous Eq. \ref{Fermi-f28}), as
\begin{equation}
n_\pi (T) = \frac{3}{2}n(T) =
0.365\frac{(kT)^3}{\hbar^3c^3},\label{Fermi-fp6b}
\end{equation}
where \citet{1950PThPh...5..570F} got also the last expression.

Substituting  Eq. (\ref{Fermi-f31}) in Eq. (\ref{Fermi-fp6}), we
get
\begin{equation}
n_\pi^{\scriptscriptstyle HE}(\sqrt{s}) = 0.0888\left(\frac{s}{\hbar c\, m_pc^2\Omega}\right)^{3/4}\, .
\label{Fermi-fp7}
\end{equation}

We can define the multiplicity of pions per collision ($N_\pi^{\scriptscriptstyle
HE}\rightarrow \#\,\rm{pions}$), for high energy, being (using Eq.
\ref{Fermi-f15})
\begin{equation}
n_\pi^{\scriptscriptstyle HE} = \frac{N_\pi^{\scriptscriptstyle HE}}{\Omega_0}\Rightarrow
N_\pi^{\scriptscriptstyle HE} = \frac{2m_pc^2}{\sqrt{s}}\Omega n_\pi^{\scriptscriptstyle HE}.\label{Fermi-fp9}
\end{equation}
Substituting Ep. (\ref{Fermi-fp7}) in Eq. (\ref{Fermi-fp9})
\begin{equation}
N_\pi^{\scriptscriptstyle HE}(\sqrt{s}) = 0.1777\left(\frac{m_pc^2\Omega^{\scriptscriptstyle
LF}s}{\hbar^3c^3}\right)^{1/4}.\label{Fermi-fp10}
\end{equation}
According Eqs. (\ref{Fermi-f2}) and (\ref{Fermi-f3}),
\begin{equation}
\Omega = \frac{4\pi a^3\hbar^3c^3}{3(m_\pi
c^2)^3}.\label{Fermi-fp11}
\end{equation}
Substituting Eq. (\ref{Fermi-fp11}) in Eq. (\ref{Fermi-fp10})
\begin{gather}
N_\pi^{\scriptscriptstyle HE}(\sqrt{s}) = 0.25422\left[\frac{m_pc^2a^3s}{(m_\pi
c^2)^3}\right]^{1/4}.\label{Fermi-fp12}
\end{gather}

The pion rest masses are different, $m_{\pi^0}c^2=0.135\,\rm{GeV}$ and $m_{\pi^\pm}c^2=0.139\,\rm{GeV}$. Fermi got that when considerer the conservation of angular momentum, it has the effect of reduction the numbers of pions and nucleons, then he found a factor obtained numerically of $0.51$. The total energy via pions is divided approximately equal among $\pi^0$, $\pi^-$ and $\pi^+$, then multiplying and dividing Eq. (\ref{Fermi-fp12}) per, respectively, $0.51$ and $3$, it gets the $\pi^0$ and $\pi^\pm$ multiplicities
\begin{gather}
N_{\pi^0}^{\scriptscriptstyle HE}(\sqrt{s}) = 0.0432\left[\frac{m_pc^2 a^3s}
{(m_{\pi^0} c^2)^3}\right]^{1/4},\label{Fermi-fp12a}\\
\end{gather}

Doing $m_{\pi^0}c^2=0.14385m_pc^2$ and
$m_{\pi^\pm}c^2=0.14875m_pc^2$ in the lest two equations,
\begin{gather}
N_{\pi^0}^{\scriptscriptstyle HE}(\sqrt{s}) = 0.185a^{3/4}\sqrt{\frac{\sqrt{s}}{m_pc^2}},\label{Fermi-fp12c}\\
N_{\pi^\pm}^{\scriptscriptstyle HE}(\sqrt{s}) = 0.180a^{3/4}\sqrt{\frac{\sqrt{s}}{m_pc^2}},\label{Fermi-fp12d}\\
N_{\pi\,total}^{\scriptscriptstyle HE}(\sqrt{s}) = 0.546a^{3/4}\sqrt{\frac{\sqrt{s}}
{m_pc^2}},\label{Fermi-fp12e}
\end{gather}
where $N_{\pi\,total}^{\scriptscriptstyle HE} = 2N_{\pi^\pm}^{\scriptscriptstyle
HE}+N_{\pi^0}^{\scriptscriptstyle HE}$, can note that the lest value is the same
in Fermi \citep{1950PThPh...5..570F}.

The center of mass energy is according Eqs.(\ref{Fermi-fx5})-(\ref{Fermi-fx7}). Substituting
in Eq. (\ref{Fermi-fp12}), we get the equation of Fermi to the
$\pi^0$ multiplicity for extreme high energies
\begin{equation}
N_{\pi^0}^{\scriptscriptstyle HE}(E_p) = 0.777a^{3/4}\left(
1+\frac{E_p}{m_pc^2}\right)^{1/4}.\label{Fermi-fp13}
\end{equation}

Doing the same procedure from Eq. (\ref{Fermi-fp9}) to (\ref{Fermi-fp12c}),
\begin{equation}
N_{\pi^0}^{\scriptscriptstyle ME}(\sqrt{s}) =
0.6533a^{3/4}\frac{(\sqrt{s}/m_pc^2-2)^{3/2}}{\sqrt{s}/m_pc^2}.\label{Fermi-fp14}
\end{equation}
Substituting Eq. (\ref{Fermi-fx5}) in Eq. (\ref{Fermi-fp14}), we get the equation of
Fermi to the multiplicity of pions in intermediate energy range,
\begin{equation}
N_{\pi^0}^{\scriptscriptstyle ME}(E_p) = 0.777a^{3/4}\frac{\left(
\sqrt{1+E_p/m_pc^2}-\sqrt{2}\right)^{3/2}}
{\sqrt{1+E_p/m_pc^2}}.\label{Fermi-fp15}
\end{equation}

\subsection{Modern approach}

\begin{figure}
\centering
\includegraphics[width=0.6\hsize]{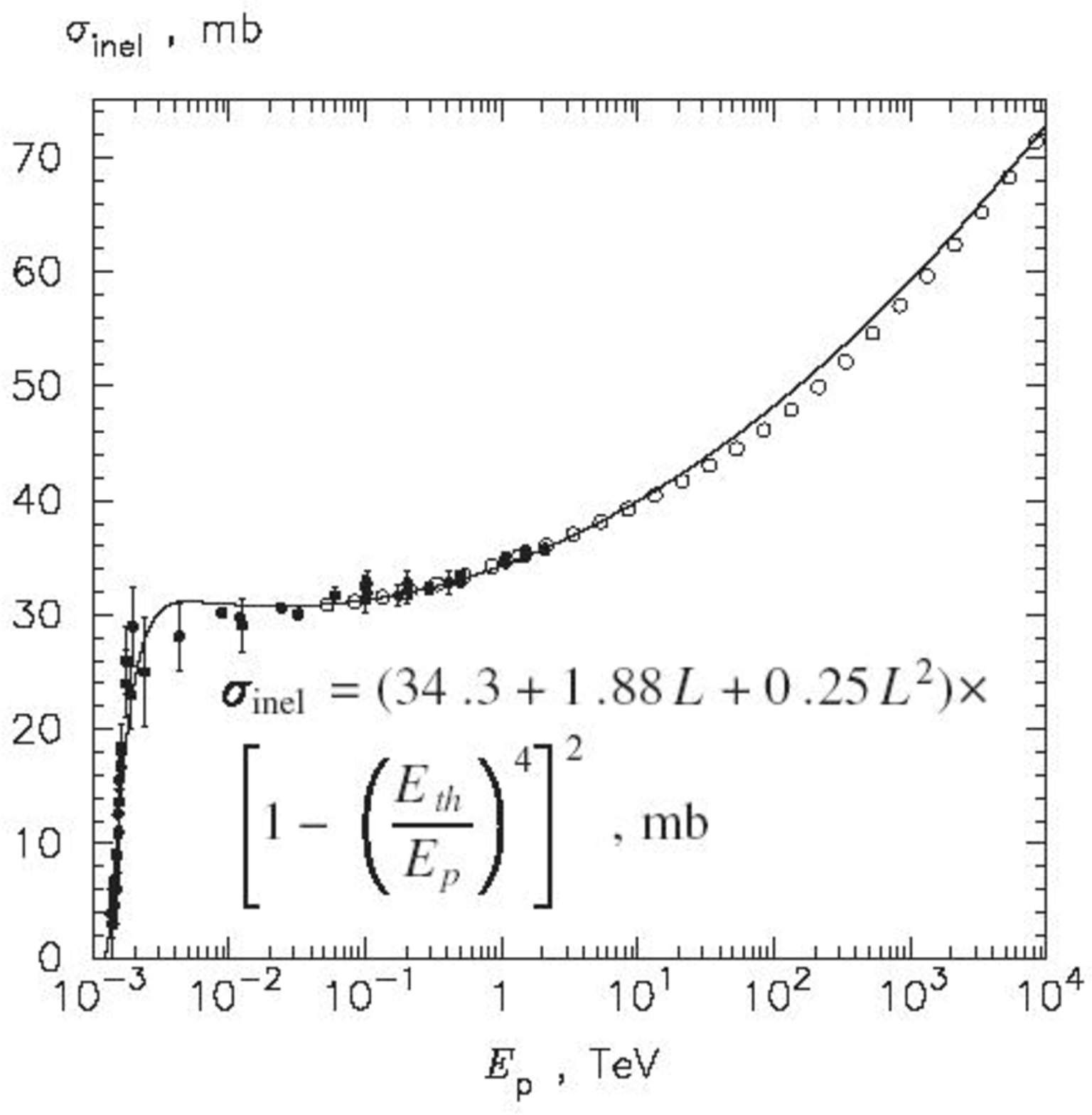}
\caption{The graphic compares the experimental data (black points)
with the SIBYLL code (open points) and with the Kelner's expression
\citep{2006PhRvD..74c4018K} for the cross section (solid curve). Reproduced from \citet{2006PhRvD..74c4018K}.}
\label{Fermi-fig1}
\end{figure}

\begin{figure}
\centering
\includegraphics[width=0.6\hsize]{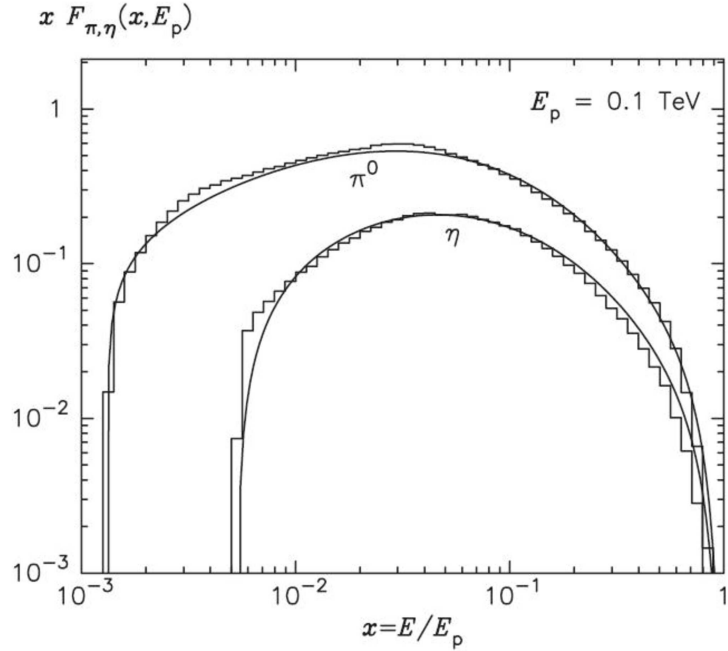}
\caption{Energy spectra of $\pi$ and $\eta$ mesons from the
numerical simulations of the SIBYLL code (histograms) and from the
analytical presentations given by Eqs. (\ref{Fermi-k2}) and (\ref{Fermi-k3}) for
energy 0.1 TeV. Reproduced from \citet{2006PhRvD..74c4018K}.}
\label{Fermi-fig2}
\end{figure}

Currently the modeling of pp interactions is done through computational codes, Monte Carlo codes
such as SIBYLL, PHYTHIA, Dpmjet. Kelner \citep{2006PhRvD..74c4018K} presented new parameterizations of energy
spectra of secondary particles, $\pi$ and $\eta$ mesons, gamma rays,
electrons, and neutrinos produced in inelastic proton-proton collisions
based on the SIBYLL code by Lipari \citep{1994PhRvD..50.5710F}. These
parameterizations have very good accuracy in the energy range above
$100$ GeV (see figure \ref{Fermi-fig1} and \ref{Fermi-fig2}).

The fit to the pp cross section obtained by \citet{2006PhRvD..74c4018K} is
\begin{equation}
\sigma_{pp} (E_p) = (34.3+1.88L+0.25L^2)\left[1-\left(\frac{E_{th}}
{E_p}\right)^4\right]^2\,\rm{mb},\label{Fermi-k1}
\end{equation}
where $E_p$ is the incident proton energy in laboratory frame (the same that $E_p$ in the last section), $L=\ln[E_p(TeV)]$ and $E_{th}$ is the minimum threshold energy of the incident proton for production of a pion ($E_{th}=1.22\,\rm{GeV}$) and $1\,\rm{mb}=1\, \rm{mbarn}=10^{-27}\,\rm{cm^2}$.

The function that Kelner got for the multiplicity of pions ``$N_\pi$'' is
given as
\begin{gather}
dN_\pi \equiv F_\pi (x,E_p)dx,\label{Fermi-k4}\\
F_\pi (x,E_p) = \frac{d}{dx}\phi(x,E_p),\label{Fermi-k2}\\
\phi_{\scriptscriptstyle SIBYLL} = -B_\pi\left(\frac{1-x^{\beta_{\scriptscriptstyle\gamma}}}{1+k_\gamma
x^{\beta_{\scriptscriptstyle\gamma}}(1-x^{\beta_{\scriptscriptstyle\gamma}})}\right)^4,\label{Fermi-k3}
\end{gather}
where $E_\pi$ is the total energy via neutral pions productions
(secondary pions mesons), $x=E_\pi /E_p$, and
$B_\pi$, $\beta^\gamma$ and $k^\gamma$ are functions dependent only
of $E_p$ in the Kelner parametrization \citep[][see graphic
\ref{Fermi-fig2}]{2006PhRvD..74c4018K}. Figure \ref{Fermi-fig2} shows the spectrum distribution
$xF_\pi(x,E_p)$ of $\pi^0$ production as a function of $x=E_\pi /E_p$,
which is the percentage of incident proton energy
transferred to the pions. Note in Fig. \ref{Fermi-fig2} shoes the small
probabilities of the $\pi^0$ productions at small energy transformation
$x\leq 0.0015$ and large energy transformation $1\geq x\geq 0.7$ $E_p$.
The maximum probability is at
$x=0.03$, indicating the approximate $3\%$ of incident proton energy is
transferred via neutral pion $\pi^0$. Since the experimental data
show that the production of neutral pions $\pi^0$ is practically the
same as the productions of positive $\pi^+$ and negative $\pi^-$
charged pions, then we get that approximate $10\%$ of incident proton
energy $E_p$ is transferred via pions ($E_\pi$).

Integrating Eq. (\ref{Fermi-k4}) over $x$, Kelner obtained \citep{2006PhRvD..74c4018K} the multiplicity
(number) of $\pi^0$ per pp collision as a function of $E_p$
\begin{equation}
N^{\scriptscriptstyle KS}_\pi (E_p) = 3.92+0.83L+0.075L^2.\label{Fermi-k4a}
\end{equation}

\subsection{Comparison between Fermi's and Kelner-SIBYLL's approaches}

We compared the multiplicities of neutral pions production ($\pi^0$)
obtained the Fermi theoretical approach (Eqs. \ref{Fermi-fp13} and \ref{Fermi-fp15})
and the Kelner analytical parametrization (Eq. \ref{Fermi-k4a}) to the SIBYLL
code \citep{1994PhRvD..50.5710F}. The analytical parametrization of Kelner has good
agreement with the SIBYLL code for energy $E_p>100GeV$.

\begin{figure}
\includegraphics[angle=270,scale=0.31]{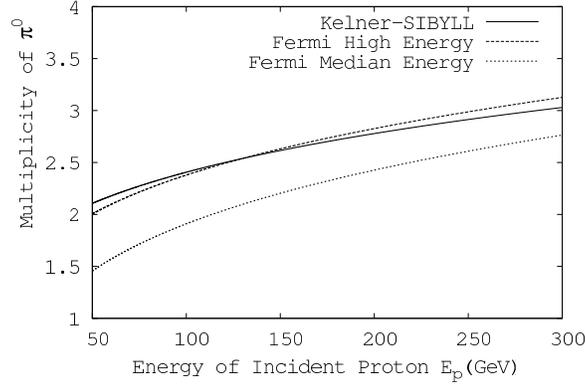}
\caption{The multiplicity of neutral pions $\pi^0$ in function of
incident proton energy $E_p$ for intermediate energy range
$50\leq E_p\leq 300$ GeV. We compared three results: Kelner-SIBYLL,
Fermi approaches in median energy and high energy for $a=5$.}\label{Fermi-fig4}
\end{figure}

In the figure \ref{Fermi-fig4} we compare the Fermi results in high-energy
region (Eq. \ref{Fermi-fp13}) and median-energy region (Eq. \ref{Fermi-fp15}) with
the description of Kelner-SIBYLL in intermediate energy range $50-300$
GeV. We used parameter $a=5$, which give a very good agreement between
Fermi result in high energy and Kelner-SIBYLL result in the range
$100-300$ GeV. But Fermi result in median energy Eq.
(\ref{Fermi-fp15}) gives lower multiplicity of $\pi^0$ in this energy range.

\begin{figure}
\includegraphics[angle=270,scale=0.31]{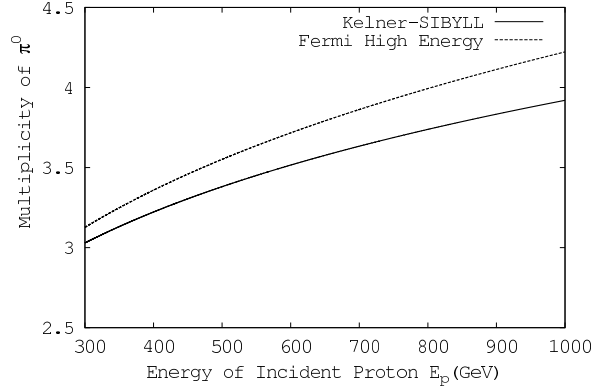}
\caption{The multiplicity of neutral pions $\pi^0$ in function of
incident proton energy $E_p$ for high energy range
$300\leq E_p\leq 1000$ GeV. We
compared two results obtained respectively by Kelner-SIBYLL
and Fermi Eq. (\ref{Fermi-fp13}) for $a=5$.}
\label{Fermi-fig5}
\end{figure}

In the figure \ref{Fermi-fig5}, we use $a=5$ and compare the Fermi result
for high energy (\ref{Fermi-fp13}) with Kelner-SIBYLL result in the energy
range $300-1000$ GeV. We find that Fermi result is about $8\%$
larger than Kelner-SIBYLL result, this indicates that the validity of the
Fermi approach in extremal high-energy range is in question.
Because the cross-section
grows with $E_p$ (see Figure 1), and the $a$ parameter by its definition
should become larger with larger as the cross-section
grows in this energy range. This makes that Fermi result further deviate from
Kelner-SIBYLL result. One of reasons
could probably be the fact that more particles, e.g.,gluons and
quarks, are exited and participate the thermalization in the Fermi
volume $\Omega$, as a result, the energy transferred to $\pi$-productions
is smaller than that estimated by Fermi with three particles
proton, neutron and pion.

\subsection{Maximum and minimum energy of the Pions}

In this section we will show the energy limits of the pions created via
pp interactions and we will apply the limits in the work of \citet{2000PhRvD..62i4030B}.

The energy of a pion in the laboratory frame (LF) in function of its
energy in center of mass (c.m.) frame, is as
\begin{equation}
 E_\pi = \gamma(E_{0\pi}+vp_{0\pi}cos{\theta}),\label{Fermi-ap1}
\end{equation}
where $\gamma$ is the Lorentz factor, ``v'' the velocity of the pion,
$E_{0\pi}$ the pion energy in the c.m. frame and
$p_{0\pi}$ the pion momentum in the c.m. frame (the index ``0'' inform
c.m. frame and without the index ``0'' give in the LF). We will
consider $c=1$. If $cos{\theta}=1$ the pion energy is maximum ($E_\pi^{max}$),
and if $cos{\theta}=-1$ the pion energy is minimum ($E_\pi^{min}$).
Then the maximum and minimum energy of the pion is,
\begin{equation}
 E_\pi^{\tfrac{max}{min}} = \gamma(E_{0\pi}\pm vp_{0\pi}).\label{Fermi-ap2}
\end{equation}

Developing the calculus we obtain the maximum and minimum energy of the
pions,
\begin{align}
&E_\pi^{max} = \frac{1}{4m_p}(2m_pE_p-2m_p^2+m_\pi^2+R_p),\label{Fermi-ap3}\\
&E_\pi^{min} =
\begin{cases}
\frac{1}{4m_p}({\scriptscriptstyle 2m_pE_p-2m_p^2+m_\pi^2-R_p}), &\rm{if}\quad E_p>E_p^*\\
m_\pi, &\rm{if}\quad E_p \leq E_p^*,
\end{cases}\label{Fermi-ap4}
\end{align}
where
\begin{gather}
 R_p = \sqrt{\scriptscriptstyle\frac{(E_p-m_p)[(2m_pE_p-2m_p^2-m_\pi^2)^2-16m_p^2m_\pi^2]}
{E_p+m_p}}.\label{Fermi-ap5}\\
E_p^* = \frac{2m_p^2+2m_pm_\pi -m_\pi^2}{2(m_p-m_\pi)}\simeq
1.242\qquad\rm{GeV}.\label{Fermi-ap6}
\end{gather}
where $E_p^*$ is the limit (threshold) of energy in the case that
$E_\pi = m_\pi$.

\begin{figure}
\centering
\parbox{8cm}{
\ifpdf
\includegraphics[width=5.5cm,height=6.2cm]{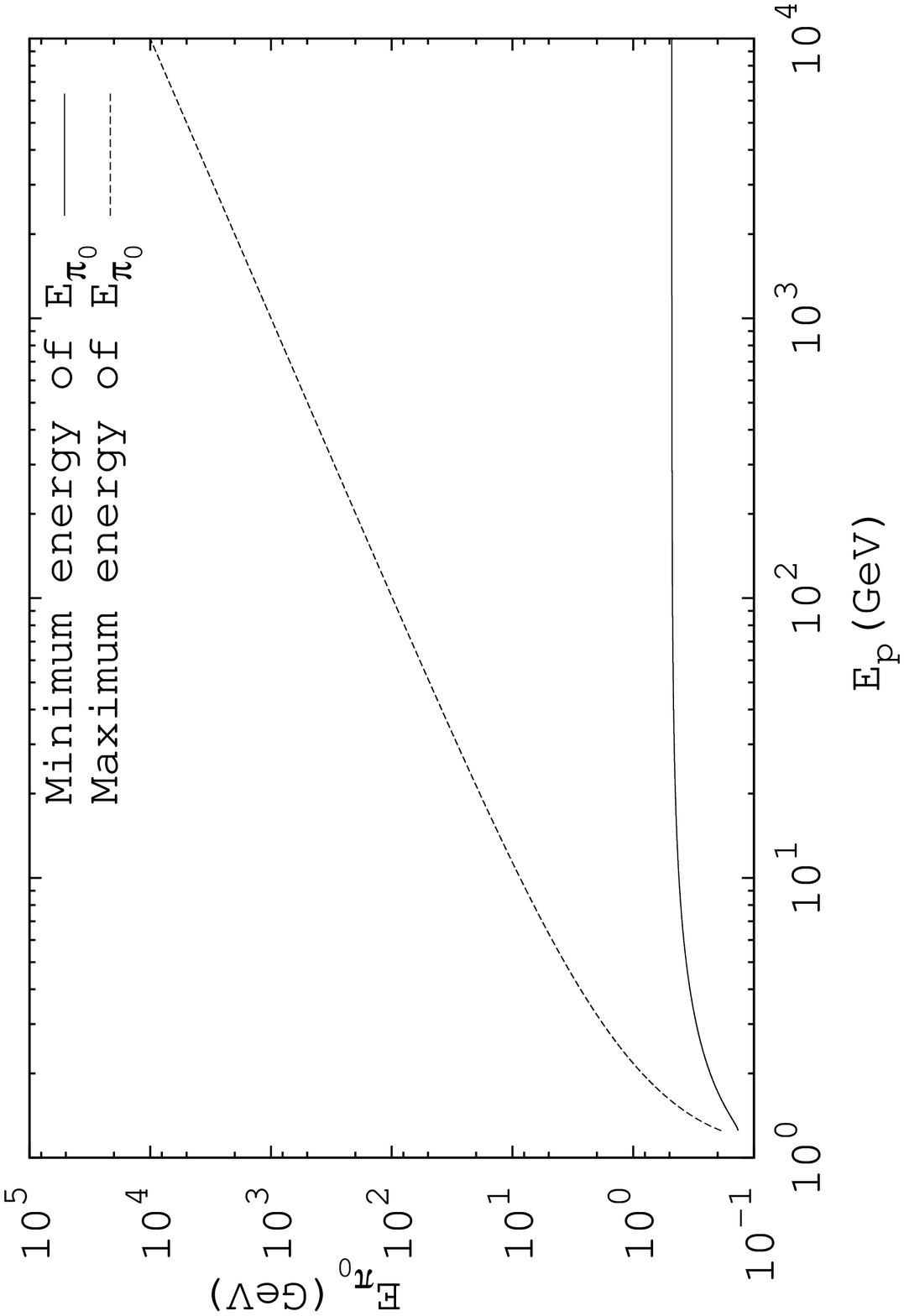}
\else
\includegraphics[width=5.5cm,height=6.2cm,angle=-90]{Fermi/th1}
\fi}
\hspace{-1.8cm}
\parbox{8cm}{
\ifpdf
\includegraphics[width=5.8cm,height=6.2cm]{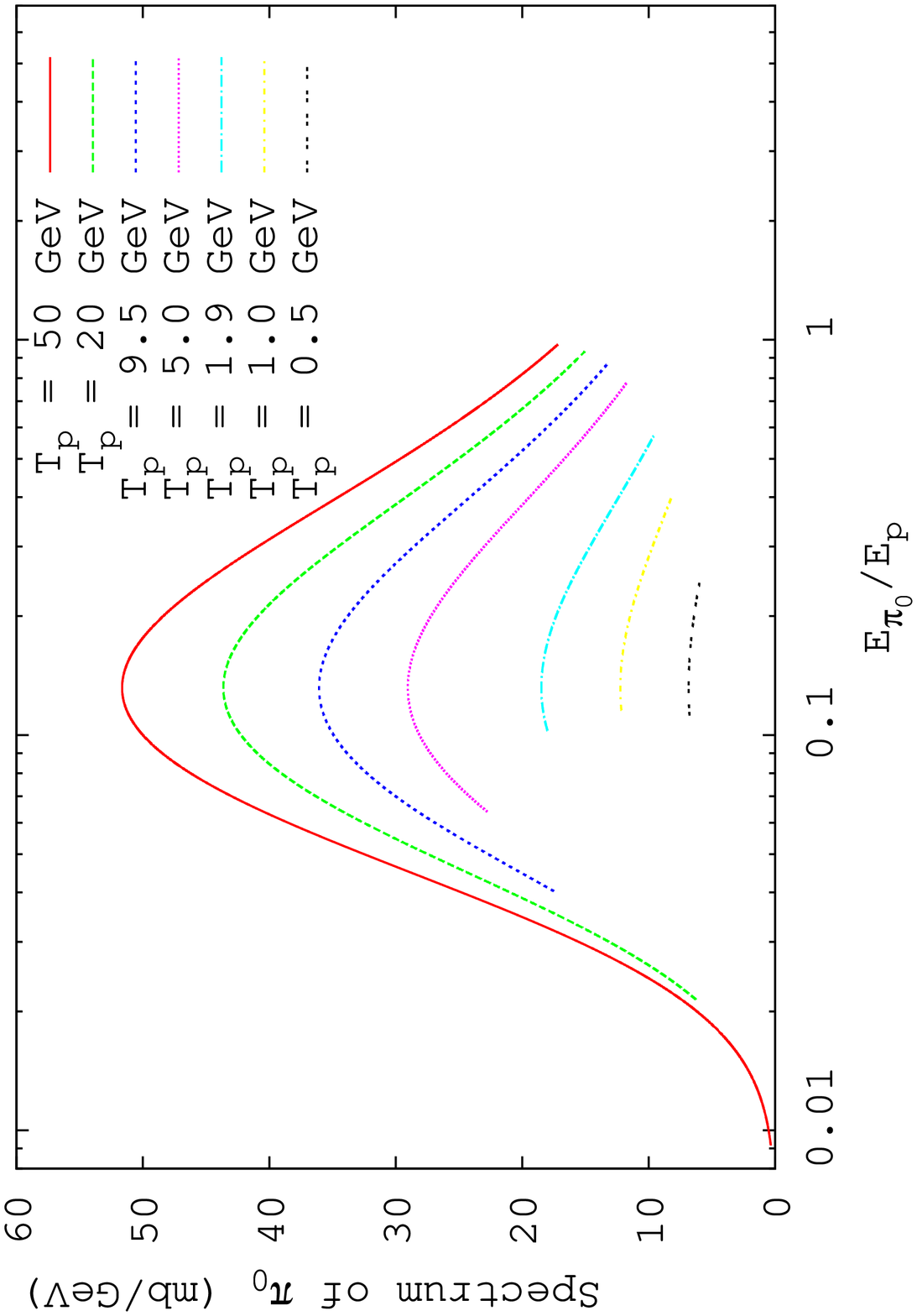}
\else
\includegraphics[width=5.8cm,height=6.2cm,angle=-90]{Fermi/Blat_mod}
\fi}
\caption{a)The confront $E_{\pi^0}^{min}\& E_{\pi^0}^{max}$ vs $E_p$.
b) Application of the minimum (\ref{Fermi-ap4}) and maximum (\ref{Fermi-ap3})
thresholds of the $\pi^0$ created in the Blattnig parametrization (\ref{Fermi-a31}).}
\label{Fermi-fig1a}
\end{figure}

In Fig. \ref{Fermi-fig1a}a, we can see the behavior of the Eqs. (\ref{Fermi-ap3}) and
(\ref{Fermi-ap4}), where they are the maximum and minimum limits of the neutral pion
energy created. We can note that the minimum limit is approximated of
$E_{\pi^0}^{min}=0.478$ GeV for energy of incident proton $E_p=10^4$ GeV,
where $E_{\pi^0}^{min}$ increase very softly. It is possible note also that the
maximum limit tends to $E_{\pi^0}^{max}=E_p-3m_p/2$.

 \citet{2000PhRvD..62i4030B} obtains the spectral distribution of $\pi^0$
[$d\sigma/dE_p(\rm{mb/GeV})$] in function of the kinetic energy of $\pi^0$
created ($T_{\pi^0}$), for energy of incident proton $E_p\le 50$ GeV. Blattnig
obtain the spectral distribution and total cross section of $\pi^0$ for seven
different kinetic energies of incident protons, $T_p =
0.5,1.0,1.9,5.0,9.5,20,50$ GeV. The analytical function that Blattnig got is
\begin{equation}
 \frac{d\sigma}{dE_p} = \exp\left(K_1+\frac{K_2}{T_p^{0.4}}+\frac{K_3}
{T_\pi^{0.2}}+\frac{K_4}{T_\pi^{0.4}}\right),\label{Fermi-a31}
\end{equation}
where $K_1=-5.8$, $K_2=-1.82$, $K_3=13.5$, and $K_4=-4.5$. The last expression
is the analytical fit of a numerical integration  \citep{2000PhRvD..62i4030B}. In Fig. 
\ref{Fermi-fig1a}b we apply the maximum (\ref{Fermi-ap3}) and minimum
(\ref{Fermi-ap4}) limit energies of the $\pi^0$ in the spectrum distribution of
 \citet{2000PhRvD..62i4030B} (Fig. \ref{Fermi-fig1a}b), and we obtained results that
contradict their limits of energy, where it shows problems in minimum and
maximum energies.

Similar conclusions are being explored using the PYTHIA code \citep{1987CoPhC..46...43B}, and results will be presented soon \citep{HA}.

	\section{Conclusions}

Our current understanding of GRBs is based on a general picture which was presented in a set of letters \citep{2001ApJ...555L.107R,2001ApJ...555L.113R,2001ApJ...555L.117R}. On that basis a canonical GRB scenario has emerged, with three distinct phases:
\begin{enumerate}
\item The vacuum polarization process occurring in the gravitational collapse to a black hole, and the consequent creation of an electron-positron plasma.
\item An optically thick fireshell characterized by the self-acceleration of such an optically thick electron-positron plasma, with the engulfed baryon loading. This phase ends with the reaching of transparency, when the P-GRB is emitted.
\item An optically thin fireshell characterized by an accelerated beam of protons and electrons with a Lorentz $\gamma$ factor roughly inversely proportional to the baryon loading, interacting with the CBM.
\end{enumerate}

This basic scenario is currently evolving in a large number of theoretical details, ranging from a) the thermalization process of electrons and positrons after their production, to b) the dynamics of the electron-positron pairs in the optically thick phase and their instabilities, to c) the probing and determination of the CBM distribution around the gravitationally collapsed object by the interaction of the ultrarelativistic baryons and electrons colliding with the CBM.

Many of the properties of the observed extended afterglow X- and $\gamma$-ray emission below $\sim 1$ MeV has been obtained by postulating a thermal spectrum of the emission process in the co-moving frame. This treatment is particularly appealing since it allowed to derive explicit analytic formulas to compute all the relativistic transformations between the co-moving, the laboratory frame and the arrival time of the signal. This approach has been very satisfactory in explaining the overall bolometric luminosity of the sources, its evolution as a function of the arrival time, and the CBM filamentary structure. With the improvement of the observational techniques, more and more time resolved spectra have been observed and our approach shows some discrepancy in the low and high energy tails of the time resolved spectra, although the time integrated spectral distribution is very well recovered. We are currently evolving this basic mechanism by assuming some departures from a pure thermal spectral shape in the co-moving frame.

This analysis has led to a clear identification of sources occurring in CBM with average density $n_{cbm} \sim 1$ particle/cm$^3$ and with $n_{cbm} \sim 10^{-3}$ particles/cm$^3$. The first ones correspond to sources occurring in star forming regions in the host galaxy, and the second ones to sources occurring in the galactic halo. This problematic has led to a new understanding of the traditional separation between long and short GRBs, as exemplified in these lectures.

We are now approaching, in view of the new data from the Fermi and AGILE satellites, an analysis of the GRB radiation over $1$ MeV. It is by now clear that the emission process previously considered of a purely thermal spectrum is not appropriate to the description of this high-energy component.

In parallel, we are currently examining how Fermi ideas \citep{1950PThPh...5..570F} have been further developed in large data analysis procedures at CERN and other accelerators all over the world (see the last section of this paper). We are going to probe, by fitting the observational data, an higher density nodule component in the CBM, which will add to the understanding of the CBM itself and will lead very likely to the explanation of GRB sources.

The major effort now is also directed to the understanding of the process of the electrodynamics of gravitational collapse.

\end{document}